\documentclass[a4paper,12pt]{report}
\usepackage[latin1]{inputenc}
\usepackage[brazil]{babel}
\usepackage{epsfig}
\usepackage{float}
\usepackage{latexsym}
\usepackage{graphicx}
\usepackage{epsfig}
\usepackage{amsmath}
\usepackage{amsfonts}
\usepackage{amssymb}

\begin{document}
\thispagestyle{empty}
\begin{center}
{\large{Universidade de São Paulo\\
Instituto de Física}}

\vspace{2cm}
{\bf \Large{Análise de Buracos Negros Esfericamente
Simétricos em modelos de Mundo Brana}}\\
\vspace{2cm}
{\bf Alan Bendasoli Pavan\\

Orientador: Prof. Dr. Élcio Abdalla}
\end{center}
\vspace{1cm}
\hfill\
\begin{flushright}
\begin{minipage}[t]{8cm}
\hrulefill

Dissertação de mestrado apresentada ao Instituto de Física da Universidade de São
Paulo para a obtenção do título de Mestre em Ciências

\hrulefill
\end{minipage}
\end{flushright}
\vspace{3cm}
{\bf Comissão Examinadora:}\\
Prof. Dr. Élcio Abdalla (IF-USP)\\
Prof. Dr. Luís Raul Weber Abramo (IF-USP)\\ 
Prof. Dr. George Emanuel Avraam Matsas (IFT-UNESP)
\vspace{2cm}
\begin{center}
São Paulo\\
2006
\end{center}
\newpage
\thispagestyle{empty}
\vspace*{2cm}
\begin{center}
{\bf
\Large{FICHA CATALOGRÁFICA} \\
\large{Preparada pelo Serviço de Biblioteca e Informação}\\
\large{do Instituto de Física da Universidade de São Paulo}\\}
\vspace{2mm}
\begin{tabular}{|p{135mm}|}
\hline
\\ 
\hspace{13mm} \large Pavan, Alan Bendasoli\\ 
\\
\hspace{13mm} \large Análise de buracos negros esfericamente simétricos em\\ 
\hspace{13mm} \large modelos de Mundo Brana.\ \ \ \ \ \ São Paulo, 2006.\\
\\
\hspace{13mm} \large Dissertação (Mestrado) - Universidade de São Paulo.\\
\hspace{13mm} \large Instituto de Física - Depto. de Física Matemática.\\
\\
\hspace{13mm} \large Orientador: Prof. Dr. Élcio Abdalla\\
\hspace{13mm} \large Área de Concentração: Física\\
\\
\hspace{13mm} \large Unitermos:\\
\\ 
\hspace{13mm} \large 1. Buracos Negros;\\
\hspace{13mm} \large 2. Relatividade-Mundo Brana;\\
\hspace{13mm} \large 3. Relatividade-Estabilidade de Buracos Negros.\\
\\
\\
\large USP/IF/SBI-017/2006
\\
\hline
\end{tabular}
\end{center}
\newpage
{\large \bf Resumo}\\
\\
\\
O estudo de objetos compactos e suas propriedades físicas tem sido alvo de constantes investigações nos últimos 90 anos. Com o surgimento da Teoria de Cordas e os modelos de Mundo Brana como alternativas de descrição do nosso Universo, cresceu o interesse no estudo de buracos negros nesses contextos. Neste trabalho é desenvolvido um estudo de buracos negros em modelos de Mundo Brana. Uma classe de buracos negros esfericamente simétricos na brana é investigada. Critérios de estabilidade são estabelecidos. Proprie-dades termodinâmicas e a existência de modos quasi-normais de oscilação também são abordadas. Os buracos negros estudados são o Buraco Negro tipo CFM e o Buraco Negro tipo SM (sem massa), obtidos por Casadio {\it et al.} e Bronnikov {\it et al.}, respectivamente. A geometria do {\it bulk} no qual a brana está imersa é desconhecida. Entretanto, o teorema de Campbell-Magaard nos garante a existência de uma solução 5-dimensional no {\it bulk} que tem como projeção sobre a brana a classe de buracos negros estudados. Eles mostraram-se estáveis quando submetidos a perturbações do tipo escalar. A existência de modos quasi-normais de oscilação foi observada tanto para o Buraco Negro tipo CFM quanto para o Buraco Negro tipo SM. As caudas das perturbações apresentaram comportamentos semelhantes. O limite superior da entropia de um corpo absorvido pelos buracos negros estudados foi calculado mostrando que esse limite independe dos parâmetros dos buracos negros, sugerindo novamente sua universalidade.
\newpage

{\large \bf Abstract}\\
\\
\\
Research on black holes and their physical proprieties has been active on last 90 years. With the appearance of the String Theory and the Braneworld models as alternative descriptions of our Universe, the interest on black holes, in these context, increased. In this work we studied black holes in Braneworld models. A class of spherically symmetric black holes is investigaded as well its stability under general perturbations. Thermodynamic proprieties and quasi-normal modes are discussed. The black holes studied are the SM (zero mass) and CFM solutions, obtained by Casadio {\it et al.} and Bronnikov {\it et al.}. The geometry of bulk is unknown. However the Campbell-Magaard Theorem guarantees the existence of a 5-dimensional solution in the bulk whose projection on the brane is the class of black holes considered. They are stable under scalar perturbations. Quasi-normal modes were observed in both models. The tail behavior of the perturbations is the same. The entropy upper bound of a body absorved by the black holes studied was calculated. This limit turned out to be independent of the black hole parameters. 
\newpage
{\large \bf Dedicatória}
\vspace{18cm}
\begin{flushright}
\begin{minipage}[t]{10cm}
``Ao Espírito Imortal que indicou o caminho pelo qual eu deveria seguir."
\end{minipage}  
\end{flushright}
\newpage
{\large \bf Agradecimentos}\\
\\
À todos os Espíritos que já caminharam  ou ainda caminham comigo durante essa etapa da minha existência.\\
\\
Ao amigo e orientador Prof Elcio Abdalla pela oportunidade de trabalho, pela paciência e pelas conversas descontraídas sobre a vida.\\
\\
Aos amigos e companheiros de trabalho, Bertha e Prof Carlos, pelo auxílio e pela atenção dedicada a este trabalho. Sem a ajuda de vocês esse trabalho não seria possível.\\
\\
À minha grande família em Ribeirão Preto e aqui em São Paulo, especialmente aos amigos e irmãos do apartamento 132, Thiago, Mineiro, Bruno e Rodrigo.\\
\\
À família do Departamento de Física-Matemática que me acolheu com tanto carinho, atenção e respeito: Amélia, Simone, Bete, Rodrigo, Jéferson, Flávio, Carlos Eduardo, Carlos, Bertha, Michele, Arlene, Cecília, David, Karlúcio e muitos outros que por lá passaram.\\
\\
Aos meus pais pela oportunidade da existência, pelo carinho e pelo desmedido amor.\\
\\
À minha irmã por tudo aquilo que ela representa em minha vida.\\
\\
Aos meus avós por serem modelos no qual pretendo sempre me espelhar.\\
\\
À FAPESP pelo apoio a Ciência no Brasil e pelo apoio a este trabalho.

\newpage

\tableofcontents
\newpage

\chapter[Introdução]{Introdução}

Eis que continua a longa jornada do homem, em busca daquilo que é
o Universo. Do início do caminho até os dias de hoje grandes
mentes contribuiram na construção, ou melhor seria dizer, no
entendimento do espaço e do tempo em que nos encontramos
inseridos.

Desvendar os mistérios do Criador para aqueles que Nele acreditam
ou simplesmente entender a Natureza, eis a nossa busca, quem sabe
nosso martí-rio, quem sabe nossa redenção. Sigamos então pela
estrada do conhecimento na tentativa de contribuir ou tornar mais
claro pequenos pontos do nosso misterioso Universo.

Nossa jornada começa em 1915. O surgimento da Teoria da
Relatividade Geral de Albert Einstein nos traz um novo
entendimento do que é o espaço e o tempo e os transforma em um
espaço-tempo.

Essa entidade na qual repousamos, adquire dinâmica, movimento, e
então as fronteiras do nosso entendimento sobre o mundo se largam.
Todo o Universo surge novo e das próprias equações que descrevem
essa dinâmica surgem mundos até então impensados, no máximo
sonhados. E o Universo começa a não ser mais algo estático,
eterno, e passa a se movimentar completamente fora do nosso
limitado controle. Então surgem novos e muitos Universos.
Universos vazios e parados, Universos cheios, de matéria e luz, a
girar, Universos que se expandem e tem começo, meio e fim. E neste
ponto até parece que nos aproximamos do Criador mostrando que o
Universo pode ter tido um início, um Grande Início; afinal não
estamos brincando de criar também.

E desses Universos, talvez aqueles, que mais nos causaram espanto
a princípio por descreverem algo novo, foram os que hoje
conhecemos pelo nome de {\it buracos negros}.

O objetivo deste trabalho é trilhar duramente a jornada que no
conduz ao entendimento desses objetos estranhos, os buracos negros,
que surgiram com a Relatividade Geral em $(3+1)$ dimensões e
aparecem até hoje como desenvolvimentos dos mais intrigantes e
motivantes em teorias mais recentes tais como a Teoria de Cordas e
o Mundo Brana com suas várias dimensões extras.

Na verdade, a idéia de um objeto com um campo gravitacional tão
forte que fosse capaz de impedir até a luz de escapar é uma idéia
antiga, anterior à Relatividade Geral, que foi estudada por Michel
\cite{Michel} e Laplace \cite{Laplace}. A princípio esse objeto
recebeu o nome de {\it estrela escura} pois se tratava de uma
estrela que não emitia luz nenhuma. Foi Wheeler, em 1968, quem
batizou tais objetos como buracos negros.

A primeira solução exata (esfericamente simétrica) das equações de
Einstein foi obtida por Schwarzschild em 1916. Além de possuir uma
singularidade física em $r=0$, essa solução possuía uma outra
singularidade em $r=2M$ devido ao sistema de coordenadas adotado.
A partir deste momento muitas outras soluções das equações de
Einstein foram obtidas por inúmeros pesquisadores.

Entres elas destacamos a solução obtida por Snyder e Oppenheimer
em 1939, que mostra que um buraco negro é obtido como resultado do
colapso gravitacional de uma estrela massiva, a solução de Reisner
(1916) e Nordstrom (1918) para um buraco negro com carga elétrica,
a solução de Kerr em 1963, para um buraco negro em rotação.

Ao longo desses desenvolvimentos uma importante questão surgiu, a
respeito dos buracos negros. Seriam eles objetos estáveis quando
submetidos a alguma perturbação?

O estudo da estabilidade de buracos negros surgiu com Regge e
Wheeler em 1957. Eles submeteram o buraco negro de Schwarzschild a
uma perturbação linear na métrica e verificaram que ele era
estável. Estudos com outros tipos de perturbações tais como
perturbações escalares, eletromagnéticas e de {\it spin} $1/2$
também foram realizados para o buraco negro de Schwarzschild,
confirmando sua estabilidade. Depois desses trabalhos muitos
outros foram realizados, complementando e extendendo os resultados
obtidos para outros buracos negros.

Nos anos 60 os primeiros indícios da possibilidade de detecção de
buracos negros surgiram no cenário astrofísico. O estudo de
acreção de matéria e sistemas binários onde um dos constituintes
era um buraco negro mostrou que tais sistemas seriam poderosas
fontes de raio-X. Medidas realizadas posteriormente mostraram
alguns possíveis candidatos a buracos negros.

Mais recentemente com a descoberta de que buracos negros emitem
ondas gravitacionais surgiu a possibilidade das primeiras medidas
através de antenas gravitacionais, de propriedades de tais
objetos. Essa descoberta motivou uma corrida entre equipes de
pesquisadores pela primeira medição de uma onda gravitacional.

Além do interesse por tais medidas, cresceu também o interesse no
estudo de modelos capazes de descrever o comportamento das ondas
gravitacionais emitidas por buracos negros.

Neste contexto é que surge o conceito de {\it freqüências
quasi-normais} de objetos massivos introduzido por Press
\cite{Press} em um trabalho de 1971. Resultado de perturbações no
exterior de um buraco negro, as freqüências quasi-normais são
freqüências complexas de oscilação do espaço-tempo no exterior do
buraco negro. Essas perturbações se propagam na forma de ondas e
tem duração limitada no tempo pois são atenuadas exponencialmente.

A característica interessante desses modos quasi-normais de
oscilação é que eles independem do tipo de perturbação inicial
dependendo apenas dos parâmetros do buraco negro.

Portanto a análise dos modos quasi-normais de buracos negros ganha
acentuada importância uma vez que através dela podemos dizer se a
onda gravitacional que medimos é provinda de um buraco negro em
rotação ou um buraco negro estático, por exemplo.

Outro grande desenvolvimento na física de buracos negros surgiu
com o trabalho de Hawking \cite{Hawking1} sobre a propagação de
campos quânticos no exterior de um buraco negro estacionário. Ele
mostrou que um buraco negro estacionário irradiava para o infinito
todas as espécies de partículas com um espectro de corpo negro
perfeito, à uma temperatura $T_{H}=\kappa/2\pi$.

Esse trabalho de certo modo, pôs fim a um empasse que existia, a
respeito da analogia entre a física de buracos negros e a
termodinâmica ordinária.

Esta analogia foi desenvolvida sistematicamente por Bardeen,
Hawking e Carter \cite{BHC} além de Bekenstein
\cite{Beken3,Beken1} que propôs uma Segunda Lei da Termodinâmica
Generalizada. Todas essas inovações impulsionaram o estudo de
buracos negros em um contexto quântico.

Com o advento da Teoria de Cordas, o interesse por buracos negros
se renovou  e abriu novas oportunidades, devido a possibilidade da
existência de dimensões espaciais extras. Iniciou-se então o
estudo de soluções do tipo buraco negro em cenários com dimensões
mais altas. Diversos cenários surgiram como alternativas de como
poderíamos enxergar o nosso Universo, entre eles os mais
conhecidos são os Modelos Randall-Sundrum (RS) \cite{RS1, RS2} e o
Modelo ADD \cite{ADD}.

Os Modelos de Mundo Brana continuam gerando grandes contribuições
no entendimento da física em altas dimensões. Avanços no contexto
de Mundo Brana tem sido feitos, tais como os realizados por
Hawking \cite{cordanegra} que obteve uma solução tipo corda negra
, Maartens \cite{Germani} que obteve uma solução que descreve uma
estrela de densidade uniforme sobre a brana, Visser \cite{Visser2}
com uma solução que descreve planetas sólidos sobre a brana,
Casadio \cite{Casadio2,Casadio1} com uma solução tipo ``cigarro
negro" e soluções tipo buracos negros e buracos de minhoca e
Bronnikov \cite{Bronnikov1,Bronnikov2} com uma classe geral de
buracos negros e buracos de minhoca. Algumas dessas soluções
apresentaram regimes instáveis quando submetidas a
perturbações \cite{gregorilaflamme}, outras ainda não foram
estudadas até agora, o que nos motiva ao estudo da estabilidade
dessas soluções.

Interessante notar que grande parte dos formalismos desenvolvidos
no contexto da Relatividade Geral funcionam sem grandes alterações
no contexto de dimensões mais altas.

Por exemplo, os modos quasi-normais, tem se mostrado uma
ferramenta eficaz no entendimento dos Mundos Brana. Em um trabalho
recente Maartens \cite{Maartens1} propõem a utilização de modos
quasi-normais como uma ferramenta na detecção de dimensões extras.
Abdalla {\it et al} \cite{elcio} utilizaram técnicas númericas
desenvolvidas no contexto de Relatividade Geral para o cálculo de
modos quasi-normais de buracos negros em branas. Neste trabalho
também foi analisada a estabilidade desses buracos negros e o
limite superior da entropia $S_{m}$ de um corpo absorvido por um
buraco negro localizado sobre a brana.

Inspirados nesta abordagem de adaptarmos um formalismo sólido
vindo da Relatividade Geral ao Mundo Brana desenvolvemos esta
dissertação.\\
\\
Este trabalho se divide em duas grandes áreas: buracos negros no
contexto da Relatividade Geral e buracos negros no contexto de
Mundo Brana. Os primeiros 5 capítulos tratam dos desenvolvimentos
realizados no estudo de buracos negros em Relatividade Geral. Os
capítulos finais tratam do estudo de buracos negros no Mundo
Brana.

No capítulo 2 apresentaremos um resumo sobre buracos negros em
(3+1) dimensões. No capítulo 3 estudaremos o formalismo aplicado
no estudo de perturbações de buracos negros esfericamente
simétricos. No capítulo 4 estudaremos os critérios de estabilidade
e os modos quasi-normais de oscilação de um buraco negro
esfericamente simétrico. No capítulo 5 estudaremos as propriedades
termodinâmicas de buracos negros esfericamente simétricos. No
capítulo 6 apresentaremos uma introdução ao Mundo Brana e o
desenvolvimento do estudo de buracos negros neste contexto. No
capítulo 7 adaptamos o formalismo desenvolvido nos capítulos 3, 4
e 5, para buracos negros localizados sobre uma brana. No capítulo
8 são apresentados os resultados obtidos no trabalho assim como o
método numérico utilizado para a solução das equações obtidas. No
capítulo 9 apresentamos as conclusões do trabalho desenvolvido
nesta dissertação. Nos Apêndices se encontram as notações adotadas
neste trabalho e algumas demonstrações de interesse.
\newpage

\chapter[Buracos Negros em 3+1 dimensões]{Buracos Negros em 3+1 dimensões}

{\it Aqui começa nossa jornada}.\\
\\

A compreensão de que espaço e tempo é uma estrutura coesa,
dinâmica e intrincada é uma conquista da Relatividade Geral. No
contexto da Relatividade Geral, espaço e tempo são tratados em pé
de igualdade formando uma única entidade chamada espaço-tempo.

Para representarmos um evento no espaço-tempo, como por exemplo um
raio que cai em um determinado ponto de São Paulo, utilizamos
quatro coordenadas, $(x,y,z,t)$. Aqui o tempo não é mais tratado
como absoluto e a cada evento associamos um tempo.

Portanto dois eventos {\bf A} e {\bf B} devem ser representados no
espaço-tempo por $(x_{a},y_{a},z_{a},t_{a})$ e
$(x_{b},y_{b},z_{b},t_{b})$ respectivamente.

Assim para medirmos o intervalo $ds$ entre os eventos {\bf A} e
{\bf B} usamos o conceito do tensor métrico $g_{ij}$ que nos diz
como esses dois eventos devem ser relacionados.

O intervalo $ds$ é dado por
\begin{eqnarray}
\label{BN01} ds^2=g_{ij}dx^{i}dx^{j}\ .
\end{eqnarray}
Outro aspecto que surge da Relatividade é que o espaço-tempo
possui uma dinâmica própria.

Como sabemos, as equações que governam a dinâmica do espaço-tempo
são as conhecidas equações de campo de Einstein. Entre muitos
outros aspectos elucidativos, as equações de Einstein contêm em si
a informação de como campos de matéria afetam a geometria em que
estão contidos. Aliás a compreensão de que a gravidade nada mais é
do que o efeito da deformação do espaço-tempo por um corpo é fruto
destas equações.

As equações de campo de Einstein escrita na forma covariante são
\cite{Weinberg}
\begin{eqnarray}
\label{BN02} R_{ij}-\frac{1}{2}g_{ij}R=-8\pi\ T_{ij}\ .
\end{eqnarray}
O termo do lado esquerdo da equação (\ref{BN02}) representa a
parte relacionada com a geometria do espaço-tempo onde
identificamos o tensor de Ricci $R_{ij}$ e o escalar de curvatura
$R$.

O termo da direita representa o conteúdo material presente no
espaço-tempo. Esse conteúdo é representado pelo tensor
momento-energia $T_{ij}$. É o tensor momento-energia que nos dirá
como o espaço-tempo será deformado. Toda vez que introduzimos
algum conteúdo no espaço-tempo devemos especificar seu tensor
momento-energia.

Tomemos um caso bem geral onde introduziremos um campo
eletromagnético livre de fonte que é governado pelas equações de
Maxwell e cujo tensor momento-energia é dado por
$F_{ij}^{Max}=\frac{1}{4\pi}\left(-g^{ab}F_{ia}F_{jb}+\frac{1}{4}g_{ij}F_{ab}F^{ab}\right)
$, um fluído de matéria que possui características de um fluído
perfeito com $T^{\ mat}_{ij}=pg_{ij}+(p+\rho)u_{i}u_{j}$ onde $p$
e $\rho$ são a pressão e a densidade e $u_{i}$ é a
quadri-velocidade das partículas que compõem o fluído e por fim um
termo chamado constante cosmológica $\lambda$.

Nesta situação as equações de campo de Einstein serão
\begin{eqnarray}
\label{BN03} R_{ij}-\frac{1}{2}g_{ij}R-\lambda g_{ij}=-8\pi\
(T_{ij}^{\ mat}+F_{ij}^{Max})\ .
\end{eqnarray}
Ao resolvermos esse conjunto de equações obteremos equações que
descrevem como esses campos deformam o espaço-tempo.\\
Podemos calcular ainda, as equações de Einstein para o vácuo.
Neste caso o termo à direita da equação (\ref{BN02}) é nulo e a
equação resultante será
\begin{eqnarray}
\label{BN04} R_{ij}-\frac{1}{2}g_{ij}R=0\ .
\end{eqnarray}
Quando nos propomos a estudar uma estrela, por exemplo, devemos
resolver as equações de Einstein para dois regimes diferentes, que
correspondem ao interior da estrela, o qual possui matéria e
portanto $T_{ij}\neq 0$, e o exterior da estrela que corresponde
ao vácuo. A solução obtida para o interior da estrela deve ser
então ``colada" à solução obtida para o vácuo para que o conjunto
tenha sentido físico e possa descrever adequadamente uma estrela.

Podemos então nos fazer uma pergunta. Qual é o significado de
calcularmos as equações de Einstein para o vácuo, sem entretanto
possuir uma ``fonte de matéria"?

Como resposta ao significado físico desta operação, foram obtidos
dentre outros objetos com diversas simetrias, os objetos mais
estranhos e exóticos da Relatividade Geral: {\it buracos negros} e
{\it buracos de minhoca}.

De um modo bem simplório podemos dizer que um buraco negro é o
resultado do colapso de uma estrela massiva quando seu
combustível nuclear se estingue e a pressão interna se torna incapaz de sustentar a
atração gravitacional fazendo a estrela implodir. Quando a estrela
implode ela gera um ponto de densidade infinita, uma {\it
singularidade}. Um observador localizado a grande distância pode
observar a implosão da estrela até que o raio da estrela seja
igual a duas vezes sua massa. Neste ponto o desvio para o vermelho
da luz emitida pela estrela é infinito e a estrela ``some" para um
observador distante. Após esse ponto a estrela continua a colapsar
até que seu raio seja zero.

O objeto remanescente do colapso gravitacional da estrela é
conhecido como buraco negro, ou seja, é um objeto com um campo
gravitacional tão intenso que nem mesmo a luz é capaz de escapar.
De fato, nem sempre o colapso gravitacional de uma estrela resulta
em um buraco negro. Se a estrela possui uma massa $M\lesssim 1.2
M_{\odot}$ o colapso da estrela resulta em uma estrela do tipo
{\it anã branca}. Esse limite é conhecido como limite de
Chandrasekhar. Se a estrela possui uma massa $M\lesssim
0.7M_{\odot}$ o colapso da estrela resulta em uma {\it estrela de
nêutrons}. Esse limite para a massa da estrela é conhecido como
limite de Oppenheimer-Volkoff. Pode acontecer ainda que a massa da
estrela seja grande o suficiente para que o seu colapso resulte em
uma supernova, ou seja, uma explosão que expulsa o excesso de
matéria e tem por resultado ou uma estrela de nêutrons ou uma anã
branca.

Um buraco de minhoca pode ser descrito como um atalho através do
espaço-tempo. Eles podem ser dividos em {\it buracos de minhoca
inter-universo} que conectam ``nosso" universo a um ``outro"
universo e {\it buracos de minhoca intra-universo} que conectam
duas regiões distintas de um único universo \cite{Visser}. Buracos
de minhoca não serão abordados neste trabalho.

Como foi dito acima, diversas soluções para a equação (\ref{BN04})
foram obtidas para algumas simetrias tais como simetria esférica,
cilíndrica e cônica. Neste trabalho nos restringiremos apenas a espaços-tempos com simetria esférica arbitrária.

Adotemos um espaço-tempo que preserve a simetria esférica podendo
conter matéria e campos físicos no seu interior.

É sabido que existe um sistema de coordenadas
$(x^0,x^1,\theta,\phi)$ em um espaço-tempo esfericamente simétrico
tal que a métrica dele tenha a forma
\begin{eqnarray}
\label{BN05}
ds^2&=&g_{00}(x^0,x^1)(dx^0)^2+2g_{01}(x^0,x^1)dx^1dx^0+\nonumber\\
\nonumber\\
&+&g_{11}(x^0,x^1)(dx^1)^2+g_{22}(x^0,x^1)(d\theta^2+sen\theta^2d\phi^2)\
.
\end{eqnarray}
Como ainda existe liberdade na escolha das coordenadas, podemos
fazer uma transformação de coordenadas para eliminar o termo
$g_{01}$ e torna a métrica diagonal. Fazendo a seguinte
transformação de coordenadas
\begin{eqnarray}
\label{BN06} \tilde{x}^0=\tilde{x}^0(x^0,x^1)\ , \qquad
\tilde{x}^1=\tilde{x}^1(x^0,x^1)\ , \qquad \tilde{\theta}=\theta\
, \qquad \tilde{\phi}=\phi\ ,
\end{eqnarray}
a métrica (\ref{BN05}) toma a forma diagonal dada por
\begin{eqnarray}
\label{BN07}
ds^2&=&g_{00}(x^0,x^1)(dx^0)^2+g_{11}(x^0,x^1)(dx^1)^2+\nonumber\\
\nonumber\\
&+&g_{22}(x^0,x^1)(d\theta^2+sen\theta^2d\phi^2)\ .
\end{eqnarray}
Podemos então, devido à liberdade de coordenadas remanescente,
escolher um novo sistema de coordenadas. Se o gradiente da função
$g_{22}(x^0,x^1)$ não se anula em alguma região, esta função pode
ser escolhida como uma nova coordenada que chamaremos $x^1$.

Neste sistema de coordenadas a métrica esfericamente simétrica
toma a forma
\begin{eqnarray}
\label{BN08}
ds^2&=&g_{00}(x^0,x^1)(dx^0)^2+g_{11}(x^0,x^1)(dx^1)^2+\nonumber\\
\nonumber\\
&+&(x^1)^2(d\theta^2+sen\theta^2d\phi^2)\ .
\end{eqnarray}
Se fixarmos os valores de $x^0$ e $x^1$ a métrica (\ref{BN08})
será a métrica de uma esfera bidimensional cuja a área da
superfície é $A=4\pi(x^1)^2$.

Portanto reescrevendo a equação (\ref{BN08}) de uma maneira mais agradável
teremos
\begin{eqnarray}
\label{BN09}
ds^2=g_{tt}(t,r)dt^2+g_{rr}(t,r)dr^2+r^2(d\theta^2+sen\theta^2d\phi^2)\
,
\end{eqnarray}
onde as componentes $g_{tt}$ e $g_{rr}$ são funções arbitrárias
dependentes das coordenadas $r$ e $t$.

Isso posto, passemos ao estudo de soluções da equação de Einstein
quando adotamos simetria esférica.\\
\\
{\bf Buraco Negro de Schwarzschild }\\
\\

Uma das soluções das equações de Einstein mais famosas foi obtida
em 1916 por Schwarzschild. Essa solução das equações de Einstein
para o vácuo descreve um espaço-tempo esfericamente simétrico,
estático e assintoticamente plano. Diversos autores demonstraram a
solução de Schwarzschild de muitas maneiras diferentes.

Nesta demonstração seguiremos de perto o método adotado por
Weinberg \cite{Weinberg}. As equações de campo de Einstein para o
vácuo são dadas por
\begin{eqnarray}
\label{BNS01} R_{ij}-\frac{1}{2}g_{ij}R=0\ .
\end{eqnarray}
O valor do escalar de curvatura, $R$, para esse caso, é obtido
através de uma contração da equação (\ref{BNS01}) com a métrica
$g^{ij}$
\begin{eqnarray}
\label{BNS02}
g^{ij}R_{ij}-\frac{1}{2}g^{ij}g_{ij}R=0\ ,\nonumber\\
\nonumber\\
R-2R=0\ ,\nonumber\\\
\nonumber\\
R=0\ .
\end{eqnarray}
Portanto devido ao fato do escalar de curvatura $R$ ser nulo a
equação (\ref{BNS01}) pode ser reescrita como
\begin{eqnarray}
\label{BNS03} R_{ij}=0\ .
\end{eqnarray}
Se impusermos que a solução da equação (\ref{BNS03}) possua
simetria esférica e seja estática podemos utilizar a métrica
(\ref{BN09}) como {\it Ansatz} para simplificar as contas.

A métrica então será
\begin{eqnarray}
\label{BNS04}
ds^2=-A(r)dt^2+B(r)dr^2+r^2(d\theta^2+sen\theta^2d\phi^2)\ .
\end{eqnarray}
Para a métrica (\ref{BNS04}) as componentes não-nulas do tensor de Ricci $R_{ij}$ serão
\begin{eqnarray}
\label{BNS05}
R_{rr}&=&\frac{A''(r)}{2A(r)}-\frac{1}{4}\left(\frac{A'(r)}{A(r)}\right)\left(\frac{B'(r)}{B(r)}+\frac{A'(r)}{A(r)}\right)-\frac{1}{r}\frac{B'(r)}{B(r)}\ ,\nonumber\\
\nonumber\\
R_{\theta\theta}&=&-1+\frac{r}{2B(r)}\left(-\frac{B'(r)}{B(r)}+\frac{A'(r)}{A(r)}\right)+\frac{1}{B(r)}\ ,\nonumber\\
\\
R_{\phi\phi}&=&sen^2\theta\ R_{\theta\theta}\ ,\nonumber\\
\nonumber\\
R_{tt}&=&-\frac{A''(r)}{2B(r)}+\frac{1}{4}\left(\frac{A'(r)}{B(r)}\right)\left(\frac{B'(r)}{B(r)}+\frac{A'(r)}{A(r)}\right)-\frac{1}{r}\frac{A'(r)}{B(r)}\
,\nonumber
\end{eqnarray}
onde $(\ '\ )$ corresponde a uma diferenciação com respeito a $r$.

As equações que irão nos interessar serão $R_{rr}=0$, $R_{tt}=0$ e
$R_{\theta\theta}=0$. Podemos ver que
\begin{eqnarray}
\label{BNS06}
\frac{R_{rr}}{B}+\frac{R_{tt}}{A}=-\frac{1}{rA}\left(\frac{B'}{B}+\frac{A'}{A}\right)\
.
\end{eqnarray}
Como as componentes $R_{rr}$ e $R_{tt}$ devem ser nulas devido à
equação (\ref{BNS03}) temos que
\begin{eqnarray}
\label{BNS07}
\frac{B'}{B}&=&-\frac{A'}{A}\ ,\nonumber\\
\nonumber\\
\frac{d}{dr}\ln\ B&=&-\frac{d}{dr}\ln\ A\ ,\nonumber\\
\nonumber\\
A(r)\ B(r)&=&\ C_{1}\ .
\end{eqnarray}
A constante $C_{1}$ será fixada pelo comportamento assintótico das
funções $A(r)$ e $B(r)$. Como queremos recuperar o espaço-tempo de
Minkowski longe do buraco negro de modo que a Relatividade
Restrita seja válida, podemos supor a seguinte condição de
contorno
\begin{eqnarray}
\label{BNS08} \lim_{r\rightarrow \infty}\ A(r)=\lim_{r\rightarrow
\infty}\ B(r) =1\ .
\end{eqnarray}
Desta forma teremos
\begin{eqnarray}
\label{BNS09} B(r)=\frac{1}{A(r)}\ .
\end{eqnarray}
Ainda precisamos fazer com que $R_{\theta\theta}$ e $R_{rr}$ sejam
nulos para termos nossa solução completa. Se usarmos a equação
(\ref{BNS09}) nas equações (\ref{BNS05}) teremos
\begin{eqnarray}
\label{BNS10}
R_{\theta\theta}=-1+A'(r)r+A(r)\ ,\nonumber\\
\nonumber\\
R_{rr}=\frac{A''(r)}{2A(r)}+\frac{A'(r)}{rA(r)}=\frac{R'_{\theta\theta}(r)}{2rA(r)}\
.
\end{eqnarray}
De modo que se tivermos $R_{\theta\theta}=0$ a componente $R_{rr}$
será nula.

Portanto
\begin{eqnarray}
\label{BNS11}
-1+A'(r)r+A(r)=0\ ,\nonumber\\
\nonumber\\
\frac{d}{dr}\left(r\ A(r)\right)=1\ .
\end{eqnarray}
Integrando a equação (\ref{BNS11}) teremos como solução
\begin{eqnarray}
\label{BNS12}
r\ A(r)= r+C_{2}, \nonumber\\
\nonumber\\
A(r)=\left(1+\frac{C_{2}}{r}\right)\ .
\end{eqnarray}
A constante $C_{2}$ é fixada utilizando o fato de que a grandes
distâncias da massa central $M$, a componente $g_{tt}=-A(r)$ deve
se aproximar de $-(1+2\phi)$ onde $\phi=-M/r$. Essa aproximação
é conhecida {\it aproximação de campo fraco}, ou seja, um
observador que está longe do buraco negro, enxerga-o como se fosse
um corpo de massa $M$.

Desta forma nossa solução final será
\begin{eqnarray}
\label{BNS13} A(r)=\left(1-\frac{2M}{r}\right)\ ,\\
\nonumber\\
B(r)=\frac{1}{\left(1-\frac{2M}{r}\right)}\ .
\end{eqnarray}
Assim temos como resultado a métrica de Schwarzschild dada por
\begin{eqnarray}
\label{BNS14}
ds^2=-\left(1-\frac{2M}{r}\right)dt^2+\frac{1}{\left(1-\frac{2M}{r}\right)}dr^2+r^2d\Omega^2\
.
\end{eqnarray}
Se realizarmos uma investigação na métrica (\ref{BNS14}) veremos
que existe dois valores de $r$ para o qual ao menos uma das
componentes da métrica diverge. Esses valores são $r= 0,2M$.
Entretanto podemos realizar uma investigação um pouco mais precisa
para saber se esses dois pontos são realmente singularidades
físicas. Pode ser que esses pontos singulares sejam apenas resultado do sistema de coordenadas adotado. Portanto precisamos avaliar
quantidades escalares as quais são invariantes por transformação
de coordenadas. Se algum desses escalares for infinito sobre algum
ponto então podemos dizer que aquele ponto é uma singularidade
física. Se eles forem finitos então é possível remover a
singularidade por um transformação de coordedanadas.

Os escalares mais relevantes são: o escalar de Ricci ou de
curvatura, o escalar dado pelo quadrado do tensor de Ricci e o
escalar dado pelo quadrado do tensor de Riemann.

Para o caso do buraco negro de Schwarzschild teremos
\begin{eqnarray}
\label{BNS15}
R&=&0\ ,\\
\nonumber\\
R_{ij}R^{ij}&=&0\ ,\\
\nonumber\\
R_{ijkl}R^{ijkl}&=&\frac{48M^2}{r^6}\ .
\end{eqnarray}
Como podemos ver, para $r=2M$ os escalares possuem valores finitos
o que nos diz que ele não é um ponto singular. A única
singularidade física na métrica de Schwarzschild se encontra em
$r=0$.

Já que $r=2M$ não é uma singularidade, de que modo podemos mudar
nosso sistema de coordenadas para que possamos remover essa
singularidade aparente? 

Um sistema de coordenadas que resolve esse
problema foi obtido por Eddington (1924) e Finkelstein (1958).

Primeiramente precisamos saber qual é o comportamento de
geodésicas nulas radiais no caso de Schwarzschild. Usando o
formalismo do princípio variacional \cite{dinverno}, teremos a
seguinte equação
\begin{eqnarray}
\label{BNS16} \frac{dt}{dr}=\pm \frac{r}{r-2M}\ .
\end{eqnarray}
Se adotarmos o sinal positivo e integrarmos a equação (\ref{BNS16}) teremos
\begin{eqnarray}
\label{BNS17}
t=r+2M\ ln|r-2M|+constante.
\end{eqnarray}
Analisando essa solução vemos que para a região onde $r>2M$, a
equação (\ref{BNS16}) é positiva. Desta forma $r$ cresce com o
aumento de $t$. Por esse motivo as curvas dadas pela equação
(\ref{BNS17}) são congruências de geodésicas radiais nulas tipo
{\it outgoing}, ou seja, elas iniciam perto de $r=2M$ e se dirigem
para um $r$ maior.

De forma análoga se tomarmos o sinal negativo na equação
(\ref{BNS16}) sua solução será
\begin{eqnarray}
\label{BNS18}
t=-(r+2M\ ln|r-2M|+constante).
\end{eqnarray}
Neste caso $r$ decresce com $t$. Portanto as curvas dadas pela
equação (\ref{BNS18}) são congruências de geodésicas radiais nulas
tipo {\it ingoing}, ou seja, elas iniciam em um $r$ grande e se
dirigem para um valor de $r$ menor.

Isto posto, a idéia de Eddington e Finkelstein consistia em
definir uma nova coordenada temporal no qual geodésicas radiais
nulas do tipo {\it ingoing} se tornassem linhas retas. Observando
a equação (\ref{BNS18}) vemos que a melhor escolha para nossos
propósitos é fazer
\begin{eqnarray}
\label{BNS19} \bar{t}=t+2M\ ln(r-2M)\ ,
\end{eqnarray}
pois para o intervalo $r \in (2M,\infty)$ no novo sistema de
coordenadas $(\bar{t},r,\theta,\phi)$ nossa nova coordenada
temporal será
\begin{eqnarray}
\label{BNS20} \bar{t}=-r+constante\ ,
\end{eqnarray}
a qual forma uma linha reta que faz $-45^{o}$ com o eixo $r$.

Diferenciando a equação (\ref{BNS19}) e susbtituindo na métrica
(\ref{BNS14}) teremos a métrica de Schwarzschild em termos das
coordenadas de Eddington-Finkelstein
\begin{eqnarray}
\label{BNS21}
ds^2=-\left(1-\frac{2M}{r}\right)d\bar{t}^2+\frac{4M}{r}d\bar{t}dr+\left(1-\frac{2M}{r}\right)dr^2+r^2d\Omega^2\
.
\end{eqnarray}
Neste sistema de coordenada o ponto $r=2M$ é regular. De fato essa
transformação extende o intervalo em que a métrica  é regular
mapeando o intervalo $0<r<\infty$.

Podemos reescrever a métrica (\ref{BNS21}) de uma outra forma  se
introduzirmos uma coordenada nula
\begin{eqnarray}
\label{BNS23} v=\bar{t}+r\ ,
\end{eqnarray}
o que resulta então, na métrica em coordenadas avançadas de Eddington-Finkelstein
\begin{eqnarray}
\label{BNS24}
ds^2=-\left(1-\frac{2M}{r}\right)dv^2+2dvdr+r^2d\Omega^2.
\end{eqnarray}
A superfície formada por $r=2M$ faz o papel de uma membrana que
conduz todas as curvas tipo luz e tipo tempo de fora para dentro
evitando que nada saia para fora desse limite. Por isso esse ponto
é conhecido como {\it horizonte de eventos}.

Podemos obter uma outra solução se introduzirmos uma nova
coordenada temporal
\begin{eqnarray}
\label{BNS22} t^{*}=t-2M\ ln(r-2M)\ ,
\end{eqnarray}
o qual torna as geodésicas nulas radiais do tipo {\it outgoing} linhas retas.
Através dela podemos escrever a coordenada nula $u$ com parâmetro de tempo retardado
\begin{eqnarray}
\label{BNS25} u=t^{*}-r\ ,
\end{eqnarray}
e escrever a métrica (\ref{BNS14}) em coordenadas retardadas de
Eddington-Finkelstein como
\begin{eqnarray}
\label{BNS26}
ds^2=-\left(1-\frac{2M}{r}\right)du^2-2dudr+r^2d\Omega^2\ .
\end{eqnarray}
\\
\\
{\bf Buraco Negro de Reissner-Nordstrom}\\
\\
Uma outra solução esfericamente simétrica para as equações de
Einstein na presença de carga elétrica é a solução de
Reissner-Nordstrom. Sendo solução das equações de Einstein-Maxwell
ela descreve um espaço-tempo esfericamente simétrico, estático,
assintoticamente plano e com um campo elétrico radial.

Como neste problema temos um campo eletromagnético inserido no
espaço-tempo o tensor momento-energia $T_{ij}$ é não-nulo. A
métrica obtida como solução também deve ser solução das equações
de Maxwell.

Assim as equações de Einstein-Maxwell serão
\begin{eqnarray}
\label{BNRN01} R_{ij}-\frac{1}{2}g_{ij}R=-8\pi \ T_{ij}\ ,
\end{eqnarray}
onde \cite{dinverno}
\begin{eqnarray}
\label{BNRN02}
T_{ij}=\frac{1}{4\pi}\left(-g^{ab}F_{ia}F_{jb}+\frac{1}{4}g_{ij}F_{ab}F^{ab}\right)\
.
\end{eqnarray}
Como solução das equações (\ref{BNRN01}) teremos a métrica de
Reissner-Nordstrom que é dada por
\begin{eqnarray}
\label{BNRN03}
ds^2=-\left(1-\frac{2M}{r}+\frac{Q^2}{r^2}\right)dt^2+\frac{1}{\left(1-\frac{2M}{r}+\frac{Q^2}{r^2}\right)}dr^2+r^2d\Omega^2\
.
\end{eqnarray}
Para discussões adicionais e desenvolvimentos recentes, a respeito
do estudo de buracos negros em $(3+1)$ dimensões, ver
\cite{Novikov}.

\newpage

\chapter[Perturbações Gerais em Buracos Negros (3+1) dimensionais]{Perturbações Gerais em Buracos Negros (3+1) dimensionais}
{\it E como em toda jornada cruzamos estradas perigosas e pontes estreitas. Ao nos depararmos com uma ponte pergutamo-nos. O quanto ela é resistente?}\\
\\
Ainda que buracos negros sejam objetos exóticos  contendo
singuralidades, o estudo da evolução de campos no
exterior de tais objetos pode revelar-se surpreendentemente
simples e bastante instrutivo.

Quando colocamos algum campo, seja ele um campo escalar ou um
campo eletromagnético ou ainda um campo gravitacional, para
evoluir no exterior de um buraco negro, o campo perturba o espaço
ao seu redor. Essas perturbações ainda que inicialmente pequenas,
que será o caso tratado neste trabalho, podem ganhar tal proporção
com o passar do tempo que podem ser capazes de tornar o objeto
instável e destruí-lo. Portanto uma compreensão da evolução de
tais campos pode nos fornecer uma grande quantidade de
informações sobre os objetos que serão perturbados.

O formalismo desenvolvido neste capítulo será estendido para o
caso de buracos negros localizados sobre branas no capítulo 7.
Serão necessárias algumas adaptações para o novo modelo mas em
essência o procedimento será o mesmo.

\section{Perturbações Escalares}
Como se sabe, campos escalares não-massivos não descrevem objetos
físicos. Entretanto alguns conceitos básicos sobre a teoria de
perturbações de buracos negros podem ser obtidos de perturbações
escalares. Uma informação interessante que pode ser obtida da
perturbação escalar é sobre a estabilidade do buraco negro. Se um
buraco negro é instável quando submetido a uma perturbação escalar
ele também será instável quando submetido a uma perturbação
gravitacional. Como os cálculos da perturbação escalar são mais
simples eles podem dar-nos dicas preciosas sem precisarmos de
grande sofisticação matemática.

A evolução de campos escalares em um espaço-tempo 4-dimensional é
regida pela equação de Klein-Gordon. Para espaços-tempos
esfericamente simétricos essa equação pode ser escrita de uma
forma mais generalizada devido ao desacoplamento das coordenadas
angulares e as coordenadas radial e temporal. A equação de
Klein-Gordon para um campo escalar não-massivo $\Psi$ é
\begin{equation}
\label{pe1} \Box \Psi(t,r,\theta,\phi)= 0\ ,
\end{equation}
que escrita explicitamente é dada por
\begin{equation}
\label{pe2}
\frac{1}{\sqrt{-g}}\partial_{i}(\sqrt{-g}g^{ij}\partial_{j})\Psi=0\
,
\end{equation}
onde $(i ,j=0,\dots,3)$ e $g=det\ g_{ij}$ é o determinante da métrica.

Tomemos uma métrica esfericamente simétrica 4-dimensional dada por
\begin{equation}
\label{pe3} ds^2=-A(r)dt^2+B(r)dr^2+r^2d\Omega^2\ ,
\end{equation}
onde as funções $A(r)$ e $B(r)$ são, a princípio, gerais.

Assim para calcularmos a perturbação, em uma métrica esfericamente
simétrica, causada por um campo $\Psi$, precisamos calcular
algumas
quantidades relevantes à operação.

O determinante da métrica (\ref{pe3}) será
\begin{equation}
\label{pe4} det\ g_{ij}=-A(r)B(r)r^4 sen^2\theta\ ,
\end{equation}
e conseqüentemente
\begin{equation}
\label{pe5} \sqrt{-g}=r^2\ sen\theta \ \sqrt{A(r)B(r)}\ .
\end{equation}
Apresentaremos abaixo as derivadas relevantes e não-nulas para os
cálculos:
\begin{eqnarray}
\label{pe6}
\partial_{1}\sqrt{-g}&=&2r sen\theta \ \sqrt{AB}\left(1+\frac{r}{4}\left[\frac{1}{A}\frac{dA}{dr}+\frac{1}{B}\frac{dB}{dr}\right]\right)\ ,\\
\nonumber\\
\frac{1}{\sqrt{-g}}\partial_{1}\sqrt{-g}&=&\frac{2}{r}\left(1+\frac{r}{4}\frac{d}{dr}\left[ln(AB)\right]\right)\ ,\\
\nonumber\\
\partial_{2}\sqrt{-g}&=&cos\theta \ r^2\sqrt{AB}\ ,\\
\nonumber\\
\frac{1}{\sqrt{-g}}\partial_{2}\sqrt{-g}&=&\frac{cos\theta}{sen\theta}\ ,\\
\nonumber\\
\partial_{2}\ g^{11}&=&-\frac{1}{B^2}\frac{dB}{dr}\ .
\end{eqnarray}
Devido às simetrias do problema em questão (esférica, sem rotação)
a nossa métrica torna-se dependente apenas das coordenadas $r$ e
$\theta$ o que reduz a equação (\ref{pe2}) à seguinte
\begin{eqnarray}
\label{pe7}
g^{00}\partial_{0}^2\Psi+
\frac{1}{\sqrt{g}}\left(\partial_{1}\sqrt{g}\right)g^{11}\partial_{1}\Psi+\left(\partial_{1}g^{11}\right)\partial_{1}\Psi
+g^{11}\partial_{1}^2\Psi&+&\nonumber
\\+g^{22}\partial_{2}^2\Psi+\frac{1}{\sqrt{g}}\left(\partial_{2}\sqrt{g}\right)g^{22}\partial_{2}\Psi+g^{33}\partial_{3}^2\Psi&=&0\
.
\end{eqnarray}
Substituindo as derivadas calculadas na equação (\ref{pe7}) teremos
\begin{eqnarray}
\label{pe8}
g^{00}\frac{\partial^2\Psi}{\partial t^2}&+&\left[\frac{2}{r}+\frac{1}{2A}\frac{dA}{dr}-\frac{1}{2B}\frac{dB}{dr}\right]g^{11}\frac{\partial \Psi}{\partial r}+g^{11}\frac{\partial^2\Psi}{\partial r^2}+\nonumber\\
&+&\frac{1}{r^2}\left[\frac{1}{sen\theta}\frac{\partial}{\partial
\theta}\left(sen\theta\frac{\partial \Psi}{\partial
\theta}\right)+\frac{1}{sen^2\theta}\frac{\partial^2\Psi}{\partial
\phi^2}\right]=0\ .
\end{eqnarray}
Utilizando a seguinte substituição
\begin{equation}
\label{pe9} \Psi(t,r,\theta,\phi)=Z(r,t)Y_{lm}(\theta,\phi)\ ,
\end{equation}
onde $Y_{lm}(\theta,\phi)$ são os chamados Harmônicos Esféricos e
solução da equação
\begin{equation}
\label{pe10} \frac{1}{sen\theta}\frac{\partial}{\partial
\theta}\left(sen\theta\frac{\partial Y_{lm}}{\partial
\theta}\right)+\frac{1}{sen^2\theta}\frac{\partial^2
Y_{lm}}{\partial \phi^2}+\ell(\ell + 1)Y_{lm}=0\ ,
\end{equation}
teremos uma equação diferencial de segunda ordem que depende
apenas de $r$ e $t$ dada por
\begin{eqnarray}
\label{pe11} -\frac{\partial^2 Z}{\partial
t^2}&+&\frac{A}{B}\frac{\partial^2 Z}{\partial
r^2}+\left[\frac{2}{r}+\frac{1}{2A}\frac{dA}{dr}-\frac{1}{2B}\frac{dB}{dr}\right]\frac{A}{B}\frac{\partial
Z}{\partial r}+\nonumber\\
&=&A\left(\frac{\ell(\ell+1)}{r^2}\right)Z\ .
\end{eqnarray}
Essa equação diferencial pode ser solucionada numericamente se
pudermos escrevê-la em um modo mais agradável. Isso pode ser feito
através da substituição da coordenada radial $r$ pela coordenada
$r_{*}$ (ver Apêndice \ref{sec:coordtartaruga}).

Para isso façamos as seguintes substituições
\begin{eqnarray}
\label{pe12} Z(r,t)=R(r,t)\ b(r)\ , \qquad f(r)=\frac{A}{B}\ ,
\qquad r=r(r_{*})\ ,
\end{eqnarray}
na equação(\ref{pe11}), resultando em
\begin{eqnarray}
\label{pe13}
&-&\frac{\partial^2 R}{\partial t^2}\ b+\left[\frac{2}{r}+\frac{1}{2}\frac{\partial r_{*}}{\partial r}\frac{\partial }{\partial r_{*}}(ln \ f)\right] f\ \frac{\partial r_{*}}{\partial r}\frac{\partial }{\partial r_{*}}(R \ b)+\nonumber\\
\nonumber\\
&+& f\ \frac{\partial r_{*}}{\partial r}\frac{\partial }{\partial
r_{*}}\left[\frac{\partial r_{*}}{\partial r}\frac{\partial
}{\partial r_{*}}(\chi \ b)\right]= A R b
\left[\frac{\ell(\ell+1)}{r^2}\right]\ .
\end{eqnarray}
Reescrevendo a equação (\ref{pe13}) e dividindo-a por $b(r)$
teremos
\begin{eqnarray}
\label{pe14}
-\frac{\partial^2 R}{\partial t^2}+\frac{f}{b}\dot{r}_{*}\dot{r}_{*}'\left[R'\ b+b'\ R\right]+\frac{f}{b}\dot{r}_{*}^2\left[R''\ b+2b'R'+b''\ R\right]+\nonumber\\
\nonumber\\
+\left[\frac{2}{r}+\frac{1}{2}\frac{\dot{r}_{*}}{f}f'\right]\frac{f}{b}\dot{r}_{*}\left[R'\
b+R\ b'\right]=A R\left[\frac{\ell(\ell+1)}{r^2}\right]\ ,
\end{eqnarray}
onde $(\ '\ )$ e $(\ \dot{}\ )$ representam diferenciação com
respeito a $r_{*}$ e $r$ respectivamente.

Escrevendo a equação (\ref{pe14}) de modo mais conveniente temos
\begin{eqnarray}
\label{pe15} -\frac{\partial^2 R}{\partial t^2}+f
\dot{r}_{*}^2\frac{\partial^2 R}{\partial r_{*}^2}+\frac{\partial
R}{\partial
r_{*}}\left(f\dot{r}_{*}\dot{r}_{*}'+2f\dot{r}_{*}^2\frac{b'}{b}+\left[\frac{2}{r}+\frac{\dot{r}_{*}f'}{2f}\right]f\dot{r}_{*}\right)+\nonumber\\
\nonumber\\
+R\left[f\dot{r}_{*}\dot{r}_{*}'\frac{b'}{b}+f\dot{r}_{*}^2\frac{b''}{b}+\left[\frac{2}{r}+\frac{\dot{r}_{*}f'}{2f}\right]f\dot{r}_{*}\frac{b'}{b}\right]=A
R\left[\frac{\ell(\ell+1)}{r^2}\right]\ .
\end{eqnarray}
Como queremos que equação (\ref{pe15}) tome a seguinte forma
\begin{eqnarray}
\label{pe16} -\frac{\partial^2 R}{\partial
t^2}(r,t)+\frac{\partial^2 R}{\partial r_{*}^2}(r,t)=
V_{ef}(r)R(r,t)\ ,
\end{eqnarray}
as seguintes condições devem ser satisfeitas:
\begin{eqnarray}
\label{pe17}
f(r)\dot{r}_{*}\dot{r}_{*}'+2f(r)\dot{r}_{*}^2\frac{b'}{b}&+&\left[\frac{2}{r}+\frac{\dot{r}_{*}f'}{2f}\right]f\dot{r}_{*}=0\ ,\\
\nonumber\\
\label{pe18} f(r)\dot{r}_{*}^2&=&1\ .
\end{eqnarray}
Solucionando a equação (\ref{pe18}) teremos
\begin{eqnarray}
\label{pe19}
\dot{r}_{*}=\frac{1}{\sqrt{f}} \qquad &\Longrightarrow & \qquad \frac{d}{dr_{*}}=\sqrt{f}\frac{d}{dr}\ ,\\
\nonumber\\
\label{pe20}
\frac{d}{dr_{*}}&=&\sqrt{\frac{A}{B}}\frac{d}{dr}\ ,\\
\nonumber\\
 r_{*}&=&\int \frac{1}{\sqrt{f}}\ dr\ .
\end{eqnarray}
Substituindo as equações (\ref{pe19}) e (\ref{pe20}) na equação
(\ref{pe17}) encontraremos a função $b(r)$
\begin{eqnarray}
\label{pe21}
\dot{r}_{*}'+2\dot{r}_{*}\frac{b'}{b}+\left[\frac{2}{r}+\frac{\dot{r}_{*}f'}{2f}\right]=0\nonumber\\
\nonumber\\
\label{pe22}
-\frac{1}{2f}\frac{df}{dr}+\frac{2}{b}\frac{db}{dr}+\frac{2}{r}+\frac{1}{2f}\frac{df}{dr}=0\nonumber\\
\nonumber\\
\label{pe22a} \frac{db}{dr}=-\frac{b}{r}\qquad \Longrightarrow
\qquad b(r)=\frac{1}{r}\ .
\end{eqnarray}
Para finalizar, o potencial efetivo será dado por
\begin{eqnarray}
\label{pe23}
V_{ef}(r)=A(r)\left[\frac{\ell(\ell+1)}{r^2}\right]-V(r)\ ,
\end{eqnarray}
onde
\begin{eqnarray}
\label{pe24}
V(r)=\left[f\dot{r}_{*}\dot{r}_{*}'\frac{b'}{b}+f\dot{r}_{*}^2\frac{b''}{b}+\left[\frac{2}{r}+\frac{\dot{r}_{*}f'}{2f}\right]f\dot{r}_{*}\frac{b'}{b}\right]\
.
\end{eqnarray}
Substituindo as equações (\ref{pe22a}) e (\ref{pe19}) na equação
(\ref{pe24}) teremos o potencial $V(r)$
\begin{eqnarray}
\label{pe25}
V(r)&=&f\dot{r}_{*}r\left[\frac{d}{dr}\left(\sqrt{f}\left(-\frac{1}{r^2}\right)\right)+\frac{2}{r}\sqrt{f}\left(-\frac{1}{r^2}\right)\right]\nonumber\\
\nonumber\\
V(r)&=&fr\left(\frac{1}{\sqrt{f}}\right)\left(-\frac{1}{2r^2\sqrt{f}}\frac{df}{dr}\right)\nonumber\\
\nonumber\\
V(r)&=&-\frac{1}{2r}\frac{df}{dr}\ .
\end{eqnarray}
Desta forma, nossa equação para a perturbação causada por um campo
escalar não-massivo em um espaço-tempo esfericamente simétrico, será
reduzida a uma equação tipo-Schroedinger dependente de $t$ e
$r=r(r_{*})$ com um potencial efetivo $V_{ef}(r)$ dados por
\begin{eqnarray}
\label{pe26}
-\frac{\partial^2 R}{\partial t^2}(r,t)+\frac{\partial^2 R}{\partial r_{*}^2}(r,t)= V_{ef}(r)R(r,t)\ ,\\
\nonumber\\
\label{pe27}
V_{ef}(r)=A(r)\left[\frac{\ell(\ell+1)}{r^2}\right]+\frac{1}{2r}\frac{df}{dr}\
.
\end{eqnarray}
A função de onda que descreve a evolução do campo escalar nesse
espaço-tempo será
\begin{equation}
\label{pe28} \Psi
(t,r,\theta,\phi)=\sum_{lm}\frac{R(t,r)}{r}Y_{lm}(\theta,\phi)\ .
\end{equation}
\newpage
\section{Perturbações Eletromagnéticas}

Um campo de grande interesse a ser estudado é o campo
eletromagnético pois ele representa um campo de matéria com algum
sentido físico. Quando estivermos estudando o caso de buracos
negros na brana a evolução do campo eletromagnético será de grande
interesse uma vez que ele pernamece confinado na brana. Mas por
enquanto nos restringiremos a buracos negros em (3+1) dimensões.

A evolução de um campo de Maxwell eletromagnético sem fonte em
um espaço-tempo esfericamente simétrico são regidas pelas equações
de Maxwell
\begin{eqnarray}
\label{pele01}
F^{ij}_{\phantom{ ij}{;j}}=0,\qquad F_{ij}=A_{ j;i}-A_{ i;j},
\end{eqnarray}
onde $A_{i}$ é potencial vetor associado ao campo.\\
Para uma métrica esfericamente simétrica dada por
\begin{eqnarray}
\label{pele02}
ds^2=-A(r)dt^2+B(r)dr^2+r^2d\Omega^2,
\end{eqnarray}
podemos expandir $A_{i}$ em termos dos harmônicos esféricos
vetoriais $(Y_{lm})_{i}(\theta,\phi)$ (ver \cite{Ruffini}):
\begin{displaymath}
A_{i}(t,r,\theta,\phi)=\sum_{l,m}\left( \left[{\small
\begin{array}{c}
0\\
0\\
\frac{a^{lm}(t,r)}{sen\theta} \ \partial_{\phi}Y_{lm}\\
-a^{lm}(t,r)\ sen\theta\ \partial_{\theta}Y_{lm}
\end{array} }\right]\right.
+\left.\left[{\small\begin{array}{c}
f^{lm}(t,r)\ Y_{lm}\\
h^{lm}(t,r)\ Y_{lm}\\
k^{lm}(t,r)\ \partial_{\theta}Y_{lm}\\
k^{lm}(t,r)\ \partial_{\phi}Y_{lm}
\end{array}}\right]\right)
\end{displaymath}
onde o primeiro termo à direita tem paridade $(-1)^{\ell+1}$ e o
segundo termo tem paridade $(-1)^{\ell}$, $m$ é o número azimutal
e $\ell$ é o número de momento angular orbital.

Se substituirmos essa expansão acima no tensor $F^{ij}$ teremos
como resultado dois conjuntos de equações. Um para a paridade
$(-1)^{\ell+1}$ e outro para a paridade $(-1)^{\ell}$. As
componentes não-nulas do tensor $F^{ij}$ de paridade
$(-1)^{\ell+1}$ são
\begin{eqnarray}
\label{pele03}
F^{t\phi}_{l,m}&=& - \frac{g^{tt}a^{lm}_{\phantom{lm}{,t}}Y^{lm}_{\phantom{lm}{,\theta}}}{r^2\ sen\theta};\qquad F^{r\phi}_{l,m}= - \frac{g^{rr}a^{lm}_{\phantom{lm}{,r}}Y^{lm}_{\phantom{lm}{,\theta}}}{r^2\ sen\theta},\nonumber\\
\nonumber\\
F^{t\theta}_{l,m}&=& \frac{g^{tt}a^{lm}_{\phantom{lm}{,t}}Y^{lm}_{\phantom{lm}{,\phi}}}{r^2\ sen\theta};\qquad \ \ F^{r\theta}_{l,m}= \frac{g^{rr}a^{lm}_{\phantom{lm}{,r}}Y^{lm}_{\phantom{lm}{,\phi}}}{r^2\ sen\theta},\nonumber\\
\nonumber\\
F^{\theta\phi}_{l,m}&=&\frac{\ell(\ell+1)\ a^{lm}\ Y^{lm}_{\phantom{lm}}}{r^4\ sen\theta}\nonumber.
\end{eqnarray}
Substituindo as componentes acima na equação (\ref{pele01}) obteremos uma equação diferencial radial
\begin{eqnarray}
\label{pele04} -\frac{1}{A}\frac{\partial^2a^{lm}}{\partial t^2}
+\frac{a^{lm}_{\phantom{lm}{,r}}}{2B}\left(ln\
A/B\right)_{,r}+\frac{a^{lm}_{\phantom{lm}{,rr}}}{B}-\frac{\ell(\ell+1)}{r^2}a^{lm}=0\
,
\end{eqnarray}
e outras equações que serão satisfeitas identicamente ou serão
equivalentes à equação (\ref{pele04}).

Como no caso da perturbação escalar, a equação (\ref{pele04}) pode
ser escrita em uma forma mais compacta se utilizarmos a coordenada
tartaruga $r_{*}$ e fizermos a substuição
\begin{eqnarray}
\label{pele09} a^{lm}(t,r)=\mathcal{A}(t,r), \qquad
f(r)=\frac{A}{B}, \qquad \frac{d}{dr_{*}}=\sqrt{f}\frac{d}{dr}.
\end{eqnarray}
Assim a equação (\ref{pele04}) pode ser reescrita como
\begin{eqnarray}
\label{pele10}
-\frac{\partial^2\mathcal{A}}{\partial t^2}+\frac{\partial^2\mathcal{A}}{\partial r_{*}^2}=V_{eletro}(r)\ \mathcal{A}
\end{eqnarray}
onde
\begin{eqnarray}
\label{pele11} V_{eletro}(r)= A(r)\frac{\ell(\ell+1)}{r^2}.
\end{eqnarray}
Essa será nossa equação para componente axial da perturbação
eletromagnética sobre um espaço-tempo esfericamente simétrico.\\
De modo semelhante as componentes não-nulas do tensor $F^{ij}$ de
paridade $(-1)^{\ell}$ são
\begin{eqnarray}
\label{pele05}
F^{r\theta}_{l,m}&=&g^{rr}(k^{lm}_{\phantom{lm}{,r}}-h^{lm})\frac{Y^{lm}_{\phantom{lm}{,\theta}}}{r^2}; \qquad F^{r\phi}_{l,m}=g^{rr}(k^{lm}_{\phantom{lm}{,r}}-h^{lm})\frac{Y^{lm}_{\phantom{lm}{,\phi}}}{r^2\ sen\theta}\nonumber\\
\nonumber\\
F^{tr}_{l,m}&=& g^{tt}\ g^{rr} (h^{lm}_{\phantom{lm}{,t}}-f^{lm}_{\phantom{lm}{,r}})Y^{lm};\qquad \ \ \ F^{t\theta}_{l,m}=g^{tt}(k^{lm}_{\phantom{lm}{,t}}-f^{lm})\frac{Y^{lm}_{\phantom{lm}{,\theta}}}{r^2},\nonumber\\
\nonumber\\
F^{t\phi}_{l,m}&=&g^{tt}(k^{lm}_{\phantom{lm}{,t}}-f^{lm})\frac{Y^{lm}_{\phantom{lm}{,\phi}}}{r^2\ sen\theta}.\nonumber
\end{eqnarray}
Substituindo essas componentes na equação (\ref{pele01}) obtemos
um conjunto de equações diferenciais radiais dados por
\begin{eqnarray}
\label{pele06}
g^{rr}[r^2(f^{lm}_{\phantom{lm}{,r}}-h^{lm}_{\phantom{lm}{,t}})]_{,r}+\ell(\ell+1)(f^{lm}-k^{lm}_{\phantom{lm}{,t}})+\nonumber\\
\nonumber\\
+\frac{(f^{lm}_{\phantom{lm}{,r}}-h^{lm}_{\phantom{lm}{,t}})}{2}g^{rr}r^2\left[(ln\ AB)_{,r}\right]=0\\
\nonumber\\
\nonumber\\
\label{pele08}
[(h^{lm}-k^{lm}_{\phantom{lm}{,r}})g^{rr}]_{,r}-(k^{lm}_{\phantom{lm}{,t}}-f^{lm})_{,t}g^{tt}-\nonumber\\
\nonumber\\
-\frac{1}{2}(ln\ AB)_{,r}g^{rr}(k^{lm}_{\phantom{lm}{,r}}-h^{lm})=0\\
\nonumber\\
\nonumber\\
\label{pele07}
\frac{1}{A}(h^{lm}_{\phantom{lm}{,t}}-f^{lm}_{\phantom{lm}{,r}})_{,t}-\frac{\ell(\ell+1)}{r^2}(k^{lm}_{\phantom{lm}{,r}}-h^{lm})=0
\end{eqnarray}
e o restante das equações ou são identicamente satisfeitas ou são
equivalentes ao conjunto acima.

Essas são as três equações da componente polar da perturbação
eletromagnética.

No caso da métrica (\ref{pele02}) ser a métrica de Schwarzschild,
ou seja, $A=\frac{1}{B}$ as equações (\ref{pele06}),
(\ref{pele08}) e (\ref{pele07}) se desacoplam, após uma
substituição conveniente, e se reduzem a uma única equação que
possuirá a mesma forma da equação (\ref{pele04}). No caso mais
geral ainda não sabemos se é possivel esse desacoplamento das
equações.

\newpage
\section{Perturbações Gravitacionais}

As perturbações gravitacionais de fato, são o caso mais
interessante a ser estudado, principalmente no que se refere à
estabilidade de soluções das equações de Einstein. Ao contrário
das perturbações escalares e eletromagnéticas, perturbações
gravitacionais, afetam levemente a métrica fundo.

A evolução de campos gravitacionais em um espaço-tempo
esfericamente simétrico é governada pelas equações de Einstein. A
perturbação causada por campos gravitacionais pode ser dividida em
axiais e polares devido à simetria esférica do problema. Essa
nomenclatura deve-se ao modo como elas se transformam sobre uma
inversão espacial em $\phi$ e foi introduzida por Chandrasekhar em
1983 \cite{Chandra}.

Perturbações {\it{axiais}}, se transformam como $(-1)^{\ell+1}$ e
induz um efeito de rotação no buraco negro.

Perturbações {\it{polares}}, se transformam como $(-1)^{\ell}$ e
como independem do sinal da coordenada $\phi$ não induzem rotação.

Nesse trabalho trataremos apenas das perturbações axiais e
consideraremos perturbações em $1^{a}$ ordem na métrica
$\breve{g}_{ij}$. Como realizado para a perturbação escalar,
calcularemos a perturbação para uma métrica esfericamente
simétrica  tão geral quanto possível e depois especificaremos caso
a caso.\\
\\
Nossa métrica esfericamente simétrica é dada por
\begin{eqnarray}
\label{pg00} ds^2=-A(r)dt^2+B(r)dr^2+r^2d\Omega^2\ ,
\end{eqnarray}
com $A(r)$ e $B(r)$ sendo duas funções em princípio arbitrárias.

As equações de Einstein 4-dimensionais para o vácuo, ou seja, o
exterior de uma estrela ou buraco negro, serão
\begin{eqnarray}
\label{pg01}
\breve{R}_{ij}-\frac{1}{2}\breve{g}_{ij}\
\breve{R}\ =\ \breve{T}_{ij}\ ,\nonumber\\
\nonumber\\
\breve{R}_{ij}-\frac{1}{2}\breve{g}_{ij}\ \breve{R}=0\ .
\end{eqnarray}
A notação $(\ \breve{} \ )$ indica que o objeto será calculado com
respeito à métrica fundo $\breve{g}_{ij}$, ou seja, o
objeto não é perturbado.

O escalar de Ricci será obtido quando contraímos o tensor de Ricci
$\breve{R}_{ij}$ com $\breve{g}^{ij}$. Entretanto ele também pode
ser calculado quando contraímos a equação (\ref{pg01}) com a
métrica. No caso do espaço-tempo vazio, o escalar de Ricci será
\begin{eqnarray}
\label{pg01a}
\breve{g}^{ij}\breve{R}_{ij}-\frac{1}{2}
\breve{g}^{ij}\breve{g}_{ij}\
\breve{R}\ =0\ ,\nonumber\\
\nonumber\\
\breve{R}-\frac{1}{2}\delta^{i}_{i}\
\breve{R}\ =0\ ,\nonumber\\
\nonumber\\
\breve{R}\ =0\ .
\end{eqnarray}
Conseqüentemente nossa equação (\ref{pg01}) se reduz a
\begin{eqnarray}
\label{pg01b} \breve{R}_{ij}=0\ .
\end{eqnarray}
A equação (\ref{pg01b}) será a equação que perturbaremos.

Quando estivermos estudando o caso de buracos negros na brana
nossa equação de Einstein será um pouco diferente devido à
influência do {\it bulk}.

Para uma perturbação em $1^{a}$ ordem a nova métrica pode ser
escrita como
\begin{eqnarray}
\label{pg02} q_{ij}=\breve{g}_{ij}+ \ h_{ij}\ , \qquad
q^{ij}=\breve{g}^{ij}-\ h^{ij}, \qquad q_{ib}\
q^{bj}&=&\delta_{i}^{j}+ \mathcal{O}(h^2),
\end{eqnarray}
onde $h_{ij} \ll \ \breve{g}_{ij}$. Como essa perturbação é
pequena os termos de $2^{a}$ ordem, $\mathcal{O}(h^{2})$, podem
ser desprezados por serem muito menores que a perturbação. As
derivadas covariantes das quantidades perturbadas serão efetuadas
com os símbolos de Christoffel $\breve{\Gamma}^{a}_{ij}$ não
perturbados. Entretanto para calcular o tensor de Ricci perturbado
precisaremos dos símbolos de Christoffel perturbados
$\Gamma^{a}_{ij}$.

Os símbolos de Christoffel perturbados são dados por
\begin{eqnarray}
\label{pg03}
\Gamma^{a}_{ij}=\frac{1}{2}q^{ab}(q_{bi,j}+q_{bj,i}-q_{ij,b})=
\breve{\Gamma}^{a}_{ij}+\delta \Gamma^{a}_{ij}+\mathcal{O}(h^2),
\end{eqnarray}
onde o termo $\delta \Gamma^{a}_{ij}$ pode ser escrito
da seguinte maneira
\begin{eqnarray}
\label{pg04} \delta
\Gamma^{a}_{ij}=\frac{1}{2}\breve{g}^{ab}(h_{bi;j}+h_{bj;i}-h_{ij;b}).
\end{eqnarray}
O tensor de Ricci perturbado é dado pelas equações
\begin{eqnarray}
\label{pg05} \delta R_{ij}= \delta \Gamma^{a}_{ia;j}-\delta
\Gamma^{a}_{ij;a}\ .
\end{eqnarray}
Assim, substituindo a nova métrica $q_{ij}$ na equação
(\ref{pg01b}), e identificando-a com o tensor de Ricci perturbado
teremos as equações de campo perturbadas para o vácuo como sendo
\begin{eqnarray}
\label{pg06}
R_{ij}=\breve{R}_{ij}+\delta R_{ij}=0\ ,\nonumber\\
\nonumber\\
\delta R_{ij}= \delta \Gamma^{a}_{ia;j}-\delta
\Gamma^{a}_{ij;a}=0\ ,\nonumber\\
\nonumber\\
\delta R_{ij}=0\ .
\end{eqnarray}
Essa é a equação que governa a propagação de campos gravitacionais
no vácuo de um espaço-tempo esfericamente simétrico.

No contexto da Relatividade Geral a única solução da equações de
Einstein que satisfaz a condições de um espaço-tempo estático,
vazio e esfericamente simétrico é a solução de Schwarzschild. Entretanto, no
contexto de branas, a influência do {\it bulk} adicionando um
termo de ``matéria"  no tensor momento-energia $T_{ij}$, permite
uma classe de soluções estáticas, vazias e esfericamente simétricas.
Tomaremos mais adiante a métrica de Schwarzschild como exemplo
para explicitarmos as equações que governam a propagação de campos
gravitacionais neste espaço-tempo. Neste trabalho restringiremos
nossa atenção à componente axial das perturbações gravitacionais.
Para o caso de Schwarzschild, por exemplo, as perturbações
gravitacionais polares
são bem explicadas e desenvolvidas em \cite{Chandra,Edelstein}.\\
\\
{\bf Perturbações gravitacionais axiais}\\
\\
O cálculo de perturbações gravitacionais é um assunto que já foi
amplamente estudado no contexto da Relatividade Geral. Entretanto
para os nossos propósitos dois formalismos são de interesse. O
primeiro deles foi desenvolvido por Regge e Wheeler \cite{Regge} e
o segundo por Chandrasekhar \cite{Chandra}.

Para o nosso cálculo das perturbações gravitacionais axiais em uma
métrica dada pela equação (\ref{pg00}), utilizaremos o método
desenvolvido por Regge e Wheeler.

Como a simetria do nosso problema é esférica podemos decompor o
nosso tensor $h_{ij}$ em uma parte angular, em termos de
harmônicos esféricos tensoriais (ver Apêndices \ref{sec:Dectensorial} e \ref{sec:ConsHET}), e uma parte dependente das
coordenadas $r$ e $t$.

Utilizando os resultados do trabalho Regge e Wheeler podemos ver
que a componente axial da perturbação gravitacional para a métrica
em estudo é dada por
\begin{displaymath}
\left[\begin{array}{cccc}
0& 0& -\frac{h_{0}(t,r)}{sen\theta}\frac{\partial Y_{lm}}{\partial \phi}&\scriptstyle h_{0}(t,r)sen\theta\frac{\partial Y_{lm}}{\partial \theta}\\
0& 0& -\frac{h_{1}(t,r)}{sen\theta}\frac{\partial Y_{lm}}{\partial
\phi}&\scriptstyle h_{1}(t,r)sen\theta\frac{\partial Y_{lm}}{\partial \theta}\\
sim&sim&\scriptstyle h_{2}(t,r)\left[\frac{1}{sen\theta}\frac{\partial^{2}}{\partial \theta
\partial\phi}-\frac{cos\theta}{sen^{2}\theta}\frac{\partial}{\partial \phi}
\right]Y_{lm}&sim\\
sim&sim&\scriptstyle \frac{1}{2}h_{2}(t,r)\left[\frac{1}{sen\theta}\frac{\partial^{2}}{\partial
\phi \partial\phi}+cos\theta \frac{\partial}{\partial \theta}-sen\theta
\frac{\partial^{2}}{\partial \theta \partial \theta}\right]Y_{lm}&\scriptstyle -h_{2}(t,r)
\left[sen\theta\ \frac{\partial^{2}}{\partial \theta \partial \phi}-cos\theta \frac{\partial}{\partial \phi}\right]Y_{lm}\\
\end{array}\right]\ .
\label{pg14}
\end{displaymath}
Trabalharemos com a condição de $m=0$ pois qualquer que seja o
valor de $m$ e $l$ teremos a mesma equação radial. A vantagem de
usarmos $m=0$ é eliminarmos a coordenada $\phi$ dos cálculos das
perturbações.

Isso se dá pois
\begin{eqnarray}
\label{pg15} L_{z}\ Y_{lm}(\theta,\phi)=\ mY_{lm}(\theta,\phi)
\qquad \Longrightarrow \qquad \frac{\partial Y}{\partial
\phi}(\theta,\phi)=0\ .
\end{eqnarray}
O nosso tensor $h_{ij}$ ainda é um tanto quanto complicado.
Portanto para simplificá-lo um pouco utilizaremos o gauge de
Regge-Wheeler para perturbações axiais.

Para isso faremos um deslocamento infinitesimal na coodernada
$x^{a}$ tal que
\begin{eqnarray}
\label{pg16} x'^{a}=x^{a}+\xi^{a}\ , \qquad (\xi^{a}\ll x^{a})
\end{eqnarray}
onde $\xi^{a}$ se transforma como um vetor.

O nosso novo tensor perturbação $h_{ij}^{\phantom{ij}{novo}}$ será
dado por
\begin{eqnarray}
\label{pg17} h_{ij}^{\phantom{ij}{novo}}=
h_{ij}+\xi_{i;j}+\xi_{j;i}\ .
\end{eqnarray}
Desta forma o gauge de Regge-Wheeler que simplifica nossa
perturbação axial para $l$ e $m$ arbitrários será
\begin{eqnarray}
\label{pg18} \xi^{0}=0,\quad \xi^{1}=0, \quad
\xi^{i}=\Lambda(r,t)\epsilon^{ij}\frac{\partial}{\partial
x^{j}}Y_{lm}(\theta,\phi), \qquad (i,j=3,4).
\end{eqnarray}
Substituindo o gauge (\ref{pg18}) na equação (\ref{pg17}),
adotando uma dependência temporal harmônica $e^{-i\omega t}$ e
fazendo $m=0$ sem perda de generalidade, nosso tensor resultante
para perturbações axiais será\\
\begin{displaymath}
h_{ij}^{\phantom{ij}{axiais}}=\left[\begin{array}{cccc}
0&0&0&h_{0}(r)\\
0&0&0&h_{1}(r)\\
0&0&0&0\\
sim&sim&0&0
\end{array}\right] \times e^{-i\omega t}\ sen\theta \frac{\partial}{\partial
\theta}P_{l}(cos\theta)\ .
\end{displaymath}
Substituindo o tensor $h_{ij}^{\phantom{ij}{axiais}}$, na equação
(\ref{pg06}), teremos como resultado três equações acopladas. Nós
utilizaremos o resultado obtido por Edelstein e Vishveshwara
\cite{Edelstein}, pela generalidade por eles adotada e porque as
equações obtidas por Regge e Wheeler não satisfaziam as equações
de Einstein.

Assim as componentes não-nulas da equação (\ref{pg06}) são
\begin{eqnarray}
\label{pg20}
\delta R_{\theta\phi}&=& -\frac{1}{2}\left(i\omega \frac{h_{0}}{A}+B\left[\frac{1}{2}\frac{d}{dr}(ln(A/B))h_{1}+\frac{dh_{1}}{dr}\right]\right)\nonumber\\
\nonumber\\
&\times& \left(cos\theta \frac{d}{d\theta}-sen\theta\frac{d^{2}}{d\theta^{2}}\right)P_{l}(cos\theta)e^{-i\omega t}=0;\\
\nonumber\\
\nonumber\\
\label{pg21}
\delta R_{r\phi}&=&\frac{1}{2}\bigg\{\bigg[\frac{\ell(\ell+1)}{r^{2}}-\frac{\omega^{2}}{A}+\frac{1}{Br}\left(\frac{d}{dr}(ln(B/A))-\frac{2}{r}\right)\bigg]h_{1}\nonumber\\
\nonumber\\
&+&\frac{i\omega}{A}\left(\frac{dh_{0}}{dr}-\frac{2h_{0}}{r}\right)\bigg\}\times sen\theta\frac{dP_{l}}{d\theta}(cos\theta)e^{-i\omega t}=0;\\
\nonumber\\
\nonumber\\
\label{pg22}
\delta R_{t\phi}&=&-\bigg\{\frac{1}{2B}\frac{d^{2}h_{0}}{dr^{2}}+i\omega\left[\frac{1}{2B}\frac{dh_{1}}{dr}+\frac{h_{1}}{Br}-\frac{1}{4B}\frac{d}{dr}(ln(AB))h_{1}\right]\nonumber\\
\nonumber\\
&-&\frac{1}{4B}\frac{d}{dr}(ln(AB))h_{1}\frac{dh_{0}}{dr}+\left(\frac{1}{Br}\frac{d}{dr}(ln(A))-\frac{\ell(\ell+1)}{2r^2}\right)h_{0}\bigg\}\nonumber\\
\nonumber\\
&\times& sen\theta\frac{dP_{l}}{d\theta}(cos\theta)e^{-i\omega
t}=0.
\end{eqnarray}
Essas equações podem ser simplificadas se levarmos em conta as propriedades dos
Polinômios de Legendre $P_{l}(cos\theta)$, ou seja, se analisarmos os índices de
multipolo $l$.

Para $l = 0$ temos que $P_{0}=1$ e por conseqüência, a parte
angular das três equações acima são identicamente nulas.

Para $l = 1$ temos que $P_{1}=-sen\theta$. Nesse caso a equação
$\delta R_{\theta \phi}$ é identicamente nula.

Como esses índices de multipolo $l=0,1$ são menores que o valor do
spin da perturbação $s=2$ eles não são modos dinâmicos, ou seja,
eles não evoluem no tempo. Eles correspondem a quantidades
conservadas.

Uma perturbação gravitacional com índice $l=0$ descreve uma
mudança na massa do buraco negro. Já uma perturbação com índice
$l=1$ corresponde a um deslocamento além de gerar um pequeno
incremento de momento angular, fazendo o buraco negro girar.

Nosso interesse se concentra naqueles índices de multipolo que são
capazes de se propagar por um intervalo de tempo suficientemente grande
de modo que possamos medir tal onda com nossos detectores.
Portanto apenas multipolos com $l\geq 2$ são relevantes.

Para $l\geq 2$ temos $P_{l\ \geq 2}\ \neq 0$. Desta forma os
fatores entre chaves nas equações devem se anular para que elas
sejam satisfeitas, resultando em três equações
radiais.

A parte radial da equação $\delta R_{t\phi}=0$ pode ser escrita
como uma combinação das outras duas equações se a seguinte condição
for satisfeita
\begin{eqnarray}
\label{pg23}
\frac{d^2}{dr^2}\left(ln\
\frac{A}{B}\right)&+&\frac{1}{2}\left(\frac{d}{dr}\left(ln\
\frac{A}{B}\right)\right)^2-\frac{1}{2}\left[\left(\frac{d}{dr}ln\
A\right)^2-\left(\frac{d}{dr}ln\ B \right)^2\right]+\nonumber\\
\nonumber\\
&+&\frac{1}{r}\left(\frac{d}{dr}\left(ln\ A-3\ ln\
B\right)\right)=0\ .
\end{eqnarray}
Se a equação (\ref{pg23}) for satisfeita poderemos eliminar
$h_{0}$ das equações de $1^{a}$ ordem e obter uma única equação
radial de $2^{a}$ ordem
\begin{eqnarray}
\label{pg24}
\frac{d^2h_{1}}{dr^2}+\left[\frac{3}{2}\frac{d}{dr}\left(ln\
\frac{A}{B}\right)-\frac{2}{r}\right]\frac{dh_{1}}{dr}&+&\Bigg[
\frac{1}{2}\frac{d^2}{dr^2}\left(ln\
\frac{A}{B}\right)+\frac{1}{2}\left(\frac{d}{dr}\left(ln\
\frac{A}{B}\right)\right)^2\nonumber\\
\nonumber\\
&-&B\frac{\ell(\ell+1)}{r^2}+\omega^2\frac{B}{A}+\frac{2}{r^2}\Bigg]h_{1}=0.
\end{eqnarray}
Caso contrário teremos duas equações de $2^{a}$ ordem acopladas
que devem ser resolvidas para $h_{0}$ e $h_{1}$.

Neste momento é necessário que especifiquemos nossa métrica para
prosseguir no desenvolvimento dos cálculos. Como a métrica de
Schwarzschild satisfaz a equação (\ref{pg23}) tomemo-a como
exemplo.

A métrica de Schwarzschild é dada por
\begin{eqnarray}
\label{pgSch} ds^2=-A(r)dt^2+B(r)dr^2+r^2d\Omega^2\ ,
\end{eqnarray}
onde
\begin{eqnarray}
\label{pgSch1}
A(r)=\left(1-\frac{2M}{r}\right), \qquad B(r)=\frac{1}{\left(1-\frac{2M}{r}\right)}.
\end{eqnarray}
Se definirmos
\begin{eqnarray}
\label{pg25} Q(r)=\left(1-\frac{2M}{r}\right)\frac{h_{1}}{r}\ ,
\end{eqnarray}
e a coordenada tartaruga como
\begin{eqnarray}
\label{pg26} dr^{*}=\frac{1}{\left(1-\frac{2M}{r}\right)}\ dr \qquad \Longrightarrow \qquad
\frac{d}{dr^{*}}=\left(1-\frac{2M}{r}\right)\frac{d}{dr},
\end{eqnarray}
e substituirmos na equação (\ref{pg24}) juntamente com os valores de $A(r)$ e $B(r)$
obteremos uma equação tipo-Schroedinger dada por
\begin{eqnarray}
\label{pg27}
\frac{d^2Q}{dr^{*}}+\left(\omega^2-V_{eff}^{axial}\right)Q=0\ ,
\end{eqnarray}
onde
\begin{eqnarray}
\label{pg28}
V_{ef}^{axial}=\left(1-\frac{2M}{r}\right)\frac{\ell(\ell+1)}{r^2}-\frac{3}{r}\frac{d}{dr^{*}}\left(1-\frac{2M}{r}\right).
\end{eqnarray}
Essa é nossa equação para perturbação gravitacional axial para o
buraco negro de Schwarzschild. Essa equação também é conhecida
como {\it equação de Regge-Wheeler}.
\newpage
\chapter[Estabilidade e Modos Quasi-normais]{Estabilidade e Modos Quasi-Normais de um Buraco
Negro}
{\it As surpresas de uma viagem tornam-na mais emocionante.
Entretanto não podemos correr o risco de cruzar uma ponte e a
mesma cair. Não podemos continuar nossa jornada sem saber se nossa
ponte é estável, segura. Que características da ponte podem nos
dizer sobre sua estabilidade? Suas cordas? Seu piso?}

\section{Estabilidade}
Como foi visto no capítulo anterior, a evolução de campos no
exterior de um buraco negro causa perturbações no espaço-tempo.
Portanto podemos nos perguntar como essas perturbações afetam o
próprio buraco negro.

Assim uma pergunta merece atenção. Buracos negros são estáveis,
quando submetidos a algum tipo de perturbação?

Para responder tal pergunta é necessário saber se soluções da
equações de Einstein tipo buraco negro são estáveis, ou seja, se
eles existem realmente e podem ser perturbados.

O início desses estudos sobre estabilidade de buracos negros
remontam ao ano de 1957 com o trabalho de Regge e Wheeler sobre a
estabilidade da singularidade de Schwarzschild \cite{Regge}.

Neste trabalho Regge e Wheeler analisaram se, perturbações
gravitacionais sobre o buraco negro de Schwarzschild, cresciam
exponencialmente com o tempo. Isso não foi observado do que eles
concluiram que o buraco negro de Schwarzschild era estável sobre
aquele tipo de perturbação. Quando estivermos estudando buracos
negros sobre a brana, utilizaremos o mesmo critério de
estabilidade, ou seja, estudaremos se as perturbações crescem
exponencialmente no tempo. Mas por enquanto voltemos nossa atenção
ao caso de buracos negros em (3+1) dimensões.\\
Como estudado anteriormente, quase todas as equações que governam
a evolução de campos no exterior de um buraco negro esfericamente
simétrico podem ser escritas na forma de uma equação diferencial
de $2^{a}$ ordem do tipo-Schroedinger quando assumimos uma
dependência temporal do tipo $e^{-i\omega t}$. Desta dependência
temporal podemos ver que se $\omega$ é puramente imaginário
$(\omega=i\alpha)$, a exponencial $e^{+\alpha t}$ cresce com o
tempo causando a instabilidade. Nessa descrição o estudo das
perturbações de um buraco negro é análogo ao estudo do problema de
propagação de onda em uma barreira de potencial em mecânica
quântica.

Quando tratamos as equações de perturbação dessa maneira o termo
$\omega^2$  faz o papel da energia da onda na equação de
Schroedinger. A quantidade resultante de
$(\omega^2-V(r))$ pode ser interpretada como o potencial efetivo.

Assim nossa atenção se reduz ao estudo da equação
\begin{eqnarray}
\label{est01}
\frac{d^2Q}{dr^{*}}+\left(\omega^2-V(r)\right)Q=0.
\end{eqnarray}
sobre um dado potencial $V(r)$, que depende do tipo da perturbação
realizada sobre o buraco negro , com condições de contorno
apropriadas. Portanto a forma do potencial $V(r)$ tem papel
decisivo na estabilidade do buraco negro.\\
Tomemos o caso do buraco negro de Schwarzschild novamente como
exemplo.\\
Para perturbações gravitacionais axiais, o potencial $V(r)$ é dado
por
\begin{eqnarray}
\label{est02}
V(r)=\left(1-\frac{2M}{r}\right)\left[\frac{\ell(\ell+1)}{r^2}-\frac{6M}{r^3}\right]\
,
\end{eqnarray}
e o comportamento da solução geral é dado por \cite{Regge}
\begin{eqnarray}
\label{est03} Q&\sim&
c_{1}e^{i\delta}\left(\frac{r}{2M}-1\right)^{2i\omega
M}+c_{1}e^{-i\delta}\left(\frac{r}{2M}-1\right)^{-2i\omega
M}\ \ r\rightarrow 2M,\\
\nonumber\\
Q&\sim& c_{2}sen(\omega r+\eta) \qquad \qquad \qquad \qquad \qquad
\qquad \qquad \ \ r \rightarrow \infty.
\end{eqnarray}
Investigando o comportamento do potencial $V(r)$ podemos ver que
ele é igual a zero sobre o raio de Schwarzschild cresce até um
máximo em $(r=3M)$ e então cai a zero novamente para grandes
valores de $r$. Em outras palavras ele é um potencial positivo
definido, ou seja, em nenhum momento ao longo do intervalo
estabelecido $r \in (2M,\infty)$, o valor do potencial muda de
sinal. Esse comportamento divide a nossa região de estudo em $3$
regimes diferentes.

No regime 1, onde $\omega^2 > V(r)$, a solução é oscilatória em
todo o espaço, ou seja, temos uma onda gravitacional se propagando
livremente.

No regime 2, $\omega^2 < V(r)$, ou seja, $\omega^2$ é ainda
positivo, mas menor que a altura da barreira. Nesse regime temos
duas condições diferentes onde as ondas gravitacionais podem
diminuir ou crescer exponencialmente na região da barreira.

A condição $2a$ corresponde a ondas gravitacionais que nunca
escapam para grandes valores de $r$, ou seja, elas passam pela
barreira mas decaem rapidamente com o aumento de $r$.

A condição $2b$ corresponde a ondas gravitacionais que atravessam
a barreira e emergem maiores do lado de fora da barreira. Estas
duas soluções são como no caso do regime 1, ou seja, são ondas
gravitacionais que se propagam livremente.

O regime 3 será quando $\omega^2 < 0$. Neste regime se $\omega$
for puramente imaginário teremos instabilidades. A solução
aceitável que cai a zero para grandes valores de $r$, também cai
a zero em $r=2M$. Portanto não há possibilidade de colá-la
suavemente a uma solução ``na outra metade da barreira de
potencial". Assim podemos concluir que não existe solução instável
para ondas geradas pelas perturbações gravitacionais axiais, ou
seja, o buraco negro de Schwarzschild é estável sobre perturbações
gravitacionais axiais.

De modo um pouco mais rigoroso podemos estabelecer um critério de
estabilidade através de uma integral tipo-energia. Para um buraco
negro com simetria esférica e componentes $g_{tt}$ e $g_{rr}$
dependentes apenas da coordenada radial $r$ podemos utilizar a
equação (\ref{pe26}). Tomemos a derivada temporal do complexo
conjugado da função de onda $\partial_{t} \bar{R}$,
 multipliquemos pela equação (\ref{pe26}) e somemos a equação
resultante ao complexo conjugado dessa equação. Teremos então
\begin{eqnarray}
\label{est04} \frac{\partial \bar{R}}{\partial
t}\left(-\frac{\partial^2 R}{\partial t^2}+ \frac{\partial^2
R}{\partial r_{*}^2}- VR\right)+ \frac{\partial R}{\partial
t}\left(-\frac{\partial^2 \bar{R}}{\partial t^2}+ \frac{\partial^2
\bar{R}}{\partial r_{*}^2}- V\bar{R}\right)=0\ .
\end{eqnarray}
Usaremos então as seguintes relações:
\begin{eqnarray}
\label{est05}
\frac{\partial}{\partial t}\left|\frac{\partial R}{\partial t}\right|^2&=&\frac{\partial R}{\partial t}\ \frac{\partial^2\bar{R}}{\partial t^2}+\frac{\partial \bar{R}}{\partial t}\ \frac{\partial^2 R}{\partial t^2};\\
\nonumber\\
\frac{\partial}{\partial t}|R|^2&=&\frac{\partial \bar{R}}{\partial t}\ R+\frac{\partial R}{\partial t}\ \bar{R};\\
\nonumber\\
\frac{\partial}{\partial t}\left|\frac{\partial R}{\partial r_{*}}\right|^2&=&\frac{\partial R}{\partial r_{*}}\  \frac{\partial^2\bar{R}}{\partial t\partial r_{*}}+\frac{\partial \bar{R}}{\partial r_{*}}\ \frac{\partial^2 R}{\partial t\partial r_{*}};\\
\nonumber\\
\frac{\partial}{\partial r_{*}}\left(\frac{\partial \bar{R}}{\partial t}\ \frac{\partial R}{\partial r_{*}}\right)&=&\frac{\partial R}{\partial r_{*}}\ \frac{\partial^2\bar{R}}{\partial t\partial r_{*}}+\frac{\partial \bar{R}}{\partial t}\ \frac{\partial^2 R}{\partial r_{*}^2};\\
\nonumber\\
\frac{\partial}{\partial r_{*}}\left(\frac{\partial R}{\partial
t}\ \frac{\partial \bar{R}}{\partial
r_{*}}\right)&=&\frac{\partial \bar{R}}{\partial r_{*}}\
\frac{\partial^2 R}{\partial t\partial r_{*}}+\frac{\partial
R}{\partial t}\ \frac{\partial^2 \bar{R}}{\partial r_{*}^2}.
\end{eqnarray}
Relacionando as equações acima e substituindo na equação
(\ref{est04}), teremos
\begin{eqnarray}
\label{est06} -\frac{\partial}{\partial
t}\left\{\left|\frac{\partial R}{\partial
t}\right|^2+\left|\frac{\partial R}{\partial
r_{*}}\right|^2+V|R|^2\right\}=-\frac{\partial}{\partial
r_{*}}\left(\frac{\partial R}{\partial t}\ \frac{\partial
\bar{R}}{\partial r_{*}}+\frac{\partial \bar{R}}{\partial t}\
\frac{\partial R}{\partial r_{*}}\right)\ .
\end{eqnarray}
Se integrarmos a equação (\ref{est06}) em relação a $r_{*}$ sobre
o intervalo de $-\infty$ a $+\infty$ o termo a direita da equação
será nulo e assim teremos a seguinte condição
\begin{eqnarray}
\label{est07} \frac{\partial}{\partial
t}\left[\int_{-\infty}^{+\infty}\left(\left|\frac{\partial
R}{\partial t}\right|^2+\left|\frac{\partial R}{\partial
r_{*}}\right|^2+V|R|^2\right)\ dr_{*}\right]&=&0\ ,\nonumber\\
\nonumber\\
\int_{-\infty}^{+\infty}\left(\left|\frac{\partial R}{\partial
t}\right|^2+\left|\frac{\partial R}{\partial
r_{*}}\right|^2+V|R|^2\right)\ dr_{*}&=&\ C_{1}\ .
\end{eqnarray}
Como queremos uma condição de estabilidade, precisamos que a
constante $C_{1}$ seja positiva e finita.

Se observarmos a equação (\ref{est07}) podemos ver que os termos
$|R|^2$ e $|\frac{\partial R}{\partial r_{*}}|^2$ são limitados e
positivos. Conseqüentemente se quisermos que $C_{1}$ seja positiva
e finita devemos avaliar o comportamento do potencial $V$.

Se $V$ for positivo definido, o termo $|\frac{\partial R}{\partial
t}|^2$ deve ser positivo e finito para que $C_{1}$ também o seja,
o que acaba excluindo todas as soluções de $R(t,r)$ que crescem
exponencialmente com o tempo. Conseqüentemente o sistema será
estável.
Se o potencial $V$ apresentar alguma região onde ele possa
assumir valores negativos nada podemos dizer a respeito da
estabilidade do
sistema pois a constante $C_{1}$ pode assumir qualquer valor.

Por exemplo, se assumirmos uma dependência temporal $e^{-i\omega
t}$ teremos
\begin{eqnarray}
\label{est08} R(r,t)&=&R(r)e^{-i\omega t}\ ,\\
\nonumber\\
\left|\frac{\partial R}{\partial t}\right|^2&=&\frac{\partial
R}{\partial t}\frac{\partial \bar{R}}{\partial
t}=(-i\omega)(i\omega)|R(r)|^2=\omega^2|R(r)|^2\ .\\
\end{eqnarray}
Sendo $V$ positivo definido, $\omega^2$ deve ser positivo para que
$C_{1}$ seja positiva e limitada. Se $\omega=i\alpha$ fosse
puramente imaginário $\omega^2=-\alpha^2$ seria negativo o que
faria com que não pudessemos garantir que $C_{1}$ fosse positiva e
limitada. Neste caso a soluçao cresce exponencialmente com o tempo
$e^{\ \alpha t}$.
\section{Modos quasi-normais}
Em nossa busca, por uma confirmação experimental da existência de
buracos negros em nosso Universo, o estudo dos modos quasi-normais
tem um papel fundamental.

Esse estudo surge como uma ferramenta extremamente útil para dar
base teórica à análise de ondas gravitacionais uma vez que as
freqüências quasi-normais de oscilação de um buraco negro estão
contidas em tais ondas.

Como os modos quasi-normais estão intimamente ligados a
propriedades tais como, a carga $Q$, a massa $M$ e o momento
angular $L$ do buraco negro, eles podem ser utilizados para
identificar se um buraco negro tem rotação ou se está
eletricamente carregado.

Os modos quasi-normais de oscilação possuem esse nome pois
diferentemente dos modos normais eles não são estacionários e
possuem duração limitada. A parte imaginária da freqüência
quasi-normal amortece a oscilação limitando-a no tempo.

Formalmente falando, modos quasi-normais são definidos como
soluções das equações de perturbações, que possuem freqüências
características complexas e satisfazem condições de contorno
específicas. Ver Nollert \cite{Nollert}, Kokkotas \cite{Kokkotas}
e Cardoso \cite{Cardoso} para maiores detalhes.

Esse comportamento de amortecimento é um tanto quanto estranho uma
vez que não existe nenhum material nem dentro nem fora do buraco
negro capaz de causar tal amortecimento.

O formalismo utilizado no estudo de modos-normais para sistemas
oscilatórios será então aplicado ao estudo dos modos quasi-normais
para observarmos em que condições tal estranho comportamento é
observado.

Na análise de modos normais de um sistema oscilatório, geralmente
temos um sistema de equações diferenciais ordinárias com condições
de contorno para que o efeito da oscilação se anule fora de uma
região finita do espaço. No nosso caso teremos a equação
diferencial
\begin{eqnarray}
\label{mqn01}
\frac{d^2Q}{dr^{*}}+\left(\omega^2-V(r)\right)Q=0,
\end{eqnarray}
governando nossas perturbações. Entretanto, perturbações em
buracos negros geram ondas gravitacionais que se propagam por todo
espaço impedindo-nos de impor um condição de modo que ela se anule
fora de uma região finita.

Portanto, devido ao caráter especial dos buracos negros, imporemos as seguintes condições de contorno:
\begin{itemize}
\item nada sai do infinito espacial.

\item nada sai do horizonte de eventos do buraco negro.
\end{itemize}
Essas condições geram soluções conhecidas como puramente
emergentes {\it (outgoing solutions)}. Dizendo de outra maneira
queremos ter apenas ondas gravitacionais, entrando no horizonte de
eventos, e entrando no infinito espacial. Essas exigências fazem
sentido pois queremos estudar a resposta da métrica fora do buraco
negro e não queremos que nenhuma onda gravitacional vinda do
infinito continue a perturbar o buraco negro.

Devemos por isso avaliar o comportamento da nossa equação
(\ref{mqn01}) nos limites quando $r_{*}\rightarrow \pm \infty$ para
obtermos o comportamento da nossa solução sobre os contornos do
problema.

Novamente, de acordo com a forma do potencial $V(r)$, nossas
condições gerarão soluções distintas.

Se aplicarmos tais condições para a equação que governa a
perturbação gravitacional axial de Schwarzschild  que pode ser
representada pela equação (\ref{mqn01}) e o potencial $V(r)$ dado
pela equação (\ref{est02}) teremos que
\begin{eqnarray}
\label{mqn02}
Q(\omega,r_{*})\sim e^{i\omega r_{*}} \qquad quando\
r_{*}\rightarrow -\infty,\\
\nonumber\\
Q(\omega,r_{*})\sim e^{-i\omega r_{*}} \qquad quando\
r_{*}\rightarrow +\infty.
\end{eqnarray}
Sob essas condições as perturbações do buraco negro de
Schwarzschild adquirem o comportamento agora não tão estranho
descrito acima que caracterizam os modos quasi-normais.

A maioria dos estudos desenvolvidos nesta área se utiliza de
soluções numéricas para a equação (\ref{mqn01}). Ching, {\it et al} \cite{Ching} obtiveram algumas soluções analíticas para comportamentos assintóticos. Recentemente Fiziev \cite{Fiziev} obteve uma solução analítica para a equação de Regge-Wheeler com o uso das funções de Heun.

Neste trabalho nós utilizaremos do cálculo númerico para obter a
solução da equação (\ref{mqn01}) para os buracos negros
estudados.

O estudo de modos quasi-normais tem se mostrado muito útil também
no estudo de propriedades quânticas e termodinâmicas de buracos
negros tais como a temperatura Hawking \cite{Hod3} e área quântica
mínima \cite{Beken4, Karlucio} do buraco negro.

Estudos de perturbações de sistemas com constante cosmológica
\cite{Abdalla,Wang} e com carga \cite{Saa} têm mostrado a existência
de modos quasi-normais de oscilação.

Em um trabalho recente Maartens \cite{Maartens1} propõem a
utilização de modos quasi-normais como uma ferramenta na detecção
de dimensões extras. Neste trabalho é realizada uma espectroscopia
da onda gravitacional emitida por uma corda negra ``{black
string}" onde é possível identificar os modos massivos do gráviton
com os picos obtidos, gerando assim um mecanismo capaz de
identificar uma ``impressão digital" da dimensão extra.
\newpage
\chapter[Termodinâmica de Buracos Negros]{Termodinâmica de Buracos Negros}
{\it Após a travessia da ponte, façamos uma merecida parada para observar a paisagem e aquecer um café. }\\
\\
A possível relação existente entre a mecânica de buracos negros e
a termodinâmica ordinária surgiu na verdade de uma possível
violação da termodinâmica por processos físicos envolvendo buracos
negros.
\section{Leis da Termodinâmica de Buracos Negros}
Wheeler notou que a existência de buracos negros na teoria clássica
da gravitação contradizia a segunda lei da termodinâmica que diz
que a entropia de uma sistema isolado nunca pode diminuir qualquer
que seja o processo físico envolvido.

Imaginemos, por exemplo, um buraco negro que acaba de absorver um
corpo quente que possui uma certa quantidade de entropia. Um
observador localizado no infinito nos dirá que a entropia total,
do mundo acessível à observação dele diminuiu, uma vez que nenhuma
informação sobre a entropia do corpo absorvido escapa do buraco
negro.

Outro problema que surgia da teoria clássica da gravitação era que
a temperatura de um buraco negro era zero absoluto. Esse fato
descartava qualquer possível relação entre uma temperatura física
e algum parâmetro do buraco negro.

Portanto utilizando apenas argumentos clássicos a relação entre a
mecâni-ca de buracos negros e a termodinâmica ordinária tornava-se
inviável. Foi necessário então, a ajuda de mecanismos quânticos
tais como efeito termal de criação de partículas originado do
estudo de propagação de campos quânticos no exterior de buracos
negros e reformulações mais precisas das leis da termodinâmica
para consolidar-se de
forma mais rigorosa a relação entre essas duas áreas.

Em 1973, Bekenstein \cite{Beken1} foi o primeiro a perceber uma
relação entre leis satisfeitas por buracos negros e a
termodinâmica ordinária. Ele notou que o teorema da área de
buracos negros \cite{Hawking2}, da Relatividade Geral clássica,
que declara que a área $A$ de um buraco negro nunca diminui por
qualquer que seja o processo físico
\begin{eqnarray}
\label{termo1} \Delta A \geq 0,
\end{eqnarray}
era análogo à declaração da segunda lei da termodinâmica que diz
que a entropia total $S$ de um sistema fechado nunca diminui
qualquer que seja o processo físico
\begin{eqnarray}
\label{termo2}
 \Delta S \geq 0.
\end{eqnarray}
Bekenstein propôs então que a área do buraco negro deveria ser
interpretada como uma entropia física.

Pouco tempo depois a analogia entre a termodinâmica e certas leis
da física de buracos negros foi desenvolvida sistematicamente por
Bardeen, Hawking e Carter \cite{BHC}. Eles mostraram que em
Relatividade Geral a gravidade superficial, $\kappa$, de um buraco
negro estacionário deve ser constante sobre o horizonte de
eventos.

Eles notaram que este resultado era análogo a declaração da Lei
Zero da termodinâmica que diz que a temperatura, $T$, deve ser
uniforme sobre um corpo em equilíbrio térmico. Eles demonstraram
também a Primeira Lei da mecânica de buracos negros que, para o
vácuo, declara que a diferença de massa, $M$, de área, $A$, e de
momento angular, $J$, de dois buracos negros estacionários
próximos, deve ser
\begin{eqnarray}
\label{termo3} \delta M=\frac{1}{8\pi}\kappa \delta A+\Omega
\delta J\ ,
\end{eqnarray}
onde $\Omega$ é a velocidade angular do horizonte de eventos. Eles
perceberam que esta lei é análoga à primeira lei da termodinâmica,
que declara que a diferença na energia, $E$, na entropia, $S$ e em
outros parâmetros de dois estados próximos do equilíbrio térmico
de um sistema é dado por
\begin{eqnarray}
\label{termo4} \delta E=T \delta S+ ``termos \ de\  trabalho".
\end{eqnarray}
Fazendo uma comparação entre a lei zero, a primeira e segunda leis
da termodinâmica e as correspondentes leis da física de buracos
negros podemos ver que as quantidades análogas são
\begin{eqnarray}
\label{termo5} E \longleftrightarrow M,\qquad T\longleftrightarrow
\alpha \kappa, \qquad S\longleftrightarrow \frac{A}{8\pi \alpha},
\end{eqnarray}
onde $\alpha$ é uma constante indeterminada. Entretanto, essa
analogia entre as quantidades pareciam ser apenas coincidências
matemáticas desprovida de qualquer significado físico pois
classicamente, a relação entre $T$ e $\kappa$ não fazia sentido, o
que acabava inviabilizando também a relação entre a entropia e a área.

Com isso a idéia dessa analogia perdeu força. Foi necessário a
utilização de conceitos quânticos para resolver esses problemas.

Em 1975 o cenário modificou-se com a importante descoberta de
Hawking \cite{Hawking1}. Em seus estudos sobre propagação de
campos quânticos no exterior de buracos negros ele descobriu que
devido ao efeito quântico de criação de partículas, um buraco
negro irradiava para o infinito todas as espécies de partículas
com um espectro de corpo negro perfeito, à uma temperatura
\begin{eqnarray}
\label{termo6} T_{H}=\frac{\kappa}{2\pi}\ ,
\end{eqnarray}
chamada Temperatura Hawking do buraco negro. Com isso, a relação
entre $T$ e $\kappa$ ganhou sentido físico pois $\kappa/2\pi$ é a
temperatura física do buraco negro.

Isso confirmou a idéia de que um buraco negro estacionário era um
estado de equilíbrio térmico.

Havia ainda o problema da diminuição da entropia quando um corpo
era absorvido pelo buraco negro. Esse problema foi contornado por
Bekenstein ao reformular a segunda lei da termodinâmica. Ele
propôs uma segunda lei generalizada \cite{Beken3,Beken1} onde a
entropia total do sistema (buraco negro $+$ objeto) nunca diminui
qualquer que seja o processo físico. Com esses desenvolvimentos
restam agora poucas dúvida de que as leis da física de buracos
negros correspondem à leis da termodinâmica aplicada a sistemas
constituídos de buracos
negros.

As leis da termodinâmica de buracos negros podem ser assim
declaradas\\
\\
{\bf Lei Zero}\\
\\
{\it A gravidade superficial $\kappa$ de um buraco negro
estacionário é constante em todo lugar, sobre a superfície do
horizonte de
eventos.}\\
\\
{\bf Primeira Lei}\\
\\
{\it Quando um sistema que possui um buraco negro muda de um
estado estacionário para outro, sua massa muda por}
\begin{eqnarray}
\label{termo7} dM=T\ dS_{BN}+\Omega_{BN}\
dJ_{BN}+\Phi_{BN}dQ+\delta q
\end{eqnarray}
{\it onde $dJ_{BN}$ e $dQ$ são as mudanças no momento angular
total e da carga elétrica do buraco negro, respectivamente. O
termo $\delta q$ é a contribuição na massa total devida à mudança
na distribuição estacionária de matéria fora do buraco negro e
$\Phi_{BN}$ é o potencial elétrico do buraco negro.}\\
\\
{\bf Segunda Lei}\\
\\
{\it Em qualquer processo clássico, a área de um buraco negro,
$A$, e conseqüentemente sua entropia $S_{BN}$}, nunca diminui:
\begin{eqnarray}
\label{termo8} \Delta S_{BN}\geq 0 \qquad onde\qquad S_{BN}=\frac{A}{4}.
\end{eqnarray}
{\it A lei de não diminuição da área do buraco negro nos diz que a
fração da energia interna do buraco negro que não pode ser
extraída aumenta com o tempo.}\\
\\
{\bf Segunda Lei Generalizada - SLG}\\
\\
{\it Qualquer que seja o processo físico envolvendo buracos
negros, a entropia total generalizada, $\widetilde{S}$, nunca
diminui:}
\begin{eqnarray}
\label{termo9} \Delta \widetilde{S}=\Delta S_{BN}+\Delta S_{m}\geq
0.
\end{eqnarray}
{\it onde $S_{BN}$ é a entropia do buraco negro e $S_{m}$ a
entropia da radiação e matéria fora do buraco negro.}\\
\\
Existe ainda uma lei análoga à terceira lei da termodinâmica, mas
omitiremos tal lei por não a utilizarmos durante o trabalho.

\section{Cálculo do limite superior da
entropia de um corpo caindo em um BN Esfericamente Simétrico}

Assumindo a validade da (SLG) uma nova condição sobre a entropia
$S_{m}$ da radiação e da matéria fora do buraco negro é estabelecida.

Quando um corpo de energia $E$ e raio efetivo $R$ é absorvido por
um buraco negro, a área da superfície do buraco negro aumenta por
um valor $8\pi ER$ \cite{Beken2}. Se quisermos que esse aumento
seja mínimo, a (SLG) será violada se a entropia do corpo $S_{m}$
for maior que $2\pi ER$. Dessa forma um limite superior para
entropia $S_{m}$ é gerado. Vamos calculá-lo formalmente.

Vamos considerar um corpo neutro com massa de repouso $m$ e raio
próprio $R$ caindo em um buraco negro esfericamente simétrico onde
a métrica é dada por
\begin{eqnarray}
\label{entropia01}
ds^2=g_{ij}dx^{i}dx^{j}=-A(r)dt^2+B(r)dr^2+r^2d\Omega^2
\end{eqnarray}
e cujo horizonte de eventos é dado pela seguinte condição
\begin{eqnarray}
\label{entropia01a} g_{tt}|_{r=r_{h}}=0 \qquad \Longrightarrow
A(r_{h})=0 \qquad para \ r=r_{h}.
\end{eqnarray}
A gravidade superficial $\kappa$ para esse buraco negro sobre o
horizonte de eventos é dada por
\begin{eqnarray}
\label{entropia01b} \left. \kappa
=\frac{1}{2\sqrt{|g_{tt}|g_{rr}}}\
\frac{d|g_{tt}|}{dr}\right|_{r=r_{h}}\ .
\end{eqnarray}
As constantes de movimento associadas
a $t$ e a $\phi$ são
\begin{eqnarray}
\label{entropia02} E=\pi_{t}=g_{tt}\dot{t},\qquad J=-\pi_{\phi}=
g_{\phi\phi}\dot{\phi}.
\end{eqnarray}
O quadrado da massa é dado por
\begin{eqnarray}
\label{entropia03} m^{2}=-\pi_{i}\pi^{i}.
\end{eqnarray}
Sem perda de generalidade consideraremos apenas o movimento
equatorial do corpo, isto é, $\theta=\pi/2$.

A equação quadrática para a energia conservada $E$ do corpo é dada
por
\begin{eqnarray}
\label{entropia04}
 - m^{2}=\frac{E^2}{g_{tt}}+\frac{J^2}{r^2},\\
\nonumber\\
\alpha E^2-2\beta E +\sigma=0\ ,
\end{eqnarray}
com
\begin{eqnarray}
\label{entropia05}
\alpha&=&1\nonumber\\
\nonumber\\
\beta&=& 0\\
\nonumber\\
\sigma&=&g_{tt}\left(\frac{J^2}{r^2}+m^2\right)\nonumber\ .
\end{eqnarray}
Como nosso corpo possui um raio próprio não-nulo, $R$ e sabendo
que a equação (\ref{entropia04}) descreve o movimento do centro de
massa do corpo precisamos calcular o novo ponto de captura do
corpo pelo buraco negro.

Isso se faz necessário pois quando parte do corpo já atravessou o
horizonte de eventos o centro de massa ainda está do lado de fora.
Portanto o novo ponto de captura será dado por $r=r_{h}+\delta$
onde $\delta$ é dado por
\begin{eqnarray}
\label{entropia06} \int_{r_{h}}^{r_{h}+\delta(R)}\ \sqrt{g_{rr}}\
dr=\ R\ .
\end{eqnarray}
Integrando a equação (\ref{entropia06}) teremos uma expressão para
$\delta$. Se estivessemos em um espaço-tempo chato, teríamos $\delta=R$.

Resolvendo a equação (\ref{entropia04}) para a energia e
calculando seu valor sobre o ponto de captura $r=r_{h}+\delta$
teremos a energia de captura que é dada por
\begin{eqnarray}
\label{entropia07} \left. E_{cap}=
\sqrt{-g_{tt}\left(\frac{J^2}{r^2}+m^2\right)}\
\right|_{r=r_{h}+\delta}\ .
\end{eqnarray}
O valor $E_{cap}$ que minimiza o aumento da área da superfície do
buraco negro é obtido quando $J=0$ de modo que
\begin{eqnarray}
\label{entropia08} \left. E_{min}=m\ \sqrt{-g_{tt}}\
\right|_{r=r_{h}+\delta}\ .
\end{eqnarray}
Da Primeira Lei da Termodinâmica de Buracos Negros sabemos que
\begin{eqnarray}
\label{entropia09} dM=\frac{\kappa}{2}\ dA_{r}
\end{eqnarray}
onde $A_{r}=A/4\pi$ e $dM=E_{min}$ é a mudança na massa do buraco
negro devido a assimilação do corpo. Assim usando a equação
(\ref{entropia01b}) nós deveremos obter que
\begin{eqnarray}
\label{entropia10} dA_{r}=2mR\ ,
\end{eqnarray}
a menos que as componentes da métrica produzam alguma correção.
Hod mostrou que para corpos carregados eletricamente esse limite é
diferente e depende da carga $Q$ \cite{Hod2} do corpo absorvido.

Assumindo a validade da Segunda Lei Generalizada,
$S_{bn}(M+dM)\geq S_{bn}(M)+S_{m}$, obtemos um limite superior
para a entropia $S_{m}$ associada a um corpo com energia própria
$E$, absorvido pelo buraco negro, que é dado por
\begin{eqnarray}
\label{entropia11}
\frac{A+dA}{4}&\geq& \frac{A}{4}+S_{m}\nonumber\\
\nonumber\\
\frac{dA}{4}&\geq& S_{m}\nonumber\\
\nonumber\\
\frac{4\pi \ dA_{r}}{4}&\geq& S_{m}\nonumber\\
\nonumber\\
S_{m}&\leq& 2\pi ER\ .
\end{eqnarray}
Portanto se a (SLG) for válida, um observador localizado no
infinito que vê um corpo caindo em um buraco negro esfericamente
simétrico dirá que houve um aumento da entropia generalizada e
nenhuma violação é constatada.

Esse limite é portanto considerado universal no sentido de que ele
depende apenas dos parâmetros do corpo absorvido sendo
independente dos parâmetros que caracterizam o buraco negro.
\newpage

\chapter[Mundo Brana]{Mundo Brana}
{\it Após a parada, sigamos agora para fora dos nossos limites, rumo ao espaço desconhecido.}\\
\\
O uso de dimensões extras na tentativa de uma descrição unificada
do nosso mundo físico não é uma idéia nova. Em meados dos anos 20
Kaluza e Klein \cite{Kaluza} conseguiram unificar o eletromagnetismo e a
gravitação utilizando-se de um dimensão extra espacial. Desde
então a utilização de dimensões extras em modelos físicos tem
ganhado força devido a sua capacidade de explicar muitos fenômenos
físicos e gerar novas descrições do mundo em que vivemos. A
maioria dos desenvolvimentos recentes em Física Teórica baseiam-se
no uso de dimensões extras, onde se destaca a Teoria de Cordas.\\
\\
{\bf Teoria de Cordas}\\
\\
Atualmente sabemos que sob altas energias, a Teoria da
Relatividade Geral falha pois perde sua capacidade de predição.
Para um entendimento de processos físicos nesse nível de energia
faz-se necessário o desenvolvimento de uma teoria quântica da
gravitação.

Uma candidata promissora, capaz de gerar uma descrição unificada
dos campos de matéria do Modelo Padrão e da gravitação, é a Teoria
de Cordas.

A Teoria de Cordas prevê que o espaço-tempo em que vivemos é
composto de 10 dimensões sendo 1 dimensão temporal e 9 espaciais.
Esse espaço-tempo 10-dimensional recebe o nome de {\it bulk}. Como
até agora, o mundo físico observável é 4-dimensional, ela assume
que as outras 6 dimensões espaciais são pequenas e compactas, ou
seja, estão enroladas de modo que um observador não seja capaz de
observá-las diretamente mas seja capaz de sentir seus efeitos no
seu mundo observável. Uma analogia interessante utilizada para
explicar essa idéia é a analogia da mangueira de jardim.

Se olharmos de perto, para uma mangueira de jardim, estendida no
chão, observaremos que ela é um objeto tridimensional com largura,
espessura e comprimento. Agora se observarmos essa mesma mangueira
de muito longe veremos apenas uma linha estendida no chão e assim
nada podemos dizer se aquele objeto possui mais dimensões.

Outra previsão da Teoria de Cordas é a existência de subvariedades
$p+1$ dimensionais do espaço-tempo 10-dimensional que recebem o
nome de $p$-branas. Um importante subconjunto das $p$-branas são
as $D$-branas.

$D$-branas são hipersuperfícies que funcionam como condições de
contorno do espaço-tempo 10-dimensional onde a extremidade de
cordas abertas são presas.

Na Teoria de Cordas os grávitons são representados por cordas
fechadas que podem se propagar pelo {\it bulk}. Já os campos de
calibre do modelo padrão são representados pelas extremidades de
cordas abertas que podem se mover apenas sobre as $p$-branas. Isto
é, os campos de calibre permanecem confinados às $p$-branas.\\
\\
{\bf Problema da Hierarquia}\\
\\
O problema da Hierarquia surge da grande diferença entre as
escalas eletrofraca $E_{ef}\sim 10^{3}\ GeV$ e a escala de Planck
4-dimensional $E_{p}\sim 10^{19}\ GeV$. Essa grande diferença
torna-se um problema quando se faz necessário um tratamento
unificado da gravitação e da mecânica quântica para um determinado
processo físico.\\
\\
Os conceitos apresentados acima inspiraram os modelos de Mundo
Brana conhecidos. Atualmente existem diversos modelos de Mundo
Brana com características bem específicas.\\
Apresentaremos na seção a seguir dois modelos bem conhecidos: ADD e Randall-Sundrum.
Estes dois modelos impulsionaram avanços na física de partículas e mesmo na cosmologia.
\section{Um olhar sobre o Mundo Brana}
Essa seção sobre Mundo Brana foi inspirada nos trabalhos
realizados por Michele F. Ferraz \cite{Michele} e Adenauer G. Casali \cite{Adenauer}.
\subsection{Modelo ADD}
A motivação do modelo ADD \cite{ADD} repousa na solução do
problema da hierarquia fazendo uso de dimensões extras.

Este modelo é constituído de um espaço-tempo $(4+d)$ dimensional,
onde campos de matéria estão confinados a uma brana 4-dimensional
de ``espessura" $E_{ef}^{\phantom{ef}{-1}}$ e apenas a gravidade é
livre para se propagar por todo o {\it bulk} $(4+d)$ dimensional.
As $d$ dimensões extras são compactas com raio $R$, ou seja,
limitada por condições de contorno que as tornam finitas, e podendo
assumir topologias diversas. Para o caso de $d=2$ por exemplo,
podemos ter uma 2-esfera ou um toros com raio interno nulo. Neste
modelo existe a possibilidade de campos de matéria sairem para o
{\it bulk} carregando energia.

Entretanto, carga elétrica, por exemplo, deve se conservar não
podendo assim escapar para o {\it bulk}.

Seguindo esses conceitos, a proposta do trabalho \cite{ADD}, é
mostrar que para um mundo com 6 dimensões o problema da hierarquia
é solucionado e pode ser confirmado experimentalmente em breve.

Podemos mostrar pela Lei de Gauss em $(4+d)$ dimensões que o
potencial em torno de uma massa pontual $M$ é dado por
\begin{eqnarray}
\label{MBADD01}
V(r)= \frac{G_{4+d}M}{r^{1+d}}
\end{eqnarray}
onde $G_{4+d}$ é a escala gravitacional em $(4+d)$ dimensões.

Tomando $r$ como a distância entre um ponto qualquer e a massa
pontual, e $R$ a largura das dimensões extras temos que se $r<R$,
então o potencial $V(r)$ será
\begin{eqnarray}
\label{MBADD02} V(r)\sim r^{-(1+d)}\ .
\end{eqnarray}
A essa distância as dimensões extras influem no comportamento do
potencial como pode ser visto.

Para uma distância $r>R$ vemos que as dimensões extras não
influenciam fortemente o potencial de modo que recuperamos o
potencial 4-dimensional
\begin{eqnarray}
\label{MBADD03} V(r)=\frac{G_{4+d}M}{rR^{d}}\ .
\end{eqnarray}
Disto podemos escrever a escala gravitacional 4-dimensional como
\begin{eqnarray}
\label{MBADD04}
G_{4}=\frac{G_{4+d}}{R^{d}}
\end{eqnarray}
ou em termos da Massa de Planck,
\begin{eqnarray}
\label{MBADD05}
M_{PL}^{2}=M_{4+d}^{d+2}R^{d}.
\end{eqnarray}
Se supusermos $M_{4+d}$ da ordem da escala eletrofraca
$(E_{ef}\sim 10^{3}\ GeV)$ e $M_{PL}\sim 10^{19}\ GeV$ teremos da
equação (\ref{MBADD05}) que
\begin{eqnarray}
\label{MBADD06}
R=\frac{M_{PL}^{2/d}}{M_{4+d}^{\frac{d+2}{d}}}=10^{(32/d)-17}cm.
\end{eqnarray}
Se tivermos $d=1$ o raio da dimensão extra será $R\sim 10^{15} cm$
o que implicaria em violações na gravitação em distâncias da ordem
do sistema solar.

Se adotarmos entretanto $d=2$ com uma topologia tipo toros o valor
do raio das dimensões extras será
\begin{eqnarray}
\label{MBADD07}
R\approx 1mm.
\end{eqnarray}
Como experimentos, que medem a validade do potencial
$V(r)\sim\frac{1}{r}$, tem precisão até a ordem de $0,1\ mm$, fica
aberta a possibilidade de termos uma nova lei do potencial. Novos
experimentos com maior precisão poderão nos dizer como se comporta
o potencial gravitacional a pequenas distâncias e assim validar ou
não as hipóteses do modelo ADD.
\subsection{Modelos Randall-Sundrum}
Como o modelo ADD, os modelos de Randall-Sundrum (RS)
\cite{RS1,RS2} foram inicialmente propostos no intuito de fornecer
um novo mecanismo para solucionar o problema da hierarquia.
Entretanto muitas outras aplicações foram dadas a esses modelos
posteriormente, tais como o estudo de cosmologia de branas,
buracos negros em branas, etc.

Diferentemente do modelo ADD os modelos RS se utilizam apenas de
uma única dimensão extra espacial. Os modelos RS constituem um
espaço-tempo 5-dimensional onde apenas a gravidade pode
propagar-se livrevemente, podendo ter duas (RS-I) ou uma brana
(RS-II), onde os campos de matéria permanecem
confinados.\\
Nestes modelos a métrica é não fatorável de modo que a métrica de
Minkowski é multiplicada por um fator de deformação que depende da
dimensão extra. É esse fator que atenua a gravidade e corrige a
escala de Planck resolvendo o Problema da Hierarquia.\\
\\
{\bf RS-I}\\
\\
O modelo RS-I \cite{RS1} consiste em um espaço 5-dimensional com
constante cosmológica $\Lambda_{5D}$ negativa, denominado {\it
bulk}, onde a quinta dimensão é espacial e compacta. A dimensão
extra é representada pela coordenada $y$ e possui raio de
compactificação $r_{c}$ de modo que $-\pi r_{c}\leq y\leq \pi
r_{c}$.

Duas branas são utilizadas neste modelo e correspondem a pontos
fixos no {\it bulk}, ou seja, são condições de contorno do espaço
5-dimensional.

Adotaremos a simetria $Z_{2}$, $y\rightarrow -y$, ou seja, as
branas funcionarão como espelhos dividindo o {\it bulk} em duas
partes idênticas. Por isso podemos restringir o intervalo que
nossa coordenada extra cobre para $0\leq y \leq \pi r_{c}$. Podemos
então localizar a brana visível, que corresponde ao nosso universo
4-dimensional, sobre $y=0$ e a segunda brana escondida sobre o
ponto $y=\pi r_{c}$. Assim teremos as métricas sobre as branas
como sendo
\begin{eqnarray}
\label{MBRSI01}
g^{ab}_{vis}&=&G^{\mu\nu}(y=0)\delta^{a}_{\mu}\delta^{b}_{\nu},\\
\nonumber\\
g^{ab}_{esc}&=&G^{\mu\nu}(y=\pi
r_{c})\delta^{a}_{\mu}\delta^{b}_{\nu},
\end{eqnarray}
com $\mu,\nu=0\dots 4$ e $a,b=0\dots 3$.\\
O cenário adotado neste modelo é descrito pela seguinte ação clássica
\begin{eqnarray}
\label{MBRSI02} S=S_{5D}+S_{vis}+S_{esc}\ ,
\end{eqnarray}
onde
\begin{eqnarray}
\label{MBRSI03}
S_{5D}&=&\int \ \sqrt{|G|}\left[-\Lambda_{5D}+2M^{3}_{5D}R\right]\ d^{4}xdy\ ,\\
\nonumber\\
S_{vis}&=&\int \ \sqrt{|g_{vis}|}\left[\mathcal{L}_{vis}-2\lambda_{vis}\right]\delta (y)\ d^{4}xdy\ ,\\
\nonumber\\
S_{esc}&=&\int \
\sqrt{|g_{esc}|}\left[\mathcal{L}_{esc}-2\lambda_{esc}\right]\delta
(y-\pi r_{c})\ d^{4}xdy\ ,
\end{eqnarray}
onde $M_{5D}$ é a escala de Planck em 5-D, $\lambda_{vis}$ e
$\lambda_{esc}$ são as tensões nas branas que atuam como origens
gravitacionais na ausência de excitações de partículas. As
Lagrangeanas das branas são dadas por $\mathcal{L}$ entretanto
seus detalhes não são relevantes para a determinação da métrica
5-dimensional.

Efetuando uma variação na ação (\ref{MBRSI02}) e lembrando que,
nas fronteiras $\delta g_{ab}=\delta
G_{\mu\nu}\delta_{a}^{\mu}\delta_{b}^{\nu}$ e o termo $\delta
R_{\mu\nu}$ pode ser deprezado, obteremos as seguintes equações de
campo
\begin{eqnarray}
\label{MBRSI04}
\sqrt{|G|}(R^{\mu\nu}-\frac{1}{2}G^{\mu\nu}R)=-\frac{1}{4M^{3}_{5D}}\left[2\sqrt{|g_{vis}|}\ g_{vis}^{ab}\ \delta_{a}^{\mu}\delta_{b}^{\nu}\ \lambda_{vis}\ \delta(y)\right.\nonumber\\
\nonumber\\
+\sqrt{|G|}G^{\mu\nu}\Lambda_{5D}+\left. 2\sqrt{|g_{esc}|}\
g_{esc}^{ab}\ \delta_{a}^{\mu}\delta_{b}^{\nu}\ \lambda_{esc}\
\delta(y-\pi r_{c})\right]\ .
\end{eqnarray}
Usando como {\it Ansatz} a métrica não fatorável
\begin{eqnarray}
\label{MBRSI05} ds^2=e^{2\sigma(y)}\eta_{ab}dx^{a}dx^{b}+dy^2
\end{eqnarray}
e aplicando-o na equação (\ref{MBRSI04}) teremos as equações
\begin{eqnarray}
\label{MBRSI06}
6\sigma'^2&=&-\frac{\Lambda_{5D}}{4M_{5D}^{3}},\\
\nonumber\\
\label{MBRSI07}
3\sigma''&=&\frac{\lambda_{vis}\delta(y)}{2M_{5D}^{3}}+\frac{\lambda_{esc}\delta(y-\pi r_{c})}{2M_{5D}^{3}}.
\end{eqnarray}
A solução consistente com a simetria do problema $y\rightarrow -y$ para a equação (\ref{MBRSI06}) é
\begin{eqnarray}
\label{MBRSI08} \sigma =\pm
|y|\sqrt{-\frac{\Lambda_{5D}}{24M_{5D}^{3}}}.
\end{eqnarray}
Nesta etapa podemos observar que a única solução fisicamente
aceitável é uma constante cosmológica $\Lambda_{5D}$ negativa o
que caracteriza um {\it bulk}  do tipo $AdS_{5}$. Portanto o
espaço-tempo entre as duas branas é uma fatia de uma geometria
$AdS_{5}$.

Derivando a equação (\ref{MBRSI08}) e lembrando que supusemos uma
condição de periodicidade para a coordenada $y$ temos
\begin{eqnarray}
\label{MBRSI09}
\sigma'' =\pm 2r_{c}\sqrt{-\frac{\Lambda_{5D}}{24M_{5D}^{3}}}\left\{\delta(y)-\delta(y-\pi r_{c})\right\}.
\end{eqnarray}
Comparando-a com a equação (\ref{MBRSI06}) vemos que, para que a
métrica (\ref{MBRSI05}) seja solução das equações de campo devemos
ter as seguintes relações \cite{Michele,Adenauer}
\begin{eqnarray}
\label{MBRSI10}
\lambda_{vis}=-\lambda_{esc}=\mp 12M_{5D}^{3}k;\qquad \Lambda_{5D}=-24 M_{5D}^{3}k^2.
\end{eqnarray}
Desta forma podemos escrever a métrica (\ref{MBRSI05}) como
\begin{eqnarray}
\label{MBRSI11} ds^2=e^{\pm 2k|y|}\eta_{ab}dx^{a}dx^{b}+dy^2\ .
\end{eqnarray}
O sinal positivo na métrica (\ref{MBRSI11}) corresponde ao modelo
RS-I onde o crescimento da exponencial na dimensão extra é barrado
pela brana escondida. Neste modelo a brana visível possui tensão
negativa e a brana escondida tensão positiva. Esse fato é
problemático, pois induz uma gravidade repulsiva na brana visível.
Esse efeito é observado pois a constante de Newton 4-dimensional
depende de $\lambda_{vis}$
\begin{eqnarray}
\label{MBRSI12} G_{4D}=\frac{\lambda_{vis}}{6M_{5D}^{6}}\ .
\end{eqnarray}
\\
{\bf RS-II}\\
\\
O modelo RS-II \cite{RS2} é construído de maneira muito similar ao
modelo RS-I exceto pelos fatos de que a dimensão extra não é mais
compacta mas sim infinita $(r_{c}\rightarrow \infty)$ e que temos
agora uma única brana com tensão positiva.

A simetria $Z_{2}$ é mantida neste modelo e o {\it bulk} ainda é
descrito por um espaço-tempo $AdS_{5}$. O método para a dedução
dessa solução é semelhante ao descrito anteriormente de modo que
omitiremos os detalhes.

Para o modelo RS-II a métrica que satisfaz as equações de campo
será
\begin{eqnarray}
\label{MBRSII01}
ds^2=e^{- 2k|y|}\eta_{ab}dx^{a}dx^{b}+dy^2,
\end{eqnarray}
que corresponde a equação (\ref{MBRSI11}) com sinal negativo. Esse
modelo apresenta algumas vantagens sobre o modelo RS-I. Neste
modelo a tensão positiva na brana fornece uma gravidade atrativa
como esperado, além de possuir um estado ligado do gráviton
4-dimensional não massivo que não se propaga para a dimensão
extra.

Uma propriedade válida aos modelos RS é que eles podem ser
escritos na forma conformalmente plana se fizermos a seguinte
transformação de coordenadas
\begin{eqnarray}
\label{MBRSII02}
z=l\  e^{\ y/l}
\end{eqnarray}
onde $l$ é o raio $AdS$.
Desta forma a métrica (\ref{MBRSII01}), por exemplo, será escrita como
\begin{eqnarray}
\label{MBRSII03}
ds^2=\frac{\ell}{z}\left[\eta_{ab}dx^{a}dx^{b}+dz^2\right].
\end{eqnarray}
Disso podemos notar que se a métrica de Minkowski for trocada por
qualquer métrica do tipo Ricci plana, ela ainda será solução das
equações (\ref{MBRSI04}).\\
Com isso podemos substituir qualquer solução da Relatividade Geral
em um cenário do tipo RS podendo assim estudar o comportamento de
buraco negros sobre a brana (ver teorema Campbell-Magaard
\cite{campbell}).

\subsection{Equação de Einstein projetada sobre a brana}
Como o intuito deste trabalho é estudar buracos negros sobre a
brana precisamos saber como a influência do {\it bulk} afeta as
equações de movimento na brana.

A projeção das equações de Einstein em uma 3-brana foi estudada
por Maeda, Sasaki e Shiromizu \cite{Maeda}.

A idéia básica é utilizar as equações de Gauss-Codazzi que projeta
a curvatura 5-dimensional ao longo da brana. Apresentaremos esses
cálculos com certo detalhe pois consideramo-lo útil no
entendimento das seções seguintes quando estudaremos buracos
negros sobre a brana.

A princípio, não suporemos nenhuma propriedade especial para o {\it
bulk}, a não ser que ele é formado por um espaço-tempo
5-dimensional, onde temos 4 dimensões espaciais e 1 dimensão
temporal.

Neste trecho do trabalho a notação será alterada. Os indices $i,j$
correm de $0\dots 4$ representando o espaço-tempo 5-dimensional.
Quando tivermos objetos 4-dimensionais os índices $i,j$ correm de
$0\dots 3$.

No Mundo Brana nosso universo 4-D é descrito por uma 3-brana,
hipersuperfície $\mathcal{M}$ com (3+1) dimensões e métrica
induzida $q_{ij}$ embebida em um espaço-tempo 5-dimensional $V$ e
métrica $g_{ij}$. O vetor unitário $n^{i}$ normal a brana
$\mathcal{M}$ é do tipo espaço de modo que a relação entre a
métrica do {\it bulk} e a métrica da brana será
\begin{eqnarray}
\label{MBEB01} q_{ij}=g_{ij}-n_{i}n_{j}\ ; \qquad
q^{ij}=g^{ij}-n^{i}n^{j}\ ;\qquad
q_{i}^{\phantom{i}{j}}=g_{i}^{\phantom{i}{j}}-n_{i}n^{j}\ .
\end{eqnarray}
Como $n^{i}$ é normal à brana temos também
\begin{eqnarray}
\label{MBEB02}n^{i}n_{i}=1;\qquad g_{ij}\ n^{i}n^{j}=1;\qquad
q_{ij}\ n^{i}=0\ .
\end{eqnarray}
As equações que governam a relação entre a brana e o {\it bulk}
são as equações de Gauss-Codazzi \cite{Maeda}
\begin{eqnarray}
\label{MBEB03}
^{4}R^{\ a}_{\phantom{a}{\ bcd}}=\ ^{5}R^{\
f}_{\phantom{f}{\ ghi}}\ q_{f}^{\phantom{f}{a}}q_{b}^{
\phantom{b}{g}}q_{c}^{\phantom{c}{h}}q_{d}^{\phantom{d}{i}}&+&
K^{a}_{\phantom{a}{c}}K_{bd}-K^{a}_{\phantom{a}{d}}K_{bc},\\
\nonumber\\
\label{MBEB04}
 K_{\phantom{g}{f;\ g}}^{g}-K_{;\ f}&=&\ ^{5}R_{hi}\
n^{i}\ q_{f}^{\phantom{f}{h}}
\end{eqnarray}
onde $K_{ij}=q_{a}^{\phantom{a}{i}}q_{b}^{\phantom{b}{j}}\nabla_{a}n_{b}$
é a curvatura extrínseca de $\mathcal{M}$, cujo o traço é dado por
$K=K_{i}^{i}$ e $(\ ;\ )$ representa a derivada covariante com respeito a
métrica induzida $q_{ij}$.

A partir dessas equações construiremos as equações de Einstein
sobre a brana de modo a levar em consideração a influência
geométrica do {\it bulk}. Para tanto precisamos do tensor de Ricci
e o escalar de Ricci 4-dimensional que podem ser calculados da
equação (\ref{MBEB03}).

Se contrairmos os índices $a$ e $c$ e lembrarmos das propriedades
(\ref{MBEB01}) e (\ref{MBEB02}) nós teremos o tensor de Ricci
4-dimensional
\begin{eqnarray}
^{4}R_{bd}&=&\ ^{5}R^{\ f}_{\phantom{f}{\ ghi}}\
q_{f}^{\phantom{f}{a}}q_{a}^{\phantom{a}{h}}q_{b}^{\phantom{b}{g}}q_{d}^{\phantom{d}{i}}+
K^{a}_{\phantom{a}{a}}K_{bd}-K^{a}_{\phantom{a}{d}}K_{ba}\nonumber\\
\nonumber\\
^{4}R_{bd}&=&\ ^{5}R^{\ f}_{\phantom{f}{\ ghi}}\
(g_{f}^{\phantom{f}{a}}-n_{f}n^{a})(g_{a}^{\phantom{a}{h}}-n_{a}n^{h})\
q_{b}^{\phantom{b}{g}}q_{d}^{\phantom{d}{i}}+ K\
K_{bd}-K^{a}_{\phantom{a}{d}}K_{ba}\nonumber\\
\nonumber\\
^{4}R_{bd}&=&\ ^{5}R^{\ f}_{\phantom{f}{\ ghi}}\
(g_{f}^{\phantom{f}{a}}g_{a}^{\phantom{a}{h}}-n_{f}n^{h})\
q_{b}^{\phantom{b}{g}}q_{d}^{\phantom{d}{i}}+ K\
K_{bd}-K^{a}_{\phantom{a}{d}}K_{ba}\nonumber\\
\nonumber\\
\nonumber\\
\label{MBEB05}
^{4}R_{bd}&=&\ ^{5}R_{gi}\
q_{b}^{\phantom{b}{g}}q_{d}^{\phantom{d}{i}}-\ ^{5}R^{\
f}_{\phantom{f}{\ ghi}}\ n_{f}n^{h}\
q_{b}^{\phantom{b}{g}}q_{d}^{\phantom{d}{i}}+ K\
K_{bd}-K^{a}_{\phantom{a}{d}}K_{ba}
\end{eqnarray}
Calculemos agora o escalar de Ricci 4-dimensional contraindo a
equação (\ref{MBEB05}) com $q^{bd}$
\begin{eqnarray}
\label{MBEB06}
 ^{4}R &=& q^{bd}\ ^{4}R_{bd}\nonumber\\
\nonumber\\
^{4}R&=&\ ^{5}R_{gi}\ q^{gi} -q^{bd}\ ^{5}R^{\ f}_{\phantom{f}{\
ghi}}\ n_{f}n^{h}\ q_{b}^{\phantom{b}{g}}q_{d}^{\phantom{d}{i}}+
q^{bd}\ K\ K_{bd}-q^{bd}\ K^{a}_{\phantom{a}{d}}K_{ba}\nonumber\\
\nonumber\\
\nonumber\\
^{4}R&=&\ ^{5}R_{gi}\ q^{gi} -q^{bd}\ ^{5}R^{\ f}_{\phantom{f}{\
ghi}}\ n_{f}n^{h}\ q_{b}^{\phantom{b}{g}}q_{d}^{\phantom{d}{i}}+
K^{2}- K^{ab}K_{ab}
\end{eqnarray}
Com isso podemos escrever o tensor de Einstein $G_{km}$
4-dimensional como
\begin{eqnarray}
\label{MBEB07}
 ^{4}G_{km}&=&\ ^{4}R_{km}-\frac{1}{2}q_{km}\ ^{4}R\nonumber\\
\nonumber\\
^{4}G_{km}&=&\ ^{5}R_{gi}\
q_{k}^{\phantom{k}{g}}q_{m}^{\phantom{m}{i}}-\ ^{5}R^{\
f}_{\phantom{f}{\ ghi}}\ n_{f}n^{h}\
q_{k}^{\phantom{k}{g}}q_{m}^{\phantom{m}{i}}+ K\
K_{km}-K^{a}_{\phantom{a}{m}}K_{ka}-\nonumber\\
\nonumber\\
&-&\frac{1}{2}q_{km}\left\{\ ^{5}R_{rs}\ q^{rs} -q^{bd}\ ^{5}R^{\
t}_{\phantom{t}{\ rls}}\ n_{t}n^{l}\
q_{b}^{\phantom{b}{r}}q_{d}^{\phantom{d}{s}}+ K^{2}-
K^{jc}K_{jc}\right\}\nonumber\\
\nonumber\\
\nonumber\\
^{4}G_{km}&=&\left(\ ^{5}R_{gi}\
q_{k}^{\phantom{k}{g}}q_{m}^{\phantom{m}{i}}-\frac{1}{2}q_{km}\
^{5}R_{rs}\ q^{rs}\right)-\widetilde{E}_{km}+K\
K_{km}-K^{a}_{\phantom{a}{m}}K_{ka}-\nonumber\\
\nonumber\\
&-&\frac{1}{2}q_{km}\left[ K^{2}- K^{jc}K_{jc}-q^{bd}\ ^{5}R^{\
t}_{\phantom{t}{\ rls}}\ n_{t}n^{l}\
q_{b}^{\phantom{b}{r}}q_{d}^{\phantom{d}{s}}\right]\nonumber\\
\end{eqnarray}
\begin{eqnarray}
^{4}G_{km}&=&\left(\
^{5}R_{gi}-\frac{1}{2}(g_{gi}-n_{g}n_{i})\left[\ ^{5}R-\
^{5}R_{rs}n^{r}n^{s} \right]
\right)q_{k}^{\phantom{k}{g}}q_{m}^{\phantom{m}{i}}-\widetilde{E}_{km}+K\
K_{km}\nonumber\\
\nonumber\\
&-&K^{a}_{\phantom{a}{m}}K_{ka}-\frac{1}{2}q_{km}\left[ K^{2}-
K^{jc}K_{jc}-q^{bd}\ ^{5}R^{\ t}_{\phantom{t}{\ rls}}\ n_{t}n^{l}\
q_{b}^{\phantom{b}{r}}q_{d}^{\phantom{d}{s}}\right]\nonumber\\
\nonumber\\
\nonumber\\
^{4}G_{km}&=&\left(\ ^{5}R_{gi}-\frac{1}{2}g_{gi}\
^{5}R\right)q_{k}^{\phantom{k}{g}}q_{m}^{\phantom{m}{i}}+\frac{1}{2}\
^{5}R_{rs}\ q_{km}\ n^{r}n^{s}-\widetilde{E}_{km}+K\
K_{km}-\nonumber\\
\nonumber\\
&-&K^{a}_{\phantom{a}{m}}K_{ka}-\frac{1}{2}q_{km}\left[ K^{2}-
K^{jc}K_{jc}-\ ^{5}R^{\ t}_{\phantom{t}{\ rls}}\ n_{t}n^{l}\ q^{rs}\right]\nonumber\\
\nonumber\\
\nonumber\\
^{4}G_{km}&=&\ ^{5}G_{gi}\
q_{k}^{\phantom{k}{g}}q_{m}^{\phantom{m}{i}}-\widetilde{E}_{km}+K\
K_{km}-K^{a}_{\phantom{a}{m}}K_{ka}-\frac{1}{2}q_{km}\left[ K^{2}-
K^{jc}K_{jc}\right]\nonumber\\
\nonumber\\
&+&\frac{1}{2}q_{km}\left[\ ^{5}R_{rs}\ n^{r}n^{s}+
\ ^{5}R_{prls}\ n^{p}n^{l}\ q^{rs}\right] \nonumber\\
\nonumber\\
\nonumber\\
^{4}G_{km}&=&\ ^{5}G_{gi}\
q_{k}^{\phantom{k}{g}}q_{m}^{\phantom{m}{i}}-\widetilde{E}_{km}+K\
K_{km}-K^{a}_{\phantom{a}{m}}K_{ka}-\frac{1}{2}q_{km}\left[ K^{2}-
K^{jc}K_{jc}\right]\nonumber\\
\nonumber\\
&+&\frac{1}{2}q_{km}\left[\ ^{5}R_{rs}\ n^{r}n^{s}+\ ^{5}R_{rs}\
n^{r}n^{s}-\ ^{5}R_{rpsl}\ n^{p}n^{l}n^{r}n^{s}\right]
\end{eqnarray}
Como $R_{rpsl}$ é um tensor anti-simétrico e
$n^{p}n^{l}n^{r}n^{s}$ é simétrico, o termo acima $\ ^{5}R_{rpsl}\
n^{p}n^{l}n^{r}n^{s}$ se anula.

Assim teremos o tensor de Einstein 4-dimensional como sendo
\begin{eqnarray}
\label{MBEB08} ^{4}G_{km}&=&\ ^{5}G_{gi}\
q_{k}^{\phantom{k}{g}}q_{m}^{\phantom{m}{i}}-\widetilde{E}_{km}+K\
K_{km}-K^{a}_{\phantom{a}{m}}K_{ka}\nonumber\\
\nonumber\\
&+&q_{km}\ ^{5}R_{rs}\ n^{r}n^{s}-\frac{1}{2}q_{km}\left[ K^{2}-
K^{jc}K_{jc}\right]
\end{eqnarray}
onde
\begin{eqnarray}
\label{MBEB09} \widetilde{E}_{km}=\ ^{5}R^{\ f}_{\phantom{f}{\
ghi}}\ n_{f}n^{h}\ q_{k}^{\phantom{k}{g}}q_{m}^{\phantom{m}{i}}\ .
\end{eqnarray}
Como pretendemos obter as equações de Einstein sobre a brana com a
menor quantidade de objetos 5-dimensionais, precisamos reescrever o
tensor de Einstein 5-dimensional. Para isso usaremos as equações de
Einstein 5-dimensionais
\begin{eqnarray}
\label{MBEB10} ^{5}R_{gi}-\frac{1}{2}g_{gi}\ ^{5}R=k_{5}^{2}\
^{5}T_{gi},
\end{eqnarray}
e do tensor de Weyl $\ ^{5}C_{abcd}$
\begin{eqnarray}
\label{MBEB11} ^{5}C_{abcd}=\
^{5}R_{abcd}&-&\frac{1}{3}\left(g_{ac}\ ^{5}R_{db}-g_{ad}\
^{5}R_{cb}-g_{bc}\ ^{5}R_{da}+g_{bd}\
^{5}R_{ca}\right)\nonumber\\
\nonumber\\
&+&\frac{1}{12}\left(g_{ac}g_{db}-g_{ad}g_{cb}\right)\ ^{5}R\ .
\end{eqnarray}
Da equação (\ref{MBEB10}) obtemos, após alguma álgebra, o escalar de
Ricci e o tensor de Ricci 5-D em termos do tensor momento-energia
$\ ^{5}T_{gi}$
\begin{eqnarray}
\label{MBEB12} ^{5}R=-\frac{2}{3}k_{5}^{2}\ ^{5}T;\qquad
^{5}R_{gi}=\frac{k_{5}^{2}}{3}\left(3\ ^{5}T_{gi}-\
^{5}Tg_{gi}\right)\ .
\end{eqnarray}
Substituindo as equações (\ref{MBEB12}) na equação (\ref{MBEB08})
teremos
\begin{eqnarray}
\label{MBEB13} ^{4}G_{km}&=&k_{5}^{2}\ ^{5}T_{gi}\
q_{k}^{\phantom{k}{g}}q_{m}^{\phantom{m}{i}}-\widetilde{E}_{km}+K\
K_{km}-K^{a}_{\phantom{a}{m}}K_{ka}\nonumber\\
\nonumber\\
&+&q_{km}\frac{k_{5}^{2}}{3}\left(3\ ^{5}T_{rs}-\
^{5}Tg_{rs}\right)n^{r}n^{s}-\frac{1}{2}q_{km}\left[ K^{2}-
K^{jc}K_{jc}\right].
\end{eqnarray}
Usando a equação (\ref{MBEB11}) podemos achar uma relação para
$\widetilde{E}_{km}$ em função do tensor de Weyl,
\begin{eqnarray}
\label{MBEB14} \widetilde{E}_{km}&=&\frac{1}{3}\left(\ ^{5}R_{rs}\
q_{k}^{\phantom{k}{r}}q_{m}^{\phantom{m}{s}}+\ ^{5}R_{ig}\
n^{i}n^{g}\ g_{rs}\
q_{k}^{\phantom{k}{r}}q_{m}^{\phantom{m}{s}}\right)\nonumber\\
\nonumber\\
&-&\frac{1}{12}\ ^{5}R\ g_{rs}\
q_{k}^{\phantom{k}{r}}q_{m}^{\phantom{m}{s}}+\ ^{5}C_{irgs}\
n^{i}n^{g}\ q_{k}^{\phantom{k}{r}}q_{m}^{\phantom{m}{s}}\ .
\end{eqnarray}
Reescrevendo $\widetilde{E}_{km}$ em termos do tensor
momento-energia e seu traço teremos
\begin{eqnarray}
\label{MBEB15} \widetilde{E}_{km}&=&\frac{k_{5}^{2}}{3}\
^{5}T_{rs}\ q_{k}^{\phantom{k}{r}}q_{m}^{\phantom{m}{s}}
-\frac{k_{5}^{2}}{6}\ ^{5}T\ q_{rs}+E_{km}\nonumber\\
\nonumber\\
&+&\frac{k_{5}^{2}}{3}\left(\ ^{5}T_{gi}\ n^{i}n^{g}\right)q_{km}
\end{eqnarray}
onde
\begin{eqnarray}
\label{MBEB16} E_{km}=\ ^{5}C_{irgs}\ n^{i}n^{g}\
q_{k}^{\phantom{k}{r}}q_{m}^{\phantom{m}{s}}\ .
\end{eqnarray}
Substituindo a equação (\ref{MBEB15}) nas equações de Einstein
4-dimensionais teremos
\begin{eqnarray}
\label{MBEB17} ^{4}G_{km}&=&k_{5}^{2}\ ^{5}T_{gi}\
q_{k}^{\phantom{k}{g}}q_{m}^{\phantom{m}{i}}\left(1-\frac{1}{3}\right)+
\frac{k_{5}^{2}}{3}\left(2\ ^{5}T_{rs}n^{r}n^{s}-\frac{\ ^{5}T}{2}
\right)q_{km}\nonumber\\
\nonumber\\
&-&E_{km}+K\
K_{km}-K^{a}_{\phantom{a}{m}}K_{ka}-\frac{1}{2}q_{km}\left[ K^{2}-
K^{jc}K_{jc}\right]\ .
\end{eqnarray}
Portanto as equações de Einstein 4-dimensionais projetadas sobre uma
brana serão
\begin{eqnarray}
\label{MBEB18} ^{4}G_{km}&=&\frac{2k_{5}^{2}}{3}\left(\
^{5}T_{rs}\
q_{k}^{\phantom{k}{r}}q_{m}^{\phantom{m}{s}}+\left[^{5}T_{rs}\
n^{r}n^{s}-\frac{\
^{5}T}{4}\right]q_{km}\right)\nonumber\\
\nonumber\\
&-&E_{km}+K\
K_{km}-K^{a}_{\phantom{a}{m}}K_{ka}-\frac{1}{2}q_{km}\left[ K^{2}-
K^{jc}K_{jc}\right].
\end{eqnarray}
Podemos observar na equação acima que temos, além de termos que
dependem do tensor momento-energia $\ ^{5}T_{km}$, termos que são
puramente geométricos fazendo papel de ``matéria" e curvando o
espaço-tempo. Esse resultado nos permite trabalhar com buracos
negros com ``matéria geométrica", ou seja, uma ``matéria" induzida pelo {\it bulk}.

Se usarmos a equação de Codazzi e a equação de Einstein
5-dimensional teremos
\begin{eqnarray}
\label{MBEB19}
 K_{\phantom{g}{f;\ g}}^{g}-K_{;\ f}&=&\frac{k_{5}^{2}}{3}
 \left(3\ ^{5}T_{hi}-\ ^{5}T\ g_{hi}\
\right)n^{i}\ q_{f}^{\phantom{f}{h}}\nonumber\\
\nonumber\\
K_{\phantom{g}{f;\ g}}^{g}-K_{;\ f}&=&k_{5}^{2}\ ^{5}T_{hi}n^{i}\
q_{f}^{\phantom{f}{h}}-\frac{k^{2}_{5}}{3}\ ^{5}T\ n_{h}\
q_{f}^{\phantom{f}{h}}\nonumber\\
\nonumber\\
K_{\phantom{g}{f;\ g}}^{g}-K_{;\ f}&=&k_{5}^{2}\ ^{5}T_{hi}\
n^{i}\ q_{f}^{\phantom{f}{h}}\ .
\end{eqnarray}
De agora em diante suporemos válidas algumas das propriedades do
mundo brana. Utilizaremos uma única brana e chamaremos a
coordenada extra de $\chi$ tal que $n_{i}dx^{i}=d\chi$ de modo que
a brana se
encontre sobre o ponto $\chi=0$.

Portanto a métrica que descreve o espaço-tempo 5-dimensional será
do tipo
\begin{eqnarray}
\label{MBEB20} ds^2=d\chi^2+q_{ij}dx^{i}dx^{j}\ .
\end{eqnarray}
Em geral, modelos de mundo brana não admitem matéria no {\it bulk}.
Portanto nosso tensor momento-energia 5-dimensional será dado por
\begin{eqnarray}
\label{MBEB21}
\ ^{5}T_{ij}&=&-\Lambda_{5}g_{ij}+S_{ij}\delta(\chi),\\
\nonumber\\
S_{ij}&=&-\lambda q_{ij}+\tau_{ij}
\end{eqnarray}
onde $\Lambda_{5}$ é a constante cosmológica do {\it bulk}, $\lambda$ e $\tau_{ij}$ são a tensão na brana e seu tensor
momento-energia respectivamente.

Como o tensor momento-energia possui um comportamento singular
precisamos saber como a brana se conecta no {\it bulk} uma vez que
ela divide o {\it bulk}  em duas partes quebrando a simetria de
translação na direção da coordenada extra. Semelhante situação
aparece no estudo de uma casca esférica, de modo que existe uma
região de transição no espaço-tempo que deve ter solução de
continuidade. Essa conexão é dada pelas condições de junção de
Israel \cite{Israel}
\begin{eqnarray}
\label{MBEB22}
[q_{ij}]&=&0\\
\nonumber\\
\left[K_{ij}\right]&=&-k_{5}^{2}\left(S_{ij}-\frac{1}{3}q_{ij}S\right)
\end{eqnarray}
onde a operação $[A]$ corresponde a
\begin{eqnarray}
\label{MBEB23} [A]:=\lim_{\chi \rightarrow +0}A-\lim_{\chi
\rightarrow -0}A= A^{+}-A^{-}\ .
\end{eqnarray}
Inspirados nos modelos RS podemos impor simetria $Z_{2}$ no nosso
problema tomando a brana como o ponto fixo. Fazendo isso obtemos
das condições de junção o comportamento da curvatura extrínseca
\begin{eqnarray}
\label{MBEB24}
K_{ij}^{+}=-K_{ij}^{-}=-\frac{1}{2}k_{5}^{2}\left(S_{ij}-\frac{1}{3}q_{ij}S\right)\
.
\end{eqnarray}
Substituindo a equação (\ref{MBEB24}) na equação (\ref{MBEB18})
obteremos as equações de campo sobre a brana na forma
\begin{eqnarray}
\label{MBEB25}
\ ^{4}G_{ij}=-\Lambda_{4}q_{ij}+8\pi G_{N}\tau_{ij}+k_{5}^{2}\pi_{ij}-E_{ij},
\end{eqnarray}
onde
\begin{eqnarray}
\label{MBEB26}
\Lambda_{4}&=&\frac{1}{2}k_{5}^{2}\left(\Lambda_{5}+\frac{1}{6}k_{5}^{2}\lambda^{2}\right),\\
\nonumber\\
G_{N}&=&\frac{k_{5}^{4}\lambda}{48\pi}\ ,\\
\nonumber\\
\pi_{ij}&=&-\frac{1}{4}\tau_{ia}\tau_{j}^{\phantom{j}{a}}+\frac{1}{12}\tau\tau_{ij}+\frac{1}{8}q_{ij}\tau_{ab}\tau^{ab}-\frac{1}{24}q_{ij}\tau^2\
.
\end{eqnarray}
Alguns comentários a respeito do tensor $E_{ij}$ merecem destaque.

O tensor $E_{ij}$ é a projeção do tensor de Weyl 5-dimensional e
carrega a informação de como o campo gravitacional fora da brana
se comporta e atua sobre ela. Portanto ele depende fortemente da
forma do espaço-tempo no {\it bulk}. Ele não é especificado
livremente mas sua divergência é vinculada ao conteúdo de matéria
na brana que modo que através das identidades de Bianchi
contraídas $G_{ij;}^{\phantom{ij;}{i}}=0$ teremos
\begin{eqnarray}
\label{MBEB27}
E_{ij;}^{\phantom{ij;}{i}}=\frac{1}{4}k_{5}^{4}\left[\tau^{ab}(\tau_{ab;j}-\tau_{ja;b})+\frac{1}{3}(\tau_{ij}-q_{ij}\tau)\tau_{;}^{\phantom{;}{i}}\right]\
.
\end{eqnarray}
Podemos decompor $E_{ij}$ em uma parte transversa sem traço
$E_{ij}^{TT}$ e uma parte longitudinal $E_{ij}^{L}$, com a parte
longitudinal sendo determinada completamente pela matéria presente
na brana. Conseqüentemente se $E_{ij}^{TT}$ não estiver presente,
as equações se fecham apenas com elementos presentes na brana.
Entretanto como a parte $E_{ij}^{TT}$ corresponde aos grávitons
nas 5 dimensões, ela vai acabar excitando a matéria presente na
brana o que por sua vez acaba excitando o {\it bulk}. Isto implica
que para resolver completamente as equações de movimento sobre a
brana é necessário resolver ao mesmo tempo as equações no {\it
bulk}.

Uma análise interessante em \cite{Maeda} mostra que o tensor
$E_{ij}$ é negligenciável a baixas energias. Isso nos mostra que
no limite de baixas energias os efeitos do {\it bulk} são
desprezíveis e assim recuperamos a equação de Einstein
convecional. Quando tratarmos de perturbações gravitacionais em
buracos negros em branas discutiremos um pouco mais a respeito do
tensor $E_{ij}$.

\section{Buracos Negros no Mundo Brana}
Como foi discutido no capítulo anterior, existem grandes chances
de que nosso universo seja uma brana imersa em um espaço-tempo de
dimensão mais alta. Portanto, tão importante quanto a verificação
desses argumentos, é o entendimento de estruturas estelares e
buracos negros neste contexto.

Como motivação extra podemos salientar um resultado, que
contribui para a solidificação da correspondência $AdS/CFT$,
obtido em 2005 \cite{CutoffADS} que apresenta evidências
significativas em favor de uma conjectura \cite{citadoemCutoffADS}
que diz que buracos negros 4-D localizados sobre a brana que são
obtidos como solução clássica de um {\it bulk} $AdS_{5}$ são na
verdade buracos negros com correções quânticas.

Diversos avanços no contexto de Mundo Brana tem sido feitos, tais
como os realizados por Hawking \cite{cordanegra} que obteve uma
solução tipo corda negra, Maartens \cite{Germani} que obteve uma
solução que descreve uma estrela de densidade uniforme sobre a
brana, Visser \cite{Visser2} com uma solução que descreve planetas
sólidos sobre a brana, Casadio \cite{Casadio2,Casadio1} com uma
solução tipo ``charuto negro" e soluções tipo buracos negros e
buracos de minhoca e Bronnikov \cite{Bronnikov1,Bronnikov2} com
uma classe geral de buracos negros e buracos de minhoca. Algumas
dessas soluções apresentaram regimes instáveis quando submetidas a
perturbações \cite{gregorilaflamme}, outras ainda não foram
estudadas até agora, o que nos motiva ao estudo da estabilidade
dessas soluções.

Examinando a equação (\ref{MBEB25}) podemos observar que ela
apresenta termos extras no tensor momento-energia que surgem da
presença do {\it bulk}. Isso faz com que nossas equações de
Einstein sobre a brana dependa do {\it bulk} e portanto não formem
um conjunto de equações fechadas se não conhecemos a extensão da
solução no {\it bulk}. Entretanto essa questão pode ser tratada
sob dois aspectos diferentes. O primeiro deles diz respeito a um
teorema sobre imersões de hipersuperfícies em espaços-tempo de
mais alta dimensão. Esse teorema é conhecido como o Teorema de
Campbell-Magaard \cite{campbell}.

Ele declara que qualquer variedade $n$-dimensional pode ser
localmente embebida em um espaço de Einstein $(n+1)$
dimensional.

Uma versão generalizada desse teorema foi obtida por Wesson
\cite{campbell}. Essa versão generalizada inclui a aplicação do
teorema a modelos de Mundo Brana e é construída sobre três
quantidades: a métrica induzida $q_{ij}$, a curvatura extrínseca
$K_{ij}$, e o tensor $E_{ij}$. As equações que descrevem a
dinâmica na brana, obtidas na seção anterior dependem dessas
quantidades. Portanto não podemos especificar arbitrariamente
$q_{ij}$ e $K_{ij}$ e resolver as equações de vínculo. Isso acaba
implicando em restrições sobre as
imersões.

Se adotarmos o modelo RS como base, o que implica em assumirmos
simetria $Z_{2}$ e confinarmos os campos de matéria na brana ,
teremos a restrição de que, especificado o tensor momento-energia
sobre a brana, a métrica induzida é obtida dinamicamente.

Portanto para a versão generalizada do teorema para Mundo Brana
temos que qualquer solução $(3+1)$ dimensional pode ser
representada por uma brana no cenário RS.

O segundo deles é extender a solução obtida na brana para o {\it
bulk}. Essa extensão é realizada de maneira análoga à decomposição
padrão em $(3+1)$ dimensões de variedades globalmente hiperbólicas
na Relatividade Geral, ou seja, o formalismo ADM \cite{Wald}.

Neste caso a solução na brana funciona como uma ``hipersuperfície
inicial" que é evoluída na direção da dimensão extra construindo a
solução 5-dimensional. Esse mecanismo foi utilizado por Casadio
\cite{Casadio2} para construir sua solução tipo ``charuto negro".
Neste trabalho abordaremos apenas o primeiro aspecto.

Portanto ignoraremos a forma dos buracos negros no {\it bulk}
invocando o teorema de Campbell-Magaard. Isso implica que não nos
importaremos com a forma explícita do tensor $E_{ij}$.

A única coisa que precisamos saber a respeito do {\it bulk} é que
ele é solução das equações de Einstein 5-dimensionais para o vácuo e
com cosntante cosmológica $\Lambda_{5}$ que é dada por
\begin{eqnarray}
\label{BNBa}
\mathcal{R}_{\mu\nu}-\frac{1}{2}g_{\mu\nu}\mathcal{R}=\Lambda_{5}\
g_{\mu\nu}
\end{eqnarray}
onde $(\mu,\nu = 0,\dots ,4)$.\\
Utilizando os resultados obtidos na seção anterior podemos
escrever as equações de Einstein projetadas na brana como
\begin{eqnarray}
\label{BNBb} G_{ij}=-\Lambda_{4} q_{ij}+8\pi
G_{N}\tau_{ij}+k_{5}^{4}\pi_{ij}-E_{ij}\ ,
\end{eqnarray}
com
\begin{eqnarray}
\label{BNBc}
\Lambda_{4}&=&\frac{1}{2}k_{5}^{2}\left(\Lambda_{5} +\frac{1}{6}k_{5}^{2}\lambda^2\right)\ ,\\
\nonumber\\
G_{N}&=&\frac{k_{5}^{4}\lambda}{48\pi}\ ,\\
\nonumber\\
\pi_{ij}&=&-\frac{1}{4}\tau_{ia}\tau_{j}^{\phantom{j}
a}+\frac{1}{12}\tau
\tau_{ij}+\frac{1}{8}q_{ij}\tau_{ab}\tau^{ab}-\frac{1}{24}q_{ij}\tau^2\
.
\end{eqnarray}
onde $\Lambda_{4}$ é a constante cosmológica 4-D, $\tau_{ij}$ é o
tensor momento-energia sobre a brana, $G_{N}$  é a constante
gravitacional de Newton, $\pi_{ij}$ são potências quadráticas do
tensor momento-energia e $E_{ij}$ é a projeção ``elétrica" do
tensor de Weyl 5-D $C_{\mu\nu\alpha\beta}$ sobre a brana.

Se introduzirmos coordenadas normais Gaussianas $x^{i}\
(i=0,\dots,3)$ e $z$ (com $z=0$ sobre a brana) e considerarmos
ausência de matéria sobre brana ($\pi_{ij}=\tau_{ij}=0$), vamos
obter como única combinação das equações de Einstein (\ref{BNBa})
que podem ser escritas sobre a brana sem ambigüidades e sem
especificar $E_{ij}$ as seguintes equações de vínculo
\begin{eqnarray}
\label{BNBd} \mathcal{R}_{iz}=0, \qquad R=4\Lambda_{4}\ .
\end{eqnarray}
Deste ponto em diante faremos $\Lambda_{4}=0$ fazendo o ajuste
apropriado entre $\Lambda_{5}$ e a tensão na brana e respeitando
as condições de junção necessárias \cite{Israel}.\\
Se supusermos que nossa solução deve ser estática e possuir
simetria esférica podemos escrever nossa métrica como
\begin{eqnarray}
\label{BNBe} ds^2=-A(r)dr^2+B(r)dr^2 +r^2d\Omega^2\ .
\end{eqnarray}
Desta forma podemos obter a métrica sobre a brana de forma completa solucionando
a equação
\begin{eqnarray}
\label{BNBf}
R=\frac{1}{2}\frac{A''}{A}-\frac{1}{4}\left(\frac{A'}{A}\right)^2
-\frac{1}{4}\frac{A'}{A}\frac{B'}{B}-\frac{1}{r}\left(\frac{B'}{B}
-\frac{A'}{A}\right)-\frac{1}{r^2}(B-1)=0.
\end{eqnarray}
Portanto, para uma dada função $A(r)$ a equação (\ref{BNBf})
sempre terá uma solução de modo que a função $B(r)$ resultante,
será especificada totalmente a menos de uma constante de
integração.

Os buracos negros estudados neste trabalho são dois: BN-CFM e
BN-SM.

A solução BN-CFM foi obtida por Casadio \cite{Casadio1} e de
maneira diferente por Maartens \cite{Germani} como soluçao
exterior de uma estrela na brana e por Bronnikov \cite{Bronnikov2}
como uma solução particular de uma classe
geral de soluções.
\subsection{Buraco Negro tipo CFM}
A métrica que descreve o BN-CFM tem a forma
\begin{eqnarray}
\label{BNB01}
ds^2=-\left(1-\frac{2M}{r}\right)dt^2+\frac{\left(1-\frac{3M}{2r}\right)}
{\left(1-\frac{2M}{r}\right)\left(1-\frac{M\gamma}{2r}\right)}\ dr^2+r^2d\Omega^2\
.
\end{eqnarray}
Apesar da constante $\gamma$ poder assumir qualquer valor positivo
ou negativo nós restringiremos seu valor neste trabalho devido à
mudança da estrutura do buraco negro causada pela mudança no
intervalo de valores de $\gamma$.

\begin{figure}[!ht]
\begin{center}
\epsfig{file = 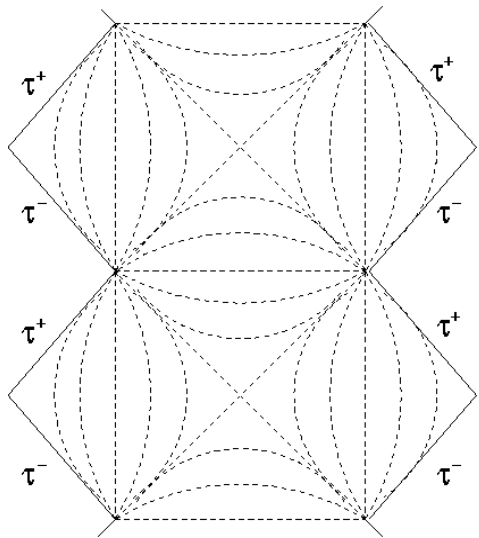, width=0.45 \linewidth, clip=}
\end{center}
\caption{Diagrama de Penrose do BN-CFM para $\gamma >3$.}
\label{fig01a}
\end{figure}

Para valores de $\gamma >4$ a solução CFM descreve um buraco negro
não-singular com um buraco de minhoca dentro do horizonte de
eventos. O diagrama de Penrose dessa solução é o mesmo de um
buraco negro de Kerr não extremo (ver figura \ref{fig01a}).

Para valores de $\gamma < 4$ a solução CFM descreve um buraco
negro tipo Schwarzschild com a singularidade física deslocada.

Deste ponto em diante assumiremos a condição $\gamma <4$.

Se analisarmos as escalares invariantes do BN-CFM, veremos que
\begin{eqnarray}
\label{BNB02}
R&=&0\ ,\\
\nonumber\\
R_{ij}R^{ij}&\sim& \frac{1}{(2r-3M)^4 r^4}\ ,\\
\nonumber\\
R_{ijkl}R^{ijkl}&\sim& \frac{1}{(2r-3M)^4 r^6}\ ,
\end{eqnarray}
ou seja, os escalares explodem para $r=0,3M/2$. Como $r=3M/2 >
r=0$ o espaço compreendido entre $0<r<3M/2$ não pertence ao nosso
universo. Portanto nosso universo é definido pelo intervalo $r \in
(3M/2,\infty)$. Sob $r=2M$ os escalares permanecem regulares.

\begin{figure}[!ht]
\begin{center}
\epsfig{file=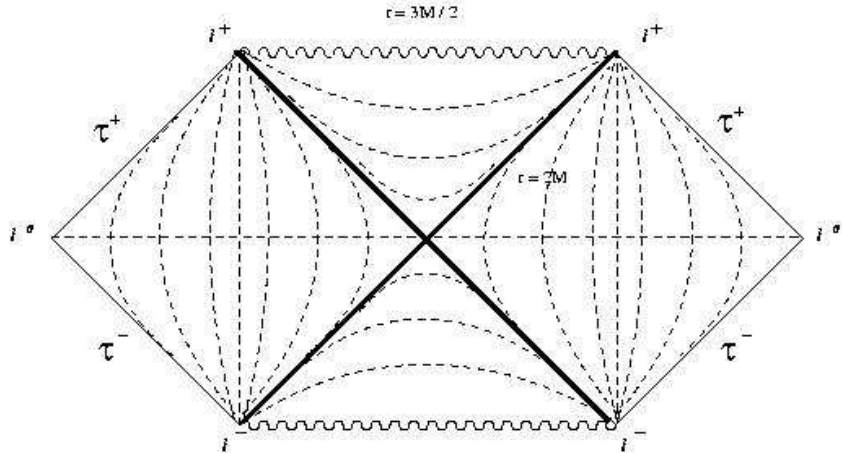, width=0.8 \linewidth, clip=}
\end{center}
\caption{Diagrama de Penrose para o BN-CFM para $\gamma <4$. A
singularidade física se encontra sobre $r=3M/2$. O horizonte de
eventos se dá sobre $r=2M$.}
\label{fig02}
\end{figure}

O diagrama de Penrose para este buraco tem a mesma forma que para
Schwarzschild, exceto pelo fato de que a singularidade física foi
deslocada de $r=0$ para $r=\frac{3M}{2}$.
O horizonte de eventos
continua sendo sobre $r=2M$ (ver figura \ref{fig02}).\\

Se fizermos $r \rightarrow \infty$ na métrica (\ref{BNB01})
observaremos que esse universo é assintoticamente plano.
\subsection{Buraco Negro tipo SM}
\begin{figure}[!ht]
\begin{center}
\epsfig{file=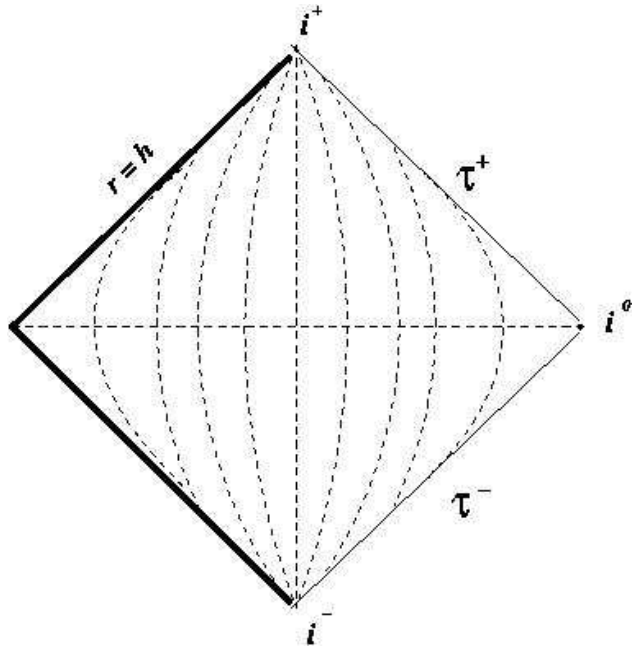, width=0.6 \linewidth, clip=}
\end{center}
\caption{Diagrama de Penrose para o BN-SM para $C>h$. A
singularidade física se encontra em $r=h/\sqrt{2}$.}
\label{fig03}
\end{figure}

A solução BN-SM foi obtida por Bronnikov \cite{Bronnikov2}. Ela representa
um buraco negro com massa zero.\\
A métrica que descreve o BN-SM tem a forma
\begin{eqnarray}
\label{BNB03}
ds^2=-\left(1-\frac{h^2}{r^2}\right)dt^2+\frac{1}{\left(1-\frac{h^2}{r^2}\right)
\left(1+\frac{C-h}{\sqrt{2r^2-h^2}}\right)}\ dr^2+r^2d\Omega^2\ .
\end{eqnarray}
onde $h$ é uma constante positiva. A constante $C$ obtida como
resultado da integração da equação (\ref{BNBf}) pode assumir
qualquer valor positivo ou negativo. Entretanto, devido aos mesmos
motivos citados para o caso do BN-CFM, assumiremos que $C>h$ daqui
pra frente.\\
Se $C>h$ a estrutura causal do buraco negro é do tipo Schwarzschild. A
figura \ref{fig03} representa o diagrama de Penrose para essa
solução.

Se observarmos os invariantes escalares desse buraco negro,
veremos que
\begin{eqnarray}
\label{BNB04}
R&=&0\ ,\\
\nonumber\\
R_{ij}R^{ij}&\sim& \frac{1}{(2r^2-h^2)^3r^8}\ ,\\
\nonumber\\
R_{ijkl}R^{ijkl}&\sim& \frac{1}{(2r^2-h^2)^3r^8}\ .
\end{eqnarray}
Deste modo vemos que para os valores $r=0,h/\sqrt{2}$ os
escalares explodem, caracterizando esses valores como
singularidades físicas. Investigando a componente $g_{tt}$ da
métrica vemos que o horizonte de eventos se dá sobre $r=h$. 

Portanto esse universo se encerra sobre $r=h/\sqrt{2}$ sendo definido pelo o intervalo $r \in (h/\sqrt{2},\infty)$.\\
Se fizermos $r \rightarrow \infty$ na métrica (\ref{BNB03})
observaremos que esse universo é assintoticamente plano.

\chapter{Evolução de Campos, Estabilidade e Termodinâmica de BN's Esfericamente Simétricos sobre a brana}
{\it No espaço onde tudo parece distante, buscamos alguma
referência. É neste momento que nos deparamos com a sua fronteira.
Continuemos pois na exploração deste universo.}\\
\\
O formalismo desenvolvido nas seções anteriores será aplicado à
dois tipos de buracos negros esfericamente simétricos localizados
sobre uma brana.

Os buracos negros estudados neste trabalho foram
\begin{itemize}
\item Buraco negro CFM

\item Buraco negro ``sem massa" SM
\end{itemize}
Apresentaremos abaixo as métricas do buracos negros estudados
assim como algumas quantidades relevantes aos cálculos realizados
nas seções posteriores.\\
\\
{\bf Buraco Negro-CFM}\\
\\
A métrica de um buraco negro tipo CFM localizado sobre uma brana é
dada por
\begin{eqnarray}
\label{BNMB01A} ds^2=
-\left(1-\frac{2M}{r}\right)dt^2+\frac{\left(1-\frac{3M}{2r}\right)}{\left(
1-\frac{2M}{r}\right)\left(1- \frac{M\gamma}{2r}\right)}\
dr^2+r^2d\Omega^2\ ,
\end{eqnarray}
com $d\Omega^2=d\theta^2+sen\theta^2d\phi^2$.

\begin{figure}[!ht]
\begin{center}
\epsfig{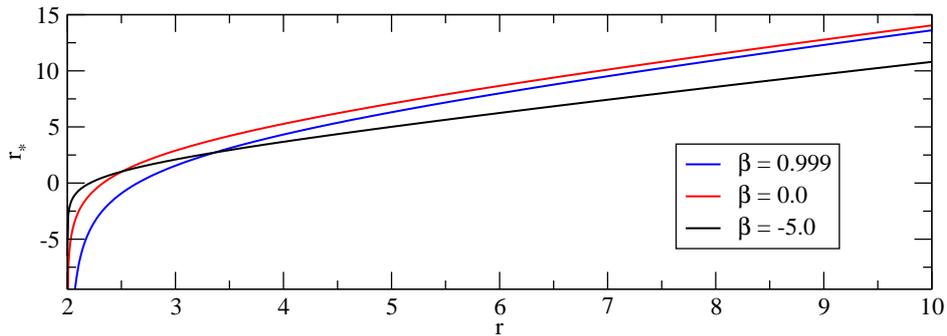}
\end{center}
\caption{Comportamento da coordenada tartaruga $r_{*}$, com $\gamma =\beta +3$ e $M=1$}
\label{figtarCFM1}
\end{figure}

A raiz quadrada do determinante da métrica covariante $g_{ij}$
será
\begin{eqnarray}
\label{BNMB02A} \sqrt{-g}= \sqrt{\frac{(2r-3M)}{(2r-M\gamma)}}\
r^2 sen\theta\ .
\end{eqnarray}
A coordenada tartaruga $r_{*}$ para esse buraco negro é dada por
\begin{eqnarray}
\label{BNMB03A}
r_{*}&=&\int \frac{1}{\sqrt{f}}\ dr\nonumber\\
\nonumber\\
r_{*}(r)&=&T_{1}(r)+T_{2}(r)+T_{3}(r),
\end{eqnarray}
onde
\begin{eqnarray}
\label{BNMB04A}
T_{1}(r)&=&\frac{\sqrt{(2r-\gamma M)(2r-3M)}}{2},\\
\nonumber\\
T_{2}(r)&=&\frac{M(5+\gamma)}{4}\ ln\left(4r-M(3+\gamma)+2\ T_{1}\ \right),\\
\nonumber\\
T_{3}(r)&=&-\frac{2M}{\sqrt{(4-\gamma)}}\ ln\left(T_{4}\right),\nonumber\\
\nonumber\\
T_{4}(r)&=&\frac{M\sqrt{4-\gamma}\ T_{1}+M(5-\gamma)r-M^2(6-\gamma)}{(r-2M)}.
\end{eqnarray}
Vejamos o comportamento da coordenada tartaruga $r_{*}$ para
alguns valores de $\beta$ na figua \ref{figtarCFM1}.\\

Portanto, com essa mudança de coordenadas mapeamos o intervalo semi-infinito
$(2M,+\infty)$ que corresponde ao exterior do BN-CFM no intervalo
infinito $(-\infty,+\infty)$.

Observando o limite quando $r\rightarrow 2M$ vemos que $r_{*}\rightarrow -\infty$. No limite de $r \gg 2M$ a coordenada $r_{*}$ cresce linearmente. Analisaremos esses limites mais detalhadamente.

Utilizando-se  do formalismo descrito no Apêndice \ref{sec:coordtartaruga} podemos
definir uma função $h(r)$ para o BN-CFM como sendo
\begin{eqnarray}
\label{BNMB05A}
h(r)=\frac{r-2M}{r}\sqrt{\frac{2r-\gamma M}{2r-3M}}.
\end{eqnarray}
Nesta forma é fácil ver que o horizonte de eventos se encontra sobre $r=r_{h}=2M$. Para que este seja um horizonte simples, ou seja, um zero simples, temos que função $P(r)$, definida abaixo, como
\begin{eqnarray}
\label{BNMB06A}
h(r)=(r-r_{h})P(r)\qquad \textrm{com}\qquad P(r)=\frac{1}{r}\sqrt{\frac{2r-\gamma M}{2r-3M}},
\end{eqnarray}
deve ser analítica e não-nula em $r=r_{h}=2M$. Isso de fato é
observado pois
\begin{eqnarray}
\label{BNMB07A}
P(r_{h})=\frac{\sqrt{4-\gamma}}{2M}.
\end{eqnarray}
Portanto, se $\gamma < 4$, $P(r_{h})$ é analítica e não-nula, e
$r=2M$ é um zero simples de $h(r)$.

Observando a equação (\ref{BNMB03A}), analisemos o comportamento de
$r_{*}$ quando $r \rightarrow \infty$. Neste caso $r_{*}(r)$
diverge linearmente, já que neste limite temos
\begin{eqnarray}
\label{BNMB08A}
T_{1}(r)&\approx& r,\\
\nonumber\\
T_{2}(r)&\approx& \frac{M(5+\gamma)}{4}\ ln(r),\\
\nonumber\\
T_{3}(r)&\approx& constante,
\end{eqnarray}
e desta forma
\begin{eqnarray}
\label{BNMB09A}
r_{*}(r)\approx r+\frac{M(5+\gamma)}{4}\ ln(r).
\end{eqnarray}
Interessante notar que a forma da função $r_{*}(r)$ no caso do BN-CFM é mesma do caso do BN-Schwarzschild
\begin{eqnarray}
\label{BNMB10A} r_{*}(r)\approx r+2M\ ln(r)\ ,
\end{eqnarray}
no limite de $r$ grande.

\begin{figure}[!ht]
\begin{center}
\epsfig{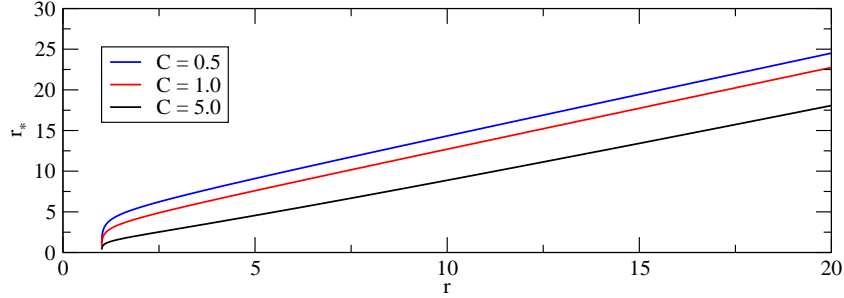}
\end{center}
\caption{{\small Comportamento da coordenada tartaruga $r_{*}$,  para $h=1$}}
\label{figtartSM1}
\end{figure}

No limite de $r\rightarrow 2M$ podemos usar os resultados obtidos
no Apêndice \ref{sec:coordtartaruga}. Como $h(r)$ possui um zero simples, a forma de
$r_{*}(r)$, próximo do horizonte de eventos, é dada por
\begin{eqnarray}
\label{BNMB11A}
r_{*}(r)=\frac{1}{P(r_{h})}\ ln(r-r_{h})\qquad \Longrightarrow \ \ r_{*}(r)=\frac{2M}{\sqrt{4-\gamma}}\ ln(r-2M).
\end{eqnarray}
Vemos então que próximo do horizonte de eventos $r_{*}$ diverge logaritimicamente.\\
\\
{\bf Buraco Negro-SM}\\
\\
A métrica de um buraco negro tipo SM localizado sobre uma brana é
dada por
\begin{eqnarray}
\label{BNMB01B} ds^2=
-\left(1-\frac{h^2}{r^2}\right)dt^2+\frac{1}{\left(
1-\frac{h^2}{r^2}\right)\left(1+
\frac{C-h}{\sqrt{2r^2-h^2}}\right)}\ dr^2+r^2d\Omega^2.
\end{eqnarray}
A raiz quadrada do determinante da métrica covariante $g_{ij}$
 neste caso será
\begin{eqnarray}
\label{BNMB02B}
 \sqrt{-g}= \sqrt{1+\frac{C-h}{\sqrt{2r^2-h^2}}}\ r^2
sen\theta\ .
\end{eqnarray}
A coordenada tartaruga $r_{*}$ para esse buraco negro é dada por
\begin{eqnarray}
\label{BNMB03B}
r_{*}=\int_{h}^{\infty}\frac{1}{\left(1-\frac{h^2}{r^2}\right)
\sqrt{1+\frac{C-h}{\sqrt{2h^2-r^2}}}}\ dr\ .
\end{eqnarray}
A equação (\ref{BNMB03B}) será integrada numericamente devido a
dificuldade em encontrarmos uma solução analítica. Vejamos o comportamento da coordenada tartaruga $r_{*}$ para alguns valores de $C$ na figura \ref{figtartSM1}.

Com essa mudança de coordenadas mapeamos o intervalo $(h,+\infty)$
no intervalo $(-\infty,+\infty)$.
Observando o limite quando $r\rightarrow h$ vemos que $r_{*}\rightarrow -\infty$. No limite de $r \gg h$ a coordenada $r_{*}$ também cresce linearmente como no caso do BN-CFM. Analisaremos esses limites mais detalhadamente.

Utilizando-se  do formalismo descrito no Apêndice \ref{sec:coordtartaruga} podemos
definir uma função $h(r)$ para o BN-SM como sendo
\begin{eqnarray}
\label{BNMB04B}
h(r)=(r-h)\frac{(r+h)}{r^2}\sqrt{1+\frac{C-h}{\sqrt{2r^2-h^2}}}.
\end{eqnarray}
Temos então que $r=r_{h}=h$ é um candidato a horizonte (simples) de eventos, desde que a função
\begin{eqnarray}
\label{BNMB05B}
P(r)=\frac{(r+h)}{r^2}\sqrt{1+\frac{C-h}{\sqrt{2r^2-h^2}}},
\end{eqnarray}
seja analítica e não-nula em $r=r_{h}=h$. Explicitamente,
\begin{eqnarray}
\label{BNMB06B}
P(r_{h})=\frac{2}{h}\sqrt{\frac{C}{h}}.
\end{eqnarray}
Desta forma, se $C>0$ então $h(r)$ possui uma raiz simples em $r=h$, e a função $r_{*}$ diverge logaritimicamente próximo ao horizonte de eventos:
\begin{eqnarray}
\label{BNMB07B}
r_{*}(r)=\frac{h}{2}\sqrt{\frac{h}{C}}\ ln(r-h).
\end{eqnarray}
Por outro lado, se $C=0$ então $P(r_{h})=0$. Neste caso, a função $h(r)$ possui zero duplo o que implica que $r_{*}$ diverge como lei de potência
\begin{eqnarray}
\label{BNMB08B}
r_{*}(r)\propto -\frac{1}{(r-h)}
\end{eqnarray}
em $r=h$. Suporemos que $C>0$ daqui para frente.

Vejamos o comportamento de $r_{*}$ para $r$ grande. Neste caso a integração explícita não é simples. Podemos entretanto expandir $h(r)$ em potências de $1/r$ obtendo\\
\begin{eqnarray}
\label{BNMB09B}
h(r)\approx 1+\left(\frac{(C-h)\sqrt{2}}{4}\right)\frac{1}{r}.
\end{eqnarray}
Usando este resultado teremos
\begin{eqnarray}
\label{BNMB10B}
r_{*}(r)\approx r-\frac{\sqrt{2}(C-h)}{4}\ ln(r).
\end{eqnarray}
A função $r_{*}$ para o BN-SM, também diverge linearmente com $r \rightarrow \infty$. Entretanto, o termo subdominante será diferente do caso do BN-CFM se, $C\geq h$.\\
Neste trabalho assumiremos a condição $C\geq h$.
Essa mudança no comportamento assintótico, acarretará alterações no comportamento das caudas nas perturbações.

\section{Evolução de campos em BN-CFM e BN-SM}
Nesta seção apresentaremos as equações que descrevem a evolução de
campos no exterior dos BN-CFM e BN-SM. A solução destas equações
foram obtidas numericamente através do método de integração com
condições iniciais características e serão apresentadas e
discutidas no capítulo 8.

\subsection{Perturbação Escalar no exterior do BN-CFM e do BN-SM}
Considerando que trabalharemos com campos escalares localizados
apenas sobre a brana, ou seja, que não se propagam para o {\it
bulk}, a evolução de um campo escalar não massivo
$\Psi(t,r,\theta,\phi)$ no exterior de um buraco negro sobre a
brana é governada pelas equações (\ref{pe26}) e (\ref{pe27}). Para
o caso do BN-CFM temos que a função $f(r)$ e sua derivada
$\frac{df}{dr}$ são
\begin{eqnarray}
\label{BNMBpertescalar01A}
f&=&\frac{(r-2M)^2(2r-M\gamma)}{r^2(2r-3M)}\\
\nonumber\\
\frac{df}{dr}&=&\frac{d}{dr}\left[\left(1-\frac{2M}{r}\right)^2\left(\frac{2r-M\gamma}{2r-3M}\right)\right]\nonumber\\
\nonumber\\
\frac{df}{dr}&=&2\left(1-\frac{2M}{r}\right)\left\{\frac{2M}{r^2}\left(\frac{2r-M\gamma}{2r-3M}\right)+\right.\nonumber\\
\nonumber\\
&+&\left.\left(1-\frac{2M}{r}\right)\left[\frac{(2r-3M)-(2r-M\gamma)}{(2r-3M)^2}\right]\right\}\
.
\end{eqnarray}
Substituindo $\gamma=\beta+3$ teremos
\begin{eqnarray}
\label{BNMBpertescalar02A}
\frac{df}{dr}&=&2\left(1-\frac{2M}{r}\right)\left\{\frac{2M}{r^2}+\frac{M\beta}{(2r-3M)^2}\left[1-\frac{6M}{r}+\frac{6M^2}{r^2}\right]\right\}.
\end{eqnarray}
Substituindo a equação (\ref{BNMBpertescalar02A}) na equação
(\ref{pe27}) obtemos o potencial efetivo $V_{esc}^{CFM}$ para o
BN-CFM. Desta forma a equação que governa a perturbação escalar de
um BN-CFM será
\begin{eqnarray}
\label{BNMBpertescalar03A}
&-&\frac{\partial^2 R}{\partial t^2}(r,t)+\frac{\partial^2 R}{\partial r_{*}^2}(r,t)= V_{esc}^{CFM}(r)R(r,t)
\end{eqnarray}
onde
\begin{eqnarray}
\label{BNMBpertescalar04A}
V_{esc}^{CFM}(r)&=&\left(1-\frac{2M}{r}\right)\left\{\frac{\ell(\ell+1)}{r^2}+\frac{2M}{r^3}+\right.\\
\nonumber\\
&+&\left.
\frac{M\beta}{r(2r-3M)^2}\left[1-\frac{6M}{r}+\frac{6M^2}{r^2}\right]\right\}\
.
\end{eqnarray}
Como podemos observar na equação acima, se fizermos, $\beta =0$,
recuperamos a equação que governa a perturbação escalar para
Schwarzschild, uma vez que para esse valor de $\beta$ a métrica do
BN-CFM se torna a métrica de Schwarzschild. O caso do BN-SM pode
ser tratado de modo análogo ao BN-CFM. Para o caso do BN-SM a
função $f(r)$ e sua derivada $\frac{df}{dr}$ serão
\begin{eqnarray}
\label{BNMBpertescalar01B}
f&=&\left(1-\frac{h^2}{r^2}\right)^2\left(1+\frac{C-h}{\sqrt{2r^2-h^2}}\right)\\
\nonumber\\
\frac{df}{dr}&=&\frac{d}{dr}\left[\left(1-\frac{h^2}{r^2}\right)^2\left(1+\frac{C-h}{\sqrt{2r^2-h^2}}\right)\right]\nonumber\\
\nonumber\\
\label{BNMBpertescalar02B}
\frac{df}{dr}&=&2\left(1-\frac{h^2}{r^2}\right)\left\{\frac{2h^2}{r^3}+\right. \nonumber\\
\nonumber\\
&+&\left.
\frac{C-h}{\sqrt{2r^2-h^2}}\left[\frac{2h^2}{r^3}-\frac{r^2-h^2}{r(2r^2-h^2)}\right]\right\}\
.
\end{eqnarray}
Agora, substituindo a equação (\ref{BNMBpertescalar02B}) na
equação (\ref{pe27}) obtemos o potencial efetivo $V_{esc}^{SM}$
para o BN-SM. A equação que governará a perturbação escalar de um
BN-SM será
\begin{eqnarray}
\label{BNMBpertescalar03B}
&-&\frac{\partial^2 R}{\partial t^2}(r,t)+\frac{\partial^2 R}{\partial r_{*}^2}(r,t)= V_{esc}^{SM}(r)R(r,t)
\end{eqnarray}
onde
\begin{eqnarray}
\label{BNMBpertescalar04B}
V_{esc}^{SM}(r)&=&\left(1-\frac{h^2}{r^2}\right)\left\{\frac{\ell(\ell+1)}{r^2}+\frac{2h^2}{r^4}\right.\nonumber\\
\nonumber\\
&+&\left.
\frac{C-h}{\sqrt{2r^2-h^2}}\left[\frac{2h^2}{r^4}-\frac{r^2-h^2}{r^2(2r^2-h^2)}\right]\right\}\
.
\end{eqnarray}

\subsection{Perturbação Eletromagnética no exterior do BN-CFM e do BN-SM}
Nesta seção trataremos da evolução de um campo de Maxwell no
exterior de um buraco negro sobre a brana. Perturbações
eletromagnéticas são de grande interesse, já que, no contexto da
conjectura $AdS/CFT$ elas podem ser vistas como perturbações para
algum campo de calibre em supergravidade. Como o campo de Maxwell
é um campo de matéria ele permanece confinado na brana. Portanto
não precisamos resolver as equações relativas ao {\it bulk}. Desta
forma a evolução de um campo de Maxwell
$\mathcal{A}(t,r,\theta,\phi)$ no exterior de um buraco negro na
brana é dada pelas equações (\ref{pele10}) e (\ref{pele11}).

A propagação de um campo de Maxwell no exterior do BN-CFM será
dada pela equação
\begin{eqnarray}
\label{BNMBperteletro01A} -\frac{\partial^2\mathcal{A}}{\partial
t^2}+\frac{\partial^2\mathcal{A}}{\partial
r_{*}^2}=V_{el}^{CFM}(r)\ \mathcal{A}
\end{eqnarray}
onde o potencial efetivo $V_{el}^{CFM}$ para o BN-CFM será
\begin{eqnarray}
\label{BNMBperteletro02A} V_{el}^{CFM}(r)=
\left(1-\frac{2M}{r}\right)\frac{\ell(\ell+1)}{r^2}\ .
\end{eqnarray}
Interessante notar que apesar do BN-CFM ter, como um caso
particular a solução de Schwarzschild, o potencial efetivo para
perturbações eletromagnéticas neste buraco negro é idêntico ao
potencial do buraco negro de Schwarzschild. Novamente o processo é
repetido para o caso do BN-SM.

Assim a propagação de um campo de Maxwell no exterior do BN-SM
será dada pela equação
\begin{eqnarray}
\label{BNMBperteletro01B}
 -\frac{\partial^2\mathcal{A}}{\partial
t^2}+\frac{\partial^2\mathcal{A}}{\partial
r_{*}^2}=V_{el}^{SM}(r)\ \mathcal{A}
\end{eqnarray}
onde o potencial efetivo $V_{el}^{SM}$ para o BN-SM será
\begin{eqnarray}
\label{BNMBperteletro02B}
 V_{el}^{SM}(r)=
\left(1-\frac{h^2}{r^2}\right)\frac{\ell(\ell+1)}{r^2}\ .
\end{eqnarray}

\subsection{Perturbações Gravitacionais no exterior do BN-CFM e do BN-SM}
O cálculo de perturbações gravitacionais, no contexto de Mundo
Brana, revela-se um pouco mais complicado do que em $(3+1)$
dimensões, devido a existência do {\it bulk}.

Diferentemente das perurbações escalares e eletromagnéticas que
permanecem confinadas na brana, as perturbações gravitacionais
além de perturbar a brana perturbam o {\it bulk}.
Conseqüentemente, o tensor $E_{ij}$ será perturbado acrescentando
alguma contribuição no tensor-momento energia perturbado.

Como foi mencionado anteriormente na seção 6.2, não conhecemos a
extensão, dos buracos negros estudados, no {\it bulk}. Isso
poderia inviabilizar o cálculo das perturbações gravitacionais
pois não poderíamos contabilizar a contribuição do tensor
$E_{ij}$.

Entretanto como estamos interessados apenas nas ondas que se
propagam sobre a brana assumiremos algumas condições adicionais
como hipótese de trabalho para contornar o problema de solucionar
as equações perturbadas no {\it bulk}. Consideraremos apenas
perturbações axiais e em $1^{a}$ ordem na métrica $q_{ij}$. Como
realizado para a perturbação escalar, calcularemos a perturbação
para uma métrica esfericamente simétrica  tão geral
quanto possível e depois especificaremos caso a caso.\\
\\
As equações de Einstein 4-dimensionais para o vácuo, sobre a brana são dadas por
\begin{eqnarray}
\label{pgb01} R_{ij}-\frac{1}{2}q_{ij}R=-\Lambda_{4} \
q_{ij}-E_{ij}\ .
\end{eqnarray}
Faremos $\Lambda_{4} =0$ daqui para frente o que implica em um ajuste ente $\Lambda_{5}$ e a tensão na brana que é dada pela equação (\ref{BNBc}). Assim a equação que devemos perturbar será
\begin{eqnarray}
\label{pgb02} R_{ij}-\frac{1}{2}q_{ij}R=-E_{ij}\ .
\end{eqnarray}
Para uma perturbação em $1^{a}$ ordem a nova métrica pode ser
escrita como
\begin{eqnarray}
\label{pgb03}
e_{ij}=q_{ij}+h_{ij} \qquad e^{ij}=q^{ij}-h^{ij}
\end{eqnarray}
onde $h_{ij}<< q_{ij}$ e $e_{ib}\ e^{bj}=\delta_{i}^{j}$.

Observando nossa equação de Einstein não-perturbada temos os
seguintes resultados
\begin{eqnarray}
\label{pgb04}
\stackrel{\circ}{R}_{ij}&=& -\stackrel{\circ}{E}_{ij}\ ,\\
\nonumber\\
\stackrel{\circ}{E}_{i}^{i}&=&\stackrel{\circ}{E}\ =0\ , \\
\nonumber\\
\stackrel{\circ}{R}&=&0\ ,
\end{eqnarray}
onde o tensor $E_{ij}$ tem traço nulo devido às simetrias do
tensor de Weyl e a notação $(\ \circ \ )$ inidica que o objeto
será calculado com respeito à métrica fundo $q_{ij}$.

A equação de Einstein perturbada será dada pela substituição da
equação (\ref{pgb03}) na equação (\ref{pgb02})
\begin{eqnarray}
\label{pgb05}
R_{ij}&-&\frac{1}{2}e_{ij}R=-E_{ij}\\
\nonumber\\
\stackrel{\circ}{R}_{ij}&+&\delta
R_{ij}-\frac{1}{2}(q_{ij}+h_{ij})(\stackrel{\circ}{R}+\delta
R)=-(\stackrel{\circ}{E}_{ij}+\delta E_{ij})\ .
\end{eqnarray}
O escalar de Ricci perturbado será dado pela contração da métrica
perturbada com o tensor de Ricci perturbado
\begin{eqnarray}
\label{pgb06}
e^{ij}R_{ij}-\frac{1}{2}e^{ij}e_{ij}R&=&-e^{ij}E_{ij}\ ,\nonumber\\
\nonumber\\
R-2R&=&-E\ ,\nonumber\\
\nonumber\\
R&=&E\ .
\end{eqnarray}
Como temos
\begin{eqnarray}
\label{pgb06a}
\stackrel{\circ}{E}=E=0,
\end{eqnarray}
devido às simetrias do tensor de Weyl, podemos reescrever a equação (\ref{pgb05}) na forma
\begin{eqnarray}
\label{pgb06b}
R_{ij}&=&-E_{ij}\ ,\\
\nonumber\\
\label{pgb06bb} \delta R_{ij}&=&-\delta E_{ij}\ .
\end{eqnarray}
Desta forma se conhecermos $\delta E_{ij}$ poderemos escrever a
equação completamente. Entretanto não conhecemos a forma exata do
{\it bulk} o que nos impede de calcular $\delta E_{ij}$ explicitamente.
Neste ponto do trabalho assumiremos algumas condições de trabalho.
Assumiremos a seguinte a hipótese de trabalho.
\begin{eqnarray}
\label{pgb06c} \delta E_{ij}=0
\end{eqnarray}
O motivo que nos leva a tal declaração repousa no fato de que
nossa perturbação gravitacional é realizada sobre a 3-brana.

Sabemos que uma parte das perturbações gravitacionais se
propagarão para o {\it bulk} perturbando-o também. Essa perturbação
afetará o tensor $E_{ij}$ fazendo com que $\delta E_{ij}$ seja
não-nulo. Entretanto o efeito do {\it bulk} sobre a brana pode ser considerado secundário.

Como a informação sobre a perturbação leva um tempo finito para
chegar ao {\it bulk} e esse também leva um tempo finito para transmitir
essa informação ao tensor de Weyl e conseqüentemente ao tensor
$E_{ij}$, podemos em primeira aproximação desconsiderar os
efeitos do {\it bulk}.

Um ponto importante a ser acrescentado é que essa condição quase
não afeta o valor das freqüências quasi-normais se essas fossem
calculadas com a presença de $\delta E_{ij}$. O maior efeito de
$\delta E_{ij}$ será observado nas caudas da perturbação pois
nesse intervalo de tempo posterior o efeito da perturbação já
alcançou a 3-brana. Esse comportamento já era esperado pois se
observarmos o comportamento da perturbação gravitacional de uma
black string veremos que o efeito dos modos massivos do gráviton
aparecem apenas na cauda.

Portanto, recuperando nossa equação (\ref{pgb06bb}) e substituindo
nela a equação (\ref{pgb06c}), teremos
\begin{eqnarray}
\label{pgb12} \delta R_{ij}=0\ ,
\end{eqnarray}
onde
\begin{eqnarray}
\label{pgb13} \delta R_{ij}=\delta \Gamma_{ij\
;b}^{\phantom{i}{b}}-\delta \Gamma_{ib\ ;j}^{\phantom{i}{b}}\ .
\end{eqnarray}
Essa é a equação a ser resolvida para uma perturbação
gravitacional de uma 3-brana embebida em um espaço 5-dimensional $AdS$.

Para o cálculo das perturbações gravitacionais axiais utilizaremos
o método desenvolvido por Chandrasekhar. A mudança no método se
deve ao fato que Chandrasekhar desenvolve um formalismo mais geral
que Regge e Wheeler por assumir condições mais fracas de
dependência das equações acopladas. Isso ficará um pouco mais
claro a seguir. A notação utilizada neste cálculo segue a notação
de Chandrasekhar \cite{Chandra}.

Assumiremos uma métrica axialmente simétrica de modo que ela
represente nossa escolha de {\it gauge}. A métrica será dada por
\begin{eqnarray}
\label{pgb14}
ds^2=-e^{2\nu}dt^2&+&e^{2\psi}(dx^{1}+\omega dt+q_{2}dx^2+q_{3}dx^3)^2\nonumber\\
\nonumber\\
&+&e^{2\mu_{2}}(dx^{2})^2+e^{2\mu_{3}}(dx^{3})^2,
\end{eqnarray}
onde temos:
\begin{eqnarray}
t&\rightarrow& 0 \qquad e^{2\nu}\ =A(r),\nonumber\\
\phi &\rightarrow& 1 \qquad e^{2\mu_{2}}=B(r),\nonumber\\
r &\rightarrow& 2 \qquad e^{2\mu_{3}}= r^2,\nonumber\\
\theta &\rightarrow& 3 \qquad e^{2\psi}\ =r^2sin(\theta)^2.\nonumber
\end{eqnarray}
As perturbações axiais na métrica são representadas pelas funções
$\omega$, $q_{1}$ e $q_{2}$ que dependem das coordenadas
$(t,r,\theta,\phi)$. Substituindo a métrica (\ref{pgb14}) na
equação (\ref{pgb12}) temos como resultado apenas 3 equações que
dependem linearmente de $\omega$, $q_{1}$ e $q_{2}$. Como estamos
interessados apenas em perturbações de $1^{a}$ ordem, elas já são
suficiente. Assim, as componentes da equação (\ref{pgb12}) que
caracterizam perturbações axiais de primeira ordem são dadas por
\begin{eqnarray}
\label{pgb15}
\delta R_{12}=0,\\
\nonumber\\
\delta R_{13}=0,\\
\nonumber\\
\delta R_{10}=0.
\end{eqnarray}
Entretanto esse sistema de equações ainda pode ser simplificado.
Podemos mostrar que a equação $\delta R_{10}=0$ é uma combinação
das duas primeiras e pode ser escrita como (ver Apêndice~\ref{sec:desacoplamento})
\begin{equation}
\label{pgb16}
\delta R_{10,0}=\delta R_{12,2}+\delta R_{13,3}
\end{equation}
Assim se as duas primeiras equações forem satisfeitas a terceira
equação é satisfeita automaticamente. Portanto basta solucionarmos as equações $\delta R_{13}=0$ e $\delta R_{12}=0$.

Nossas duas equações podem ser explicitadas como
\begin{eqnarray}
\label{pgb17}
(e^{3\psi +\nu-\mu_{2}-\mu_{3}}Q_{23})_{,3}&=&-(e^{3\psi-\nu+\mu_{3}-\mu_{2}}Q_{02})_{,0}\ ,\\
\nonumber\\
(e^{3\psi
+\nu-\mu_{2}-\mu_{3}}Q_{23})_{,2}&=&(e^{3\psi-\nu-\mu_{3}+\mu_{2}}Q_{03})_{,0}\
,
\end{eqnarray}
onde
\begin{eqnarray}
\label{pgb18} Q_{AB}=q_{A,B}-q_{B,A} \qquad
Q_{A0}=q_{A,0}-\omega_{,A}\ .\nonumber
\end{eqnarray}
Como todas as componentes da métrica dependem apenas de $r$ e $\theta$ podemos reescrever as equações acima como:
\begin{eqnarray}
\label{pgb19}
(e^{3\psi +\nu-\mu_{2}-\mu_{3}}Q_{23})_{,3}&=& -e^{3\psi-\nu+\mu_{3}-\mu_{2}} Q_{02,0}\\
\nonumber\\
(e^{3\psi +\nu-\mu_{2}-\mu_{3}}Q_{23})_{,2}&=&
e^{3\psi-\nu-\mu_{3}+\mu_{2}} Q_{03,0}\ .
\end{eqnarray}
Reescrevendo as equações e substituindo os valores de $Q_{02}$ e $Q_{03}$ temos
\begin{eqnarray}
\label{pgb20}
(e^{3\psi +\nu-\mu_{2}-\mu_{3}}Q_{23})_{,3} e^{-3\psi+\nu-\mu_{3}+\mu_{2}}&=& - (\omega_{,2}-q_{2,0})_{,0}\\
\nonumber\\
\label{pgb21}
(e^{3\psi
+\nu-\mu_{2}-\mu_{3}}Q_{23})_{,2}e^{-3\psi+\nu+\mu_{3}-\mu_{2}}&=&
(\omega_{,3}-q_{3,0})_{,0}\ .
\end{eqnarray}
Se chamarmos
\begin{eqnarray}
\label{pgb22}
Q(r,\theta,t)= e^{3\psi +\nu-\mu_{2}-\mu_{3}}Q_{23}\nonumber\\
\nonumber\\
Q(r,\theta,t)= r^2 sin(\theta)^3\sqrt{\frac{A(r)}{B(r)}} Q_{23}
\end{eqnarray}
e substituirmos nas equações (\ref{pgb20}) e (\ref{pgb21}) temos
\begin{eqnarray}
\label{pgb23}
\frac{\partial Q}{\partial \theta}e^{-3\psi+\nu-\mu_{3}+\mu_{2}}&=&- (\omega_{,2}-q_{2,0})_{,0}\\
\nonumber\\
\frac{\partial Q}{\partial r}e^{-3\psi+\nu+\mu_{3}-\mu_{2}}&=&
(\omega_{,3}-q_{3,0})_{,0}\ .
\end{eqnarray}
Nossas equações serão então
\begin{eqnarray}
\label{pgb24}
\frac{1}{r^4 sin(\theta)^3} \sqrt{A(r)B(r)} \frac{\partial Q}{\partial \theta}&=&- (\omega_{,2}-q_{2,0})_{,0}\ ,\\
\nonumber\\
\label{pgb25}
\frac{1}{r^2 sin(\theta)^3} \sqrt{\frac{A(r)}{B(r)}}\frac{\partial
Q}{\partial r}&=& (\omega_{,3}-q_{3,0})_{,0}\ .
\end{eqnarray}
Derivando a equação (\ref{pgb24}) em relação a $\theta$ e equação (\ref{pgb25}) em relação a $r$ temos
\begin{eqnarray}
\label{pgb26}
\frac{\sqrt{A(r)B(r)}}{r^4} \frac{\partial}{\partial \theta}\left(\frac{1}{sin(\theta)^3}\frac{\partial Q}{\partial \theta}\right)&=&- \omega_{,2,0,3}+q_{2,0,0,3}\ ,\\
\nonumber\\
\label{pgb27}
\frac{1}{sin(\theta)^3}\frac{\partial}{\partial
r}\left(\frac{1}{r^2} \sqrt{\frac{A(r)}{B(r)}}\frac{\partial
Q}{\partial r}\right)&=& \omega_{,3,0,2}-q_{3,0,0,2}\ .
\end{eqnarray}
Somando as duas equações teremos
\begin{eqnarray}
\label{pgb28}
\frac{\sqrt{A(r)B(r)}}{r^4} \frac{\partial}{\partial \theta}\left(\frac{1}{sin(\theta)^3}\frac{\partial Q}{\partial \theta}\right)&+&\frac{1}{sin(\theta)^3}\frac{\partial}{\partial r}\left(\frac{1}{r^2}\sqrt{\frac{A(r)}{B(r)}}\frac{\partial Q}{\partial r}\right)=\nonumber\\
\nonumber\\
&=& (q_{2,3}-q_{3,2})_{,0,0}=(Q_{23})_{,0,0}\ .
\end{eqnarray}
Substituindo $Q_{23}$ ficaremos com
\begin{eqnarray}
\label{pgb29}
\frac{\sqrt{A(r)B(r)}}{r^4} \frac{\partial}{\partial \theta}\left(\frac{1}{sin(\theta)^3}\frac{\partial Q}{\partial \theta}\right)&+&\frac{1}{sin(\theta)^3}\frac{\partial}{\partial r}\left(\frac{1}{r^2}\sqrt{\frac{A(r)}{B(r)}}\frac{\partial Q}{\partial r}\right)=\nonumber\\
\nonumber\\
&=&\frac{1}{r^2
sin(\theta)^3}\sqrt{\frac{A(r)}{B(r)}}\frac{\partial^2 Q}{\partial
t^2}.
\end{eqnarray}
Rearranjando os termos da equação (\ref{pgb29}) teremos
\begin{eqnarray}
\label{pgb30}
\frac{r^4}{\sqrt{A(r)B(r)}}\frac{\partial}{\partial r}\left(\frac{1}{r^2}\sqrt{\frac{A(r)}{B(r)}}\frac{\partial Q}{\partial r}\right)-\frac{r^2}{A(r)}\frac{\partial^2 Q}{\partial t^2}=\nonumber\\
\nonumber\\
=-sin(\theta)^3\frac{\partial}{\partial
\theta}\left(\frac{1}{sin(\theta)^3}\frac{\partial Q}{\partial
\theta}\right)\ .
\end{eqnarray}
As variáveis $r$, $t$, $\theta$ da equação (\ref{pgb30}) podem ser
separadas fazendo a substituição
\begin{equation}
\label{pgb31}
Q(r,t,\theta)= Q(r,t)C^{-3/2}_{l+2}(\theta)
\end{equation}
onde $C^{\nu}_{n}$ é a função de Gegenbauer que é solução da equação
\begin{equation}
\label{pgb32}
\left[\frac{d}{d\theta}\left(sin(\theta)^{2\nu}\frac{d}{d\theta}\right)+n(n+2\nu)sin(\theta)^{2\nu}\right]C^{\nu}_{n}(\theta)=0\
.
\end{equation}
Em nosso caso
\begin{equation}
\label{pgb33}
sin(\theta)^3\frac{d}{d\theta}\left(\frac{1}{sin(\theta)^3}\frac{dC^{-3/2}_{l+2}(\theta)}{d\theta}\right)=
-(l+2)(l-1)C^{-3/2}_{l+2}(\theta)\ .
\end{equation}
Substituindo (\ref{pgb31}) e (\ref{pgb33}) em (\ref{pgb30}) teremos a seguinte equação
\begin{eqnarray}
\label{pgb34}
\frac{r^4}{\sqrt{A(r)B(r)}}\frac{\partial}{\partial r}\left(\frac{1}{r^2}\sqrt{\frac{A(r)}{B(r)}}\frac{\partial Q}{\partial r}\right)C^{-3/2}_{l+2}(\theta)-\frac{r^2}{A(r)}\frac{\partial^2 Q}{\partial t^2}C^{-3/2}_{l+2}(\theta)=\nonumber\\
\nonumber\\
=C^{-3/2}_{l+2}(\theta)(l+2)(l-1)Q,\\
\nonumber\\
\frac{r^4}{\sqrt{A(r)B(r)}}\frac{\partial}{\partial
r}\left(\frac{1}{r^2}\sqrt{\frac{A(r)}{B(r)}}\frac{\partial
Q}{\partial r}\right)-\frac{r^2}{A(r)}\frac{\partial^2 Q}{\partial
t^2}-Q(l+2)(l-1) = 0.
\end{eqnarray}
Rearranjando mais uma vez a equação temos
\begin{eqnarray}
\label{pgb35}
r^2\sqrt{\frac{A(r)}{B(r)}}\frac{\partial}{\partial r}\left(\frac{1}{r^2}\sqrt{\frac{A(r)}{B(r)}}\frac{\partial Q}{\partial r}\right)-\frac{\partial^2 Q}{\partial t^2}-\nonumber\\
\nonumber\\
-\frac{A(r)}{r^2}Q(l+2)(l-1) = 0\ .
\end{eqnarray}
Neste ponto faremos a seguinte mudança de variável
\begin{equation}
\label{pgb36}
Q(r,t)=\chi(r,t)b(r), \qquad r=r(r_{*}), \qquad \Delta =
r^2\sqrt{\frac{A(r)}{B(r)}}\ .
\end{equation}
Dessa forma nossa equação (\ref{pgb35}) ficará da seguinte maneira
\begin{eqnarray}
\label{pgb37}
\Delta \frac{\partial}{\partial r}\left(\frac{\Delta}{r^4}\frac{\partial}{\partial r}(\chi b)\right)-\frac{A(r)}{r^2}(l+2)(l-1)\chi b - b\frac{\partial^2 \chi}{\partial t^2}=0\\
\nonumber\\
\frac{\Delta}{b} \frac{\partial r_{*}}{\partial
r}\frac{\partial}{\partial
r_{*}}\left(\frac{\Delta}{r^4}\frac{\partial r_{*}}{\partial
r}\frac{\partial}{\partial r_{*}}(\chi
b)\right)-\frac{A(r)}{r^2}(l+2)(l-1)\chi - \frac{\partial^2
\chi}{\partial t^2}=0\ .
\end{eqnarray}
Usando a notação
\begin{equation}
\label{pgb38} \frac{\partial a}{\partial r}=\dot{a}, \qquad
\frac{\partial a}{\partial r_{*}}=a'\ .
\end{equation}
Nossa equação pode ser reescrita como
\begin{eqnarray}
\label{pgb39}
\frac{\Delta}{b} \dot{r}_{*}\frac{\partial}{\partial
r_{*}}\left(\frac{\Delta}{r^4}\dot{r}_{*}(\chi'
b+b'\chi)\right)-\frac{A(r)}{r^2}(l+2)(l-1)\chi - \frac{\partial^2
\chi}{\partial t^2}=0\ .
\end{eqnarray}
Realizando as derivadas e reagrupando os termos temos
\begin{eqnarray}
\label{pgb40}
\frac{\Delta^2}{r^4}\dot{r}_{*}^2 \chi''+\chi'\left(\Delta \dot{r}_{*}^2\left[\frac{\Delta}{r^4}\right]'+\frac{\Delta^2}{r^4}\dot{r}_{*}\dot{r}_{*}'+\frac{2\Delta^2}{r^4}\dot{r}_{*}^2\frac{b'}{b}\right)+\nonumber\\
\nonumber\\
+ \chi \left[\frac{\Delta}{b} \dot{r}_{*}^2\left[\frac{\Delta}{r^4}\right]'b'+\frac{\Delta^2}{r^4}\dot{r}_{*}\dot{r}_{*}'\frac{b'}{b}+\frac{\Delta^2}{r^4}\dot{r}_{*}^2\frac{b''}{b}\right]-\nonumber\\
\nonumber\\
-\frac{A(r)}{r^2}(l+2)(l-1)\chi - \frac{\partial^2 \chi}{\partial
t^2}=0\ .
\end{eqnarray}
Queremos transformar a equação (\ref{pgb40}) em uma equação do
tipo
\begin{equation}
\label{pgb41} -\frac{\partial^2 \chi}{\partial
t^2}+\frac{\partial^2 \chi}{\partial r_{*}^2}= V(r_{*})\chi\ .
\end{equation}
Para tal, a equação (\ref{pgb40}) deve satisfazer as seguintes
condições
\begin{equation}
\label{pgb42}
\frac{\Delta^2}{r^4}\dot{r}_{*}^2 =1\ ,
\end{equation}
\begin{equation}
\label{pgb43}
\left(\Delta
\dot{r}_{*}^2\left[\frac{\Delta}{r^4}\right]'+\frac{\Delta^2}{r^4}\dot{r}_{*}\dot{r}_{*}'+\frac{2\Delta^2}{r^4}\dot{r}_{*}^2\frac{b'}{b}\right)=0\
.
\end{equation}
Resolvendo a equação (\ref{pgb42}) teremos
\begin{eqnarray}
\label{pgb44}
\dot{r}_{*}=\frac{r^2}{\Delta} \qquad &\Longrightarrow &\qquad dr_{*}=\frac{r^2}{\Delta}dr\ ,\\
\nonumber\\
\label{pgb45}
\frac{d}{dr_{*}}=\frac{\Delta}{r^2}\frac{d}{dr}\qquad
&\Longrightarrow &\qquad
\frac{d}{dr_{*}}=\sqrt{\frac{A(r)}{B(r)}}\frac{d}{dr}\ .
\end{eqnarray}
Resolvendo a equação (\ref{pgb43}) teremos
\begin{eqnarray}
\label{pgb46}
\left(\Delta \dot{r}_{*}^2\left[\frac{\Delta}{r^4}\right]'+\frac{\Delta^2}{r^4}\dot{r}_{*}\dot{r}_{*}'+\frac{2\Delta^2}{r^4}\dot{r}_{*}^2\frac{b'}{b}\right)=0\nonumber\\
\nonumber\\
r^2\left[\frac{1}{r^4}\frac{d\Delta}{dr}-\frac{4\Delta}{r^5}\right]+\frac{\Delta^2}{r^4}\left[\frac{2r}{\Delta}-\frac{r^2}{\Delta^2}\frac{d\Delta}{dr}\right]+2\frac{\Delta}{r^2}\frac{db}{dr}\frac{1}{b}=0\nonumber\\
\nonumber\\
-\frac{2\Delta}{r^3}+\frac{2\Delta}{r^2}\frac{db}{dr}\frac{1}{b}=0\nonumber\\
\nonumber\\
\frac{db}{dr}=\frac{b}{r}\qquad \Longrightarrow \qquad b(r)=r\ .
\end{eqnarray}
Dessa maneira nossa função de onda será
\begin{equation}
\label{pgb47} Q(r,\theta,t)=r\ \chi(r,t)\ C^{-3/2}_{l+2}(\theta)\ .
\end{equation}
Nossa equação será então
\begin{eqnarray}
\label{pgb48}
-\frac{\partial^2\chi}{\partial t^2}&+&\frac{\partial^2\chi}{\partial r_{*}^2}+
\chi \left[\frac{\Delta}{b} \dot{r}_{*}^2\left[\frac{\Delta}{r^4}\right]'b'+\frac{\Delta^2}{r^4}\dot{r}_{*}\dot{r}_{*}'\frac{b'}{b}+\frac{\Delta^2}{r^4}\dot{r}_{*}^2\frac{b''}{b}\right]-\nonumber\\
\nonumber\\
&-&\frac{A(r)}{r^2}(l+2)(l-1)\chi=0\ .
\end{eqnarray}
Chamemos
\begin{eqnarray}
\label{pgb49}
\widetilde{V}(r)= \left[\frac{\Delta}{b}
\dot{r}_{*}^2\left[\frac{\Delta}{r^4}\right]'b'+\frac{\Delta^2}{r^4}\dot{r}_{*}\dot{r}_{*}'\frac{b'}{b}+\frac{\Delta^2}{r^4}\dot{r}_{*}^2\frac{b''}{b}\right]-\frac{A(r)}{r^2}(l+2)(l-1).
\end{eqnarray}
Substituindo (\ref{pgb44}) e (\ref{pgb45}) na equação (\ref{pgb49}) teremos
\begin{eqnarray}
\label{pgb50}
\widetilde{V}(r)=\frac{\Delta}{b}\frac{d}{dr}\left(\frac{\Delta}{r^4}\right)\frac{db}{dr}&+&\frac{\Delta^3}{r^6}\frac{1}{b}\frac{d}{dr}\left(\frac{r^2}{\Delta}\right)\frac{db}{dr}+\frac{\Delta}{br^2}\frac{d}{dr}\left(\frac{\Delta}{r^2}\frac{db}{dr}\right)-\nonumber\\
\nonumber\\
&-&\frac{A(r)}{r^2}(l+2)(l-1)\ .
\end{eqnarray}
Usando o fato de que $b(r)=r$ temos
\begin{eqnarray}
\label{pgb51}
\widetilde{V}(r)=\frac{\Delta}{r}\left[\frac{1}{r^4}\frac{d\Delta}{dr}-\frac{4\Delta}{r^5}\right]&+&\frac{\Delta^3}{r^6}\frac{1}{r}\left[\frac{2r}{\Delta}-\frac{r^2}{\Delta^2}\frac{d\Delta}{dr}\right]+\frac{\Delta}{r^3}\left[\frac{1}{r^2}\frac{d\Delta}{dr}-\frac{2\Delta}{r^3}\right]-\nonumber\\
\nonumber\\
&-&\frac{A(r)}{r^2}(l+2)(l-1)\ .
\end{eqnarray}
Rearranjando os termos, nossa equação final para a perturbação gravitacional axial sobre a brana será
\begin{equation}
\label{pgb52}
-\frac{\partial^2 \chi}{\partial t^2}+\frac{\partial^2 \chi}{\partial r_{*}^2}+ \widetilde{V}(r)\chi=0
\end{equation}
com
\begin{equation}
\label{pgb53} \widetilde{V}(r)=
\frac{\Delta}{r^5}\frac{d\Delta}{dr}-\frac{4\Delta^2}{r^6}-\frac{A(r)}{r^2}(l+2)(l-1)\
.
\end{equation}
Especificaremos agora cada caso estudado.\\
\\
\\
Para o caso do BN-CFM temos que a função $\Delta$ é dada por
\begin{eqnarray}
\label{pgbCFM01}
\Delta = r(r-2M)\sqrt{\frac{2r-M\gamma}{2r-3M}}.
\end{eqnarray}
A função de onda que descreve a evolução do campo gravitacional será
\begin{eqnarray}
\label{pgbCFM02}
Q(r,t,\theta)=\chi(r,t)r\ C^{-3/2}_{\ell+2}(\theta),
\end{eqnarray}
onde $C^{-3/2}_{\ell+2}$ é a função de Gegenbauer.

O potencial $\widetilde{V}(r)^{CFM}$ fica
\begin{eqnarray}
\label{pgbCFM04}
\widetilde{V}(r)&=&\frac{(r-2M)}{r^4}\left\{\left(\frac{2r-M\gamma}{2r-3M}\right)(6M-2r)-\right.\nonumber \\
\nonumber \\
&-&\left.\left[(\ell+2)(\ell-1)r+\frac{r(r-2M)M(3-\gamma)}{(2r-3M)^2}\right]\right\}\
.
\end{eqnarray}
Substituindo $\gamma=\beta+3$ teremos
\begin{eqnarray}
\label{pgbCFM05}
\widetilde{V}(r)&=&\left(1-\frac{2M}{r}\right)\left\{\frac{6M-(\mu+2)r}{r^3}-\right.\nonumber\\
\nonumber\\
&-&\left.
\frac{M\beta}{r^3(2r-3M)}\left[\frac{(2r-3M)(6M-2r)-r(r-2M)}{(2r-3M)}\right]\right\}
\end{eqnarray}
onde $\mu =(\ell+2)(\ell-1)$ e $\ell$ são os índices de
multipolos.

Assim a equação que governa a evolução do campo gravitacional no
exterior do BN-CFM é dada por
\begin{equation}
\label{pgbCFM03} -\frac{\partial^2 \chi}{\partial
t^2}+\frac{\partial^2 \chi}{\partial r_{*}^2}+
\widetilde{V}(r)\chi=0\ ,
\end{equation}
com
\begin{eqnarray}
\label{pgbCFM06}
\widetilde{V}(r)&=&\left(1-\frac{2M}{r}\right)\left\{\frac{6M-(\mu+2)r}{r^3}-\right.\nonumber\\
\nonumber\\
&-&\left.
\frac{M\beta}{r^3(2r-3M)}\left[\frac{(2r-3M)(6M-2r)-r(r-2M)}{(2r-3M)}\right]\right\}
\end{eqnarray}
\\
\\
Para o caso do BN-SM a função $\Delta$ é dada por
\begin{eqnarray}
\label{pgbSM01}
\Delta = (r^2-h^2)\sqrt{1+\frac{C-h}{\sqrt{2r^2-h^2}}}.
\end{eqnarray}
A função de onda que descreve a evolução do campo gravitacional será
\begin{eqnarray}
\label{pgbSM02}
Q(r,t,\theta)=\chi(r,t)r\ C^{-3/2}_{\ell+2}(\theta),
\end{eqnarray}
onde $C^{-3/2}_{\ell+2}$ é a função de Gegenbauer.

O potencial $\widetilde{V}(r)^{SM}$ torna-se
\begin{eqnarray}
\label{pgbSM03}
\widetilde{V}(r)&=&\frac{(r^2-h^2)}{r^5}\sqrt{1+\frac{C-h}{\sqrt{2r^2-h^2}}}\left\{\ \sqrt{1+\frac{C-h}{\sqrt{2r^2-h^2}}}\left(\frac{h^2-2r^2}{r}\right)-\right.\nonumber \\
\nonumber \\
&-&\left.\left(1+\frac{C-h}{\sqrt{2r^2-h^2}}\right)^{-1/2}\left(\mu r+\frac{r(C-h)(r^2-h^2)}{(2r^2-h^2)^{3/2}}\right)\ \right\}\ ,\nonumber\\
\nonumber\\
\nonumber\\
\widetilde{V}(r)&=&\frac{(r^2-h^2)}{r^5}\left\{\ \left(1+\frac{C-h}{\sqrt{2r^2-h^2}}\right)\left(\frac{h^2-2r^2}{r}\right)-\right.\nonumber\\
\nonumber\\
&-&\left. \left(\mu
r+\frac{r(C-h)(r^2-h^2)}{(2r^2-h^2)^{3/2}}\right)\ \right\}\ ,
\end{eqnarray}
onde $\mu =(\ell+2)(\ell-1)$ e $\ell$ são os índices de
multipolos.

Portanto a equação que governa a evolução do campo gravitacional
no exterior do BN-SM é dada por
\begin{equation}
\label{pgbSM04}
-\frac{\partial^2 \chi}{\partial t^2}+\frac{\partial^2 \chi}{\partial r_{*}^2}+ \widetilde{V}(r)\chi=0
\end{equation}
com
\begin{eqnarray}
\label{pgbSM05}
\widetilde{V}(r)&=&\frac{(r^2-h^2)}{r^5}\left\{\ \left(1+\frac{C-h}{\sqrt{2r^2-h^2}}\right)\left(\frac{h^2-2r^2}{r}\right)-\right.\nonumber\\
\nonumber\\
&-&\left.\left(\mu
r+\frac{r(C-h)(r^2-h^2)}{(2r^2-h^2)^{3/2}}\right)\ \right\}\ .
\end{eqnarray}

\newpage

\section{Estabilidade de um BN-CFM e BN-SM}
Nesta seção faremos uma análise sobre a estabilidade do BN-CFM e
do BN-SM quando submetidos à perturbação de um campo  escalar
$\Psi$ não-massivo. Para tal usaremos o mesmo procedimento adotado no exemplo dado no
capítulo 4. A pergunta a ser respondida por nossa análise será se a
perturbação sobre o buraco negro cresce arbitrariamente com o
tempo. Se a resposta for positiva o buraco negro é instável, ou
seja, uma leve perturbação é capaz de destruí-lo.\\
\\
{\bf BN-CFM}\\
\\
Para o estudo da estabilidade do BN-CFM utilizaremos duas análises diferentes. Se for possível encontrar valores de $\beta$ para
o qual o potencial efetivo seja positivo definido podemos garantir
a estabilidade dos buracos negros pelos argumentos apresentados no
capítulo 4. Se o potencial efetivo apresentar alguma região onde ele possa
assumir valores negativos, qualquer que seja o valor de $\beta$,
precisaremos utilizar os resultados da solução numérica para
observar se o campo decai com o tempo, o que indicaria a
estabilidade do buraco negro.

A equação radial que governará a evolução da perturbação escalar
no exterior do BN-CFM será
\begin{eqnarray}
\label{BNMBest02A} \frac{\partial^2 R}{\partial
r_{*}^2}+(\omega^2- V_{esc}^{CFM}(r))R=0
\end{eqnarray}
onde
\begin{eqnarray}
\label{BNMBest03A}
V_{esc}^{CFM}(r)&=&\left(1-\frac{2M}{r}\right)\left\{\frac{\ell(\ell+1)}{r^2}+\frac{2M}{r^3}+\right.\\
\nonumber\\
&+&\left.
\frac{M\beta}{r(2r-3M)^2}\left[1-\frac{6M}{r}+\frac{6M^2}{r^2}\right]\right\}\ .
\end{eqnarray}
Investigando o potencial $V^{CFM}_{esc}$ para $M=1$ por exemplo, veremos que ele apresenta
três regimes distintos dependendo dos valores de $\beta$.

\begin{figure}[!ht]
\begin{center}
\epsfig{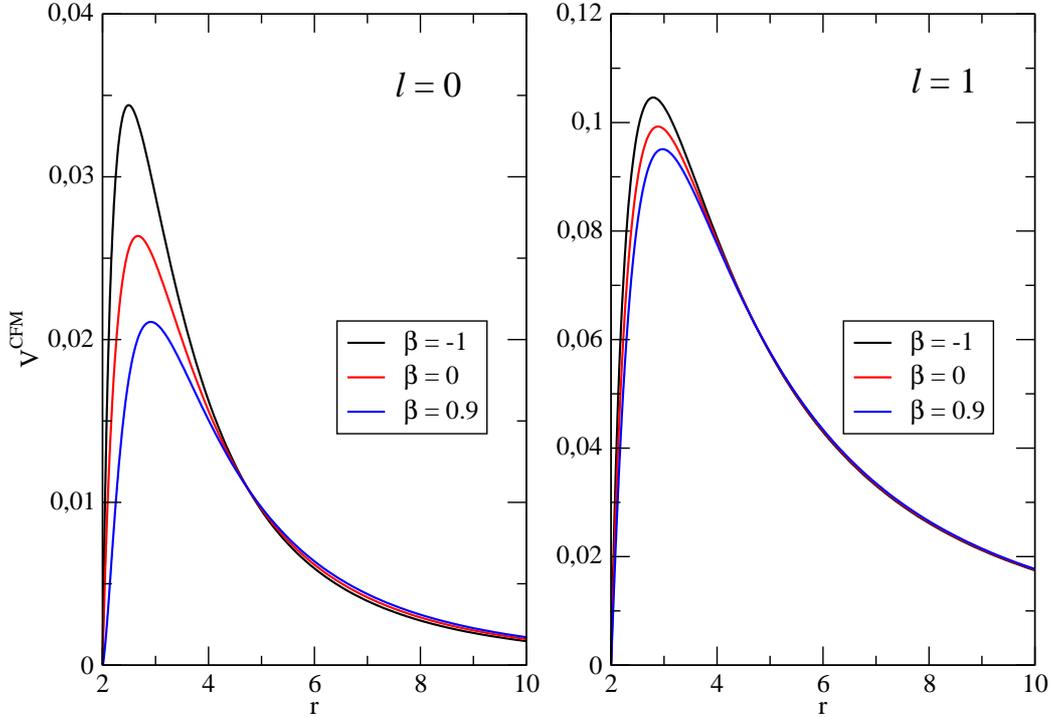}
\end{center}
\caption{{\small Evolução do campo escalar no BN-CFM com $M=1$
sobre $r_{*}=0$.}} \label{figestCFM3}
\end{figure}

O primeiro acontece quando ele é positivo definido,
independentemente do índice de multipolo $\ell$. Verificamos, via inspeção direta que este regime se dá quando $ 1 > \beta \gtrsim -8$, ou seja, para valores pequenos de $\beta$.\\
\newpage
Observemos o comportamento do potencial $V_{esc}^{CFM}$ na figura (\ref{figestCFM3}) para valores pequenos de $\beta$

Podemos ver que o potencial é nulo sobre o horizonte de eventos,
cresce até um máximo e vai a zero para valores de $r$ grandes. Portanto para esse intervalo de valores de $\beta$ podemos
garantir a estabilidade dos buracos negros devido ao fato do
potencial ser positivo definido. Os resultados obtidos numericamente no capítulo 8 concordam com
nosso critério de estabilidade uma vez que para esse intervalo de
valores de $\beta$ o campo decai com tempo.

O segundo regime acontece quando o potencial é inicialmente
positivo, mas apresenta um comportamento do tipo poço para valores
mais altos de $r$. Neste caso o potencial vai a zero para $r$ grande
por baixo. Esse comportamento se dá quando $\beta \lesssim -8$
dependendo do valor de $\ell$.

\begin{figure}[!ht]
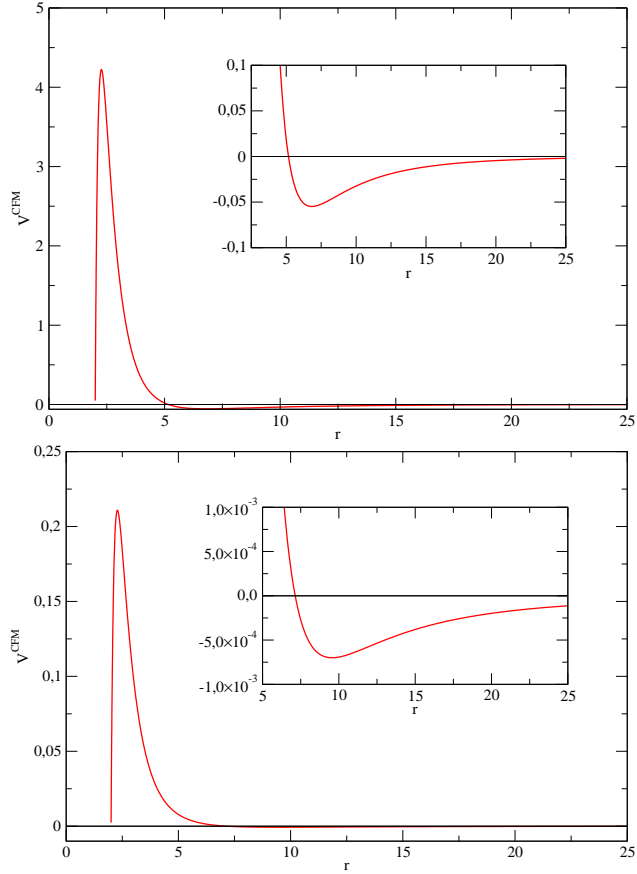

\begin{center}
\epsfig{file = g2.eps, width=0.6 \linewidth, clip=}\\
\epsfig{file =g3.eps, width=0.6 \linewidth, clip=}
\end{center}
\caption{{\small Evolução do campo escalar no BN-CFM com $M=1$,
$\beta=-390$ e $\ell=1$ sobre $r_{*}=0$ (acima) e $M=1$,
$\beta=-15$ e $\ell=0$ sobre $r_{*}$=0 (embaixo).}}
\label{figestCFM1}
\end{figure}

Neste caso precisaremos dos resultados numéricos para sabermos se
o campo decai com o tempo.
\newpage
Vejamos na figura (\ref{figestCFM1}) exemplos onde, o potencial
apresenta o comportamento referido anteriormente.

O potencial é zero sobre o horizonte de eventos, crescendo até um
máximo, caindo até um ponto de minímo e indo a zero para valores
de $r$ grande. Nos detalhes das figuras podemos ver que os poços
de potenciais são bem rasos quando comparados com os máximos dos
potenciais. Para os respectivos casos acima mostrados temos as
seguintes relações de proporção entre os máximos e mínimos dos
potenciais:
\begin{eqnarray}
\label{BNMBest04A}
\left|\frac{V_{min}^{CFM}}{V_{max}^{CFM}}\right|\approx 10^{-3} \qquad
\left|\frac{V_{min}^{CFM}}{V_{max}^{CFM}}\right|\approx 10^{-2}\ .
\end{eqnarray}
Essa diferença de magnitude explica porque, mesmo apresentando
regiões onde os potenciais são negativos, os buracos negros
estudados devem ser estáveis. Como os poços são muito rasos os potenciais podem ser encarados
como ``quase" positivos definidos entrando assim no critério de
estabilidade. Mais uma vez, a evolução numérica do campo mostrada
no capítulo 8 confirma a estabilidade dos buracos negros em
questão uma vez que os campos decaem a zero com o tempo.

Há ainda um terceiro regime quando o potencial parte do zero
crescendo até um ponto de máximo primário, apresentando em seguida
uma região intermediária do tipo poço que cresce novamente até
alcançar um ponto de máximo secundário caindo a zero para valores
altos de $r$. Esse regime se dá quando, $\beta$ assume valores
grandes e negativos e um certo valor de $\ell$ é assumido. Neste
caso também precisaremos dos resultados numéricos para sabermos se
o campo decai com o tempo.

\begin{figure}[!h]
\begin{center}
\epsfig{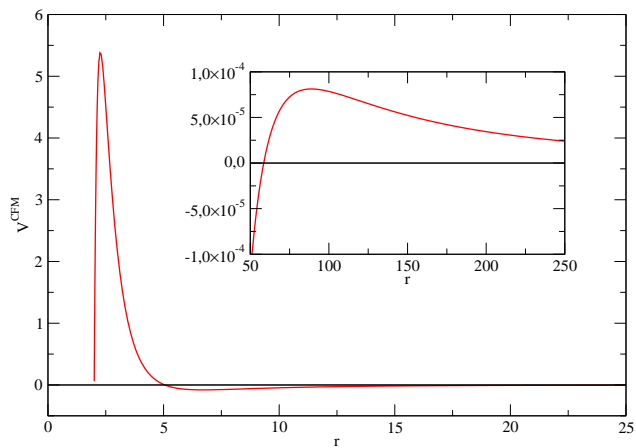}
\end{center}
\caption{{\small Evolução do campo escalar no BN-CFM com $M=1$,
$\beta=-500$ e $\ell=5$ sobre $r_{*}=0$}}
\label{figestCFM2}
\end{figure}

Vejamos na figura (\ref{figestCFM2}) um exemplo onde, o potencial
apresenta o comportamento referido anteriormente.

Observamos no detalhe da figura que o poço de potencial é raso em
comparação com o valor do máximo primário. Neste caso a relação de
proporção entre o máximo e o mínimo é
\begin{eqnarray}
\label{BNMBest05A}
\left|\frac{V_{min}^{CFM}}{V_{max}^{CFM}}\right|\approx 10^{-2}.
\end{eqnarray}
Entretanto, como declarado no regime anterior, nada podemos dizer
sobre a estabilidade desse buraco negro sem avaliar os resultados
numéricos da evolução do campo escalar.\\
\\
{\bf BN-SM}\\
\\
Para o estudo da estabilidade do BN-SM utilizaremos novamente o
mesmo procedimento adotado para o BN-CFM.

Se for possível valores de $C$ para
o qual o potencial efetivo seja positivo definido podemos garantir
a estabilidade dos buracos negros pelos argumentos apresentados no
capítulo 4.

Se o potencial efetivo apresentar alguma região onde ele possa
assumir valores negativos, qualquer que seja o valor de $C$,
precisaremos utilizar os resultados da solução numérica para
observar se o campo decai com o tempo, o que indicaria a
estabilidade do buraco negro.\\
A equação radial que governará a evolução da perturbação escalar
no exterior do BN-SM será
\begin{eqnarray}
\label{BNMBest02B} \frac{\partial^2 R}{\partial
r_{*}^2}+(\omega^2- V_{esc}^{SM}(r))R=0
\end{eqnarray}
onde
\begin{eqnarray}
\label{BNMBest03B}
V_{esc}^{SM}(r)&=&\left(1-\frac{h^2}{r^2}\right)\left\{\frac{\ell(\ell+1)}{r^2}+\frac{2h^2}{r^4}\right.\nonumber\\
\nonumber\\
&+&\left.
\frac{C-h}{\sqrt{2r^2-h^2}}\left[\frac{2h^2}{r^4}-\frac{r^2-h^2}{r^2(2r^2-h^2)}\right]\right\}\ .
\end{eqnarray}
Investigando o potencial $V^{SM}_{esc}$ veremos que ele também
apresenta três regimes distintos dependendo dos valores de $C$.

Diferentemente do potencial $V_{esc}^{CFM}$, o comportamento do
potencial $V_{esc}^{SM}$ depende mais fortemente dos valores
adotados para $h$ e $\ell$. Essa influência pode ser observada no último termo do potencial.
Dependendo do ajuste entre $C$ e $h$ esse termo pode ser negativo
ou positivo podendo alterar o comportamento assintótico de
$V_{esc}^{SM}$ quando $r\rightarrow \infty$.

Portanto precisamos estudar caso a caso para sabermos se algum
buraco negro dessa classe é estável.

Apresentaremos apenas um exemplo para ilustrar o comportamento dos
potenciais. Estudaremos o caso para $h=1$.

O primeiro regime acontece quando o potencial é positivo definido.
No caso desse buraco negro esse regime só aparece para $C=h=1$ de
modo independente do valor de $\ell$.
\newpage
Observemos o comportamento do potencial $V_{esc}^{SM}$ nesta
condição.
\begin{figure}[!ht]
\begin{center}
\epsfig{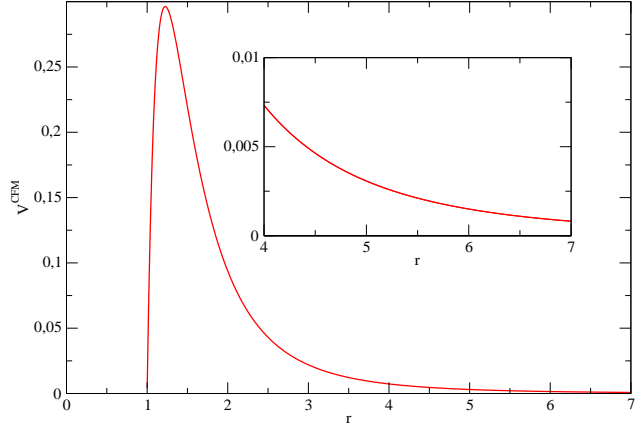}
\end{center}
\caption{{\small Evolução do campo escalar no BN-SM com $h=1$, $\ell=0$, $C=1$
sobre $r_{*}=0$.}} \label{figestSM1}
\end{figure}

Podemos ver que o potencial é nulo sobre o horizonte de eventos $h$,
cresce até um máximo e vai a zero para valores de $r$ grandes.

Como o potencial é positivo definido, o buraco negro com $C=1$ pode ser considerado estável. Os resultados numéricos da evolução do campo
obtida no capítulo 8 confirmam a estabilidade do buraco negro para
esses valores de $h$, $C$ e $\ell$.

O segundo regime acontece quando o potencial é inicialmente
positivo, mas apresenta um comportamento do tipo poço para valores
mais altos de $r$. Neste caso o potencial vai a zero para $r$ grande
por baixo. Ele se dá para valores de $C>1$ dependendo do valor de $\ell$. 

\begin{figure}[!ht]
\begin{center}
\epsfig{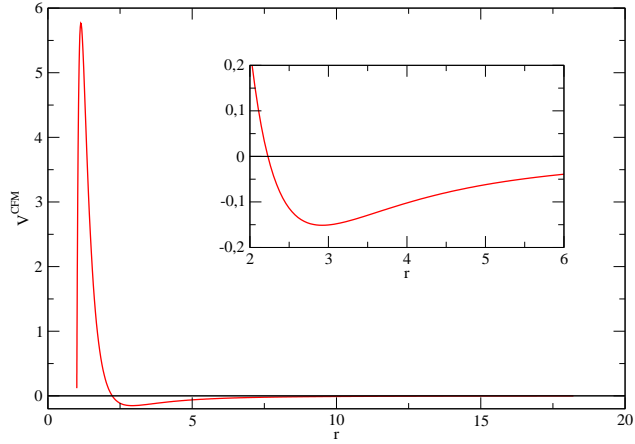}
\end{center}
\caption{{\small Evolução do campo escalar no BN-SM com $h=1$, $\ell=0$, $C=30$
sobre $r_{*}=0$.}} \label{figestSM2}
\end{figure}

Esse regime é ilustrado na figura (\ref{figestSM2}).

Podemos ver que o potencial é zero sobre o horizonte de eventos,
crescendo até um máximo, caindo em seguida até um ponto de mínimo
e indo a zero para valores de $r$ grande. No detalhe da figura
(\ref{figestSM2}) podemos ver, novamente como no caso do BN-CFM,
que o poço de potencial é raso quando comparado com o máximo do
potencial.

Para este caso temos a seguinte relação de proporção entre o máximo e mínimo do potencial:
\begin{eqnarray}
\label{BNMBest04B}
\left|\frac{V_{min}^{CFM}}{V_{max}^{CFM}}\right|\approx 10^{-2}\ .
\end{eqnarray}
Como o poço é muito raso o potencial pode ser encarado como
``quase" positivo definido entrando assim no critério de
estabilidade.

Entretanto apenas uma investigação mais profunda da evolução do
campo escalar com tempo pode definir se esses buracos negros são
estáveis. Neste caso precisaremos dos resultados numéricos para
sabermos se o campo cai com o tempo.

Há ainda um terceiro regime quando o potencial é nulo sobre o horizonte de eventos crescendo até um ponto de máximo primário, apresentando em seguida
uma região intermediária do tipo poço que cresce novamente até
alcançar um ponto de máximo secundário caindo a zero para valores
altos de $r$. Esse regime se dá quando, $C > 34$ e um certo valor de $\ell$ é assumido.

Neste caso também precisaremos dos resultados numéricos para
sabermos se o campo cai com o tempo.

\begin{figure}[!h]
\begin{center}
\epsfig{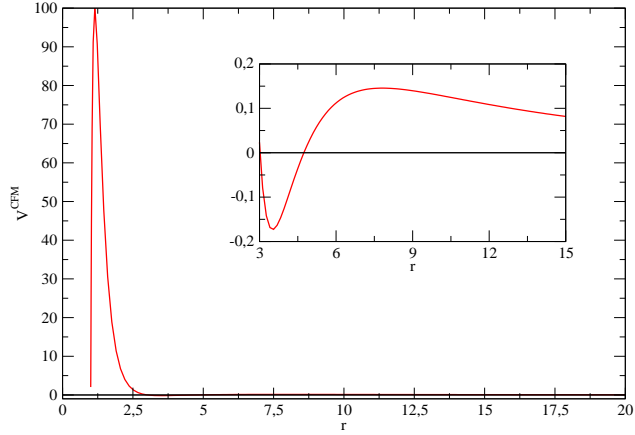}
\end{center}
\caption{{\small Evolução do campo escalar no BN-CFM com $h=1$, $C=500$ e $\ell=5$ sobre $r_{*}=0$}}
\label{figestSM3}
\end{figure}

Vejamos na figura (\ref{figestSM3}) um exemplo onde, o potencial
apresenta o comportamento referido anteriormente.

Novamente observamos que o poço de potencial é raso em comparação
com o valor do máximo primário. Neste caso a relação de proporção
entre o máximo e o mínimo é
\begin{eqnarray}
\label{BNMBest05B}
\left|\frac{V_{min}^{CFM}}{V_{max}^{CFM}}\right|\approx 10^{-3}.
\end{eqnarray}
Entretanto, como declarado no regime anterior, nada podemos dizer
sobre a estabilidade desse buraco negro sem avaliar os resultados
numéricos da evolução do campo escalar.

\section{Termodinâmica de BN-CFM e BN-SM}
O tratamento termodinâmico apresentado no capítulo 5 será
utilizado nesta seção para o cálculo do limite superior da
entropia $S_{m}$ de um corpo arbitrário caindo no BN-CFM e  no
BN-SM que estão localizados sobre a brana. O formalismo
desenvolvido não necessita de nenhuma adaptação no contexto de
branas diferentemente da evolução de
campos.\\
\\
{\bf Buraco Negro-CFM}\\
\\
Seja um corpo de raio próprio $R$ e massa própria $m$ caindo livremente no BN-CFM.
O horizonte de eventos $r_{h}$ para o BN-CFM é dado por
\begin{eqnarray}
\label{BNMBtermo00A}
g_{tt}|_{r=r_{h}}=0 \qquad \Longrightarrow \left(1-\frac{2M}{r}\right)=0 \qquad \Longrightarrow r_{h}=2M\ .
\end{eqnarray}
A gravidade superficial $\kappa$ para esse buraco negro sobre seu
horizonte de eventos será
\begin{eqnarray}
\label{BNMBtermo01A} \kappa=\frac{1}{2M}\sqrt{1-\frac{\gamma}{4}}.
\end{eqnarray}
As constantes de movimento associadas a $t$ e $\phi$ quando
consideramos o movimento do corpo no plano equatorial
$\theta=\pi/2$ serão
\begin{eqnarray}
\label{BNMBtermo02A}
E&=&-\left(1-\frac{2M}{r}\right)\dot{t},\\
\nonumber\\
J&=&-r^2\dot{\phi}.
\end{eqnarray}
Portanto a equação quadrática para a energia conservada $E$ do
corpo que está caindo no buraco negro será
\begin{eqnarray}
\label{BNMBtermo03A}
\alpha E^2-2\beta E +\sigma=0
\end{eqnarray}
com
\begin{eqnarray}
\alpha&=&r^4,\nonumber\\
\beta&=&0,\nonumber\\
\sigma&=&-r(r-2M)(m^2r^2+J^2)\ .\nonumber
\end{eqnarray}
Como nosso corpo não é uma partícula, ele possui dimensão e
conseqüentemente um centro de massa. Portanto obtemos o novo ponto
de captura $r=2M+\delta$ calculando $\delta$ como
\begin{eqnarray}
\label{BNMBtermo04A} \int_{2M}^{2M+\delta}\
\sqrt{\frac{r(r-3M/2)}{(r-2M)(r-M\gamma/2)}}\ dr=\ R\ .
\end{eqnarray}
Integrando a equação (\ref{BNMBtermo04A}) e realizando uma expansão em série teremos
\begin{eqnarray}
\label{BNMBtermo05A}
\delta=\frac{R^2}{4M^2}\left(2M-\frac{M\gamma}{2}\right)\ .
\end{eqnarray}
Igualando a energia $E$ sobre o ponto de captura $r=2M+\delta$
teremos após uma expansão até ordem de $\delta^{1/2}$, o valor da
energia de captura $E_{cap}$
\begin{eqnarray}
\label{BNMBtermo06A}
E_{cap}=\left[\frac{(m^24M^2+J^2)(2M-M\gamma/2)}{(2M)^5}\right]^{1/2}R\ .
\end{eqnarray}
Conseqüentemente a energia mínima $E_{min}$, que é obtida quando
$J=0$, será
\begin{eqnarray}
\label{BNMBtermo07A}
E_{min}=\left[\frac{m^2(2M-M\gamma/2)}{(2M)^3}\right]^{1/2}R\ .
\end{eqnarray}
Da Primeira Lei da termodinâmica de buracos negros temos
\begin{eqnarray}
\label{BNMBtermo08A}
dM=\frac{\kappa}{2}\ dA_{r}\ .
\end{eqnarray}
Lembrando que da equação de Einstein $E=mc^2$ temos, $dM=E_{min}$
e substituindo o valor de $\kappa$ sobre o horizonte de eventos na
equação acima teremos
\begin{eqnarray}
\label{BNMBtermo09A}
E_{min}&=&\frac{1}{4M}\sqrt{1-\frac{\gamma}{4}}\ dA_{r}\ ,\nonumber\\
\nonumber\\
\label{BNMBtermo10A}
\left[\frac{m^2(2M-M\gamma/2)}{(2M)^3}\right]^{1/2}R&=&\frac{1}{4M}\sqrt{1-\frac{\gamma}{4}}\ dA_{r}\ ,\nonumber\\
\nonumber\\
dA_{r}&=&2mR\ .
\end{eqnarray}
Por fim assumindo ser válida a Segunda Lei Generalizada,
$S_{bn}(M+dM)\geq S_{bn}(M)+S_{m}$, obtemos um limite superior
para a entropia $S_{m}$ associada a um corpo com energia própria
$E$, absorvido pelo BN-CFM, que é dado por
\begin{eqnarray}
\label{BNMBtermo11A}
\frac{A+dA}{4}&\geq& \frac{A}{4}+S_{m}\ ,\nonumber\\
\nonumber\\
\frac{dA}{4}&\geq& S_{m}\ ,\nonumber\\
\nonumber\\
\frac{4\pi \ dA_{r}}{4}&\geq& S_{m}\ ,\nonumber\\
\nonumber\\
\label{BNMBtermo12A}
S_{m}&\leq& 2\pi ER\ .
\end{eqnarray}
Como podemos observar, a influência do {\it bulk} não se manifesta
no limite superior da entropia $S_{m}$, uma vez que o limite
independe do parâmetro $\beta$. Isso confirma a universalidade
desse limite no sentido de que o limite depende apenas dos
parâmetros do corpo em queda, independendo dos parâmetros do
buraco negro.\\
\\
{\bf Buraco Negro-SM}\\
\\
Seja um corpo de raio próprio $R$ e massa própria $m$ caindo livremente no BN-SM.
Calculemos o horizonte de eventos $r_{h}$ para o BN-SM
\begin{eqnarray}
\label{BNMBtermo00B}
g_{tt}|_{r=r_{h}}=0 \qquad \Longrightarrow \left(1-\frac{h^2}{r^2}\right)=0 \qquad \Longrightarrow r_{h}= h\ .
\end{eqnarray}
A gravidade superficial $\kappa$ para BN-SM sobre seu
horizonte de eventos será
\begin{eqnarray}
\label{BNMBtermo01B}
\kappa=\frac{1}{h}\sqrt{\frac{C}{h}}\ .
\end{eqnarray}
As constantes de movimento associadas a $t$ e $\phi$ quando
consideramos o movimento do corpo no plano equatorial
$\theta=\pi/2$ serão
\begin{eqnarray}
\label{BNMBtermo02B}
E&=&-\left(1-\frac{h^2}{r^2}\right)\dot{t}\ ,\\
\nonumber\\
J&=&-r^2\dot{\phi}\ .
\end{eqnarray}
Portanto a equação quadrática para a energia conservada $E$ do
corpo que está caindo no BN-SM será
\begin{eqnarray}
\label{BNMBtermo03B}
\alpha E^2-2\beta E +\sigma=0\
\end{eqnarray}
com
\begin{eqnarray}
\alpha&=&r^4\ ,\nonumber\\
\beta&=&0\ ,\nonumber\\
\sigma&=&-(r^2-h^2)(m^2r^2+J^2)\ .\nonumber
\end{eqnarray}
Como nosso corpo não é uma partícula, ele possui dimensão e
conseqüentemente um centro de massa. Portanto o ponto de captura
do corpo será $r=r_{h}+\delta$. O valor de $\delta$ é dado por
\begin{eqnarray}
\label{BNMBtermo04B} \int_{h}^{h+\delta}\
\sqrt{\frac{1}{\left(1-\frac{h^2}{r^2}\right)\left(1+\frac{C-h}{\sqrt{2r^2-h^2}}\right)}}\ dr=\ R\nonumber\\
\end{eqnarray}
onde após uma integração e uma expansão série até a ordem de $\delta^{1/2}$ teremos
\begin{eqnarray}
\label{BNMBtermo05B}
\delta=\frac{CR^2}{2h^2}\ .
\end{eqnarray}
Igualando a energia $E$ sobre o ponto de captura $r=h+\delta$
teremos o valor da energia de captura $E_{cap}$ que foi calculado
de maneira análoga ao buraco negro anterior após uma expansão em
série
\begin{eqnarray}
\label{BNMBtermo06B}
E_{cap}=\left[\frac{(m^2h^2+J^2)CR^2}{h^5}\right]^{1/2}\ .
\end{eqnarray}
Conseqüentemente a energia mínima $E_{min}$, que é obtida quando
$J=0$, será
\begin{eqnarray}
\label{BNMBtermo07B}
E_{min}=\left[\frac{CR^2m^2}{h^3}\right]^{1/2}\ .
\end{eqnarray}
Da Primeira Lei da termodinâmica de buracos negros temos
\begin{eqnarray}
\label{BNMBtermo08B} dM=\frac{\kappa}{2}\ dA_{r}\ .\nonumber\\
\end{eqnarray}
Lembrando que $dM=E_{min}$ e substituindo o valor de $\kappa$ sobre
o horizonte de eventos na equação acima teremos
\begin{eqnarray}
\label{BNMBtermo09B}
E_{min}&=&\frac{1}{2h}\sqrt{\frac{C}{h}}\ dA_{r}\ ,\nonumber\\
\nonumber\\
\label{BNMBtermo10B}
\sqrt{\frac{m^2CR^2}{h^3}}&=&\frac{1}{2h}\sqrt{\frac{C}{h}}\ dA_{r}\ ,\nonumber\\
\nonumber\\
dA_{r}&=&2mR\ .
\end{eqnarray}
Por fim supondo ser válida a Segunda Lei Generalizada,
$S_{bn}(M+dM)\geq S_{bn}(M)+S_{m}$, obtemos um limite superior
para a entropia $S_{m}$ associada a um corpo com energia própria
$E$, absorvido pelo BN-SM, que é dado por
\begin{eqnarray}
\label{BNMBtermo11B}
\frac{A+dA}{4}&\geq& \frac{A}{4}+S_{m}\ ,\nonumber\\
\nonumber\\
\frac{dA}{4}&\geq& S_{m}\ ,\nonumber\\
\nonumber\\
\frac{4\pi \ dA_{r}}{4}&\geq& S_{m}\ ,\nonumber\\
\nonumber\\
\label{BNMBtermo12B}
S_{m}&\leq& 2\pi ER\ .
\end{eqnarray}
Apesar deste buraco negro possuir propriedades um tanto quanto
exóticas como não possuir uma massa de Schwarzschild, ele também
obedece ao limite superior sendo independente das características
do {\it bulk}.
\newpage

\chapter[Resultados Númericos]{Resultados Numéricos}
{\it Encerrando nossa jornada observamos que, a máquina bem dirigida, aliada ao espírito de investigação, nos proporciona grandes desenvolvimentos no entendimento do Universo e permite-nos  trilhar novos caminhos com alguma segurança.}\\
\\
Neste capítulo apresentaremos os resultados obtidos para a
evolução dos campos no exterior do BN-CFM e do BN-SM e o método numérico utilizado.

\section{Apresentação do Método Numérico}
O método númerico utilizado para a solução das equações
diferenciais que descrevem a evolução dos campos no exterior dos
buracos negros estudados é o mesmo utilizado em \cite{Carlos}.

\subsection{Problema de Condições Iniciais Características}
Trata-se do método de integração com condições iniciais características, baseado na especificação de condições iniciais em hipersuperfícies nulas.\\
Neste trabalho utilizamos um esquema específico que é conhecido como ``problema das duplas coordenadas nulas".

Como as equações que governam a evolução dos campos no exterior do buraco negro se reduziram a uma equação em $(1+1)$ dimensões devemos utilizar um esquema compatível com essas condições.

\begin{figure}[h]
\begin{center}
\epsfig{file = 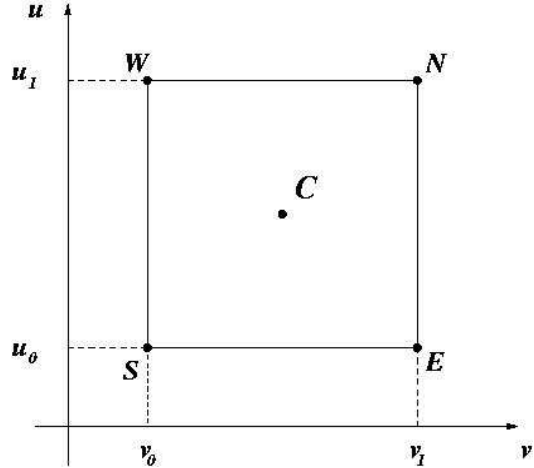, width=0.5 \linewidth, clip=}
\end{center}
\caption{Ilustração do esquema de duplas coordenadas nulas.}
\label{fignm1}
\end{figure}

A equação de movimento, escrita usando as coordenadas nulas $u=t-r_{*}$ e $v=t+r_{*}$, será
\begin{eqnarray}
\label{MN01}
\frac{\partial^2\psi(u,v)}{\partial u\partial v}=-\frac{1}{4}V(r(u,v))\psi(u,v).
\end{eqnarray}
Esse esquema consiste em especificarmos o campo $\psi$ na fronteira de um ângulo delimitado pelas semi-retas $u=u_{0}(v\geq v_{0})$ e $v_{0}=v_{0}(u \geq u_{0})$, que se interceptam no ponto $(u_{0},v_{0})$ conforme pode ser visto na figura \ref{fignm1}
\subsection{Discretização}
Passemos ao método para a discretização do plano $u-v$.

Uma possível forma de obtenção deste método é transformar a equação (\ref{MN01}) em uma equação integral. Integrando a equação (\ref{MN01}) em um retângulo nulo do plano $u-v$ delimitado pelos pontos $N$, $S$, $E$, $W$ teremos
\begin{eqnarray}
\label{MN02}
\int \frac{\partial^2\psi(u,v)}{\partial u\partial v}\ du\ dv =-\frac{1}{4}\int V(u,v)\psi(u,v)\ du\ dv.
\end{eqnarray}
Escrevemos as integrais de superfície como
\begin{eqnarray}
\label{MN03}
\int_{v_{0}}^{v_{1}} \left[\int_{u_{0}}^{u_{1}}\frac{\partial^2\psi(u,v)}{\partial u\partial v}\ du\right]\ dv =-\frac{1}{4}\int_{v_{0}}^{v_{1}} \left[\int_{u_{0}}^{u_{1}}  V(u,v)\psi(u,v)\ du\right]\ dv.
\end{eqnarray}
Resolvendo o lado esquerdo da equação (\ref{MN03}) teremos
\begin{eqnarray}
\label{MN04}
\int_{v_{0}}^{v_{1}} \left[\int_{u_{0}}^{u_{1}}\frac{\partial^2\psi(u,v)}{\partial u\partial v}\ du\right]\ dv &=&\int_{v_{0}}^{v_{1}} \frac{\partial}{\partial v}\left[\int_{u_{0}}^{u_{1}}\frac{\partial \psi(u,v)}{\partial u}\ du\right]\ dv \nonumber\\
\nonumber\\
&=& \psi(u_{1},v_{1})-\psi(u_{1},v_{0})-\psi(u_{0},v_{1})+\psi(u_{0},v_{0})\nonumber\\
\nonumber\\
&=& \psi(N)-\psi(W)-\psi(E)+\psi(S)\ .
\end{eqnarray}
Assim a equação (\ref{MN02}) se reduz a
\begin{eqnarray}
\label{MN05}
\psi(N)-\psi(W)-\psi(E)+\psi(S)=-\frac{1}{4}\int_{v_{0}}^{v_{1}}\ \int_{u_{0}}^{u_{1}}  V(u,v)\psi(u,v)\ du\ dv.
\end{eqnarray}
Definiremos uma função $p(u,v)$ como
\begin{eqnarray}
\label{MN06}
p(u,v)=V(u,v)\psi(u,v)
\end{eqnarray}
de forma que
\begin{eqnarray}
\label{MN07}
\int_{v_{0}}^{v_{1}}\ \int_{u_{0}}^{u_{1}}  V(u,v)\psi(u,v)\ du\ dv=\int_{v_{0}}^{v_{1}}\ \int_{u_{0}}^{u_{1}}  p(u,v)\ du\ dv.
\end{eqnarray}
Sem perda de generalidade, escolheremos a origem do plano $u-v$ de modo que o ponto $(0,0)$ esteja sobre o centro do retângulo. Desta forma a integral (\ref{MN07}) fica
\begin{eqnarray}
\label{MN08}
\int_{v_{0}}^{v_{1}}\ \int_{u_{0}}^{u_{1}}  p(u,v)\ du\ dv=\int_{-\Delta v/2}^{+\Delta v/2}\ \int_{-\Delta u/2}^{+\Delta u/2}  p(u,v)\ du\ dv.
\end{eqnarray}
Usando o teorema do valor médio teremos
\begin{eqnarray}
\label{MN09}
\int_{-\Delta v/2}^{+\Delta v/2}\ \int_{-\Delta u/2}^{+\Delta u/2}  p(u,v)\ du\ dv = p(u',v')\Delta u\Delta v,
\end{eqnarray}
onde $(u',v')\in [-\Delta u/2,+\Delta u/2]\otimes[-\Delta v/2,+\Delta v/2]$. Supondo que $p(u,v)$ seja pelo menos diferenciável duas vezes em $u$ e $v$, podemos expandir $p(u',v')$ em termos de $p(0,0)$, de forma que
\begin{eqnarray}
\label{MN10}
p(u',v')=p(0,0)+\left. \frac{\partial p}{\partial u}\right|_{(0,0)} u'+\left. \frac{\partial p}{\partial v}\right|_{(0,0)} v'+\ o(\Delta^2)\ .
\end{eqnarray}
Então podemos reescrever a equação (\ref{MN09}) na forma
\begin{eqnarray}
\label{MN11}
\int_{-\Delta v/2}^{+\Delta v/2}\ \int_{-\Delta u/2}^{+\Delta u/2}  p(u,v)\ du\ dv &=& p(0,0)\Delta u\ \Delta v+\Delta v \int_{-\Delta u/2}^{+\Delta u/2}  u' du\nonumber\\
&+&\Delta u \int_{-\Delta v/2}^{+\Delta v/2}  v'\ dv+O(\Delta^4)\nonumber\\
\nonumber\\
&=&p(0,0)\Delta u\ \Delta v+o(\Delta^4).
\end{eqnarray}
Substituindo a equação (\ref{MN11}) na equação integral de movimento (\ref{MN05}) teremos
\begin{eqnarray}
\label{MN12}
\psi(N)-\psi(W)-\psi(E)+\psi(S)=-\frac{1}{4}\ V(C)\psi(C)\Delta u\ \Delta dv + o(\Delta^4).
\end{eqnarray}
Por outro lado, podemos fazer a seguinte consideração
\begin{eqnarray}
\label{MN13}
\psi(C)=\frac{1}{2}\left[\psi(E)+\psi(W)\right]+ o(\Delta^4).
\end{eqnarray}
Lembrando que o potencial depende apenas de $r$ e que conseqüentemente ele tem o mesmo valor nos pontos $C$ e $S$
\begin{eqnarray}
\label{MN14}
V(C)=V(S),
\end{eqnarray}
podemos escrever nossa equação de movimento de forma discretizada como
\begin{eqnarray}
\label{MN15}
\psi(N)=\psi(E)+\psi(W)-\psi(S)-\frac{V(S)\Delta u\Delta v}{8}\left[\psi(E)+\psi(W)\right]+ o(\Delta^4)
\end{eqnarray}
A equação (\ref{MN15}) é a base do método numérico utilizado neste
trabalho assim como do trabalho \cite{elcio}.
\newpage
\section{Resultados}
\begin{figure}[!ht]
\begin{center}
\epsfig{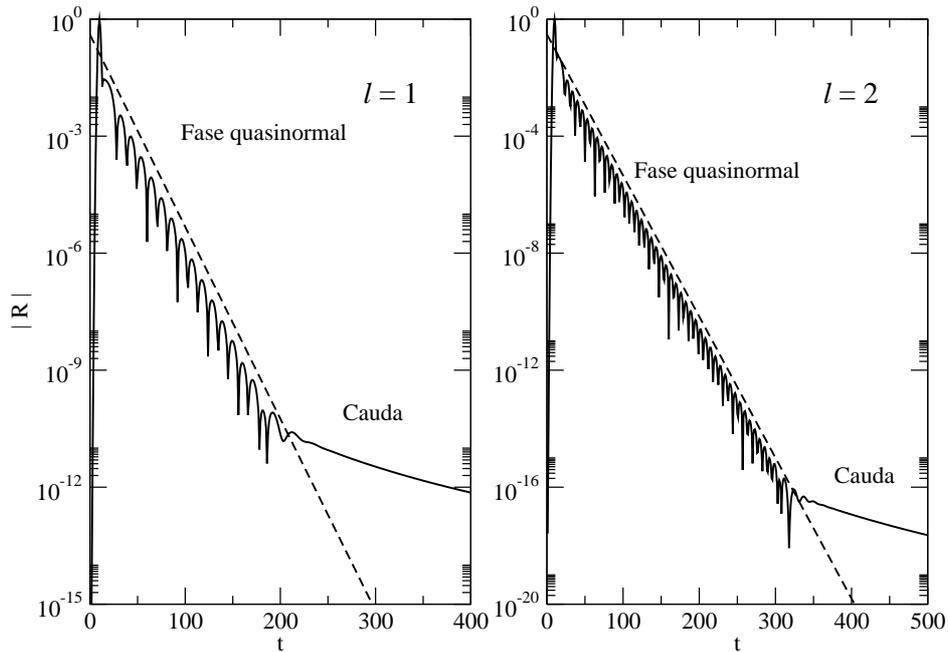}
\end{center}
\caption{Evolução temporal do campo escalar dado por $R(r,t)$ em
$r_{*}=0$ com $\beta =-1$.} \label{figevCFM2}
\end{figure}

Nesta seção apresentaremos a evolução dos campos, obtidas das
soluções das equações de perturbação através do cálculo
numérico. A estabilidade dos buracos negros, os regimes quasi-normais e os
comportamentos assintóticos são discutidos.\\
\\
{\bf BN-CFM}\\
\\
A função de onda que representa a evolução do campo escalar não
massivo $\Psi $ no exterior do BN-CFM é dada por
\begin{eqnarray}
\label{resultCFM01}
\Psi(r,t,\theta,\phi)=\frac{R(r,t)}{r}Y_{\ell m}(\theta,\phi)\ .
\end{eqnarray}
O comportamento de $R(r,t)$, que é solução da parte radial da
equação de Klein-Gordon, no exterior do BN-CFM é apresentado na figura \ref{figevCFM2} para valores de
$\ell=1,2$.

Como podemos observar, a evolução do campo escalar possui três regimes distintos.

Inicialmente o campo apresenta uma fase transiente. Logo após temos a fase quasi-normal, sendo
atenuada exponencialmente. Essa atenuação exponencial é
representada pela reta decrescente tracejada no gráfico. Nesta fase o campo
oscila com freqüência quasi-normal $\omega=\omega_{R}+i\ \omega_{i}$
onde a parte imaginária da freqüência $\omega_{i}$ é responsável
pelo atenuamento.

Após essa fase o campo apresenta um decaimento na forma de lei de
potências. Essa fase posterior é conhecida como a cauda da
perturbação. Como podemos observar, a influência do {\it bulk} não alterou a
forma do comportamento do campo. Diferentemente da perturbação
escalar de uma corda negra, o comportamento assintótico, para
tempos suficientemente grandes, do BN-CFM, é idêntico ao buraco
negro de Schwarzschild $(3+1)$ dimensional.

\begin{figure}[!ht]
\begin{center}
\epsfig{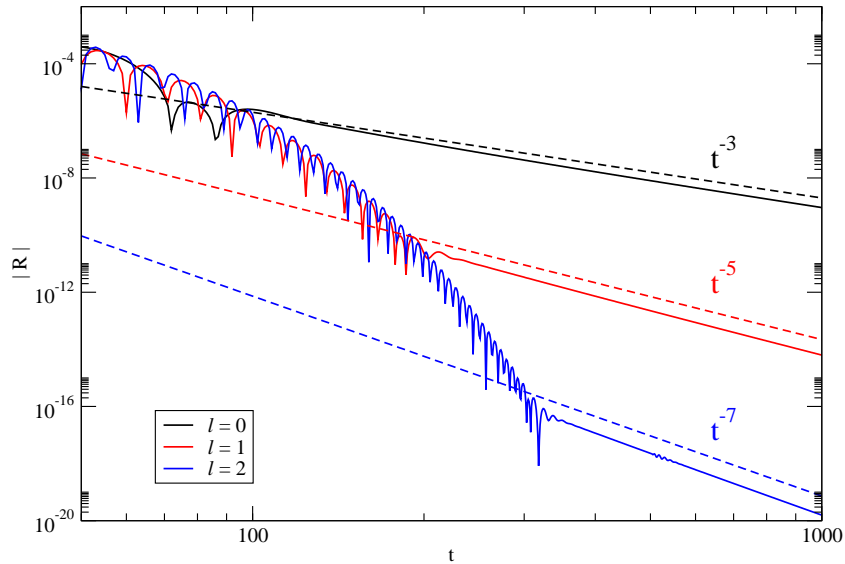}
\end{center}
\caption{\small{Caudas para o BN-CFM com $M=1$, $\beta=-1$, $r_{*}=0$ e multipolos $\ell=0,1,2$.}}
\label{figevCFM1}
\end{figure}

O comportamento das caudas da perturbação escalar para o BN-CFM
com $\beta =-1$ sobre o horizonte de eventos é apresentado na
figura \ref{figevCFM1}\\

Esse decaimento como lei de potência é devido à forma do potencial
$V^{CFM}_{esc}$.\\
Ching \cite{Ching} demonstrou analiticamente que potenciais que
possuem a forma assintótica, do tipo
\begin{eqnarray}
\label{resultCFM2} V(r)\approx
\frac{\ell(\ell+1)}{r_{*}^2}+\frac{K_{1}}{r_{*}^{\alpha}}\
log\left(\frac{r_{*}}{a}\right)\qquad \textrm{quando}\ r_{*}
\rightarrow \infty,
\end{eqnarray}
onde $K_{1}$ e $a$ são constantes, geram uma cauda que decai na
forma de $t^{-(2\ell+\alpha)}$ para valores de $\alpha<2\ell +3$. No caso do potencial
$V_{esc}^{CFM}$ a forma assintótica satisfaz a equação
(\ref{resultCFM2}) para $\alpha=3$ e $\ell\neq 0$. Neste mesmo trabalho Ching mostra também que se o potencial tem a forma
\begin{eqnarray}
\label{resultCFM2a}
 V(r)\approx
\frac{K_{1}}{r_{*}^{\alpha}}\qquad \textrm{quando}\ r_{*}
\rightarrow \infty,
\end{eqnarray} 
a cauda decai na forma de $t^{-\alpha}$ se $\alpha >2$ e real. No caso do potencial $V_{esc}^{CFM}$ a forma assintótica satisfaz a equação (\ref{resultCFM2a}) para $\alpha=3$ e $\ell=0$.

Portanto as caudas da perturbação escalar decaem como 
\begin{eqnarray}
\label{resultCFM2b}
t^{-(2\ell+3)}\qquad &\textrm{quando}& \ell \neq 0\\
\nonumber\\
t^{-3}\qquad &\textrm{quando}& \ell = 0
\end{eqnarray}
Como podemos ver na figura \ref{figevCFM1} e nas equações acima, esse decaimento depende do índice
de multipolo $\ell$. O decaimento é mais rápido para valores de
$\ell$ mais altos.

As figuras \ref{figevCFM2} e \ref{figevCFM1} mostram que o
campo escalar decai com o tempo o que nos sugere que o BN-CFM é
estável quando submetido a perturbações escalares. Essa estabilidade é preservada mesmo para os casos onde o
potencial $V_{esc}^{CFM}$ apresenta regiões negativas.\\
\\
{\bf BN-SM}\\
\\
Novamente apresentaremos apenas um exemplo para ilustrar a evolução do campo escalar. Estudaremos o caso para $h=1$.

\begin{figure}[!ht]
\begin{center}
\epsfig{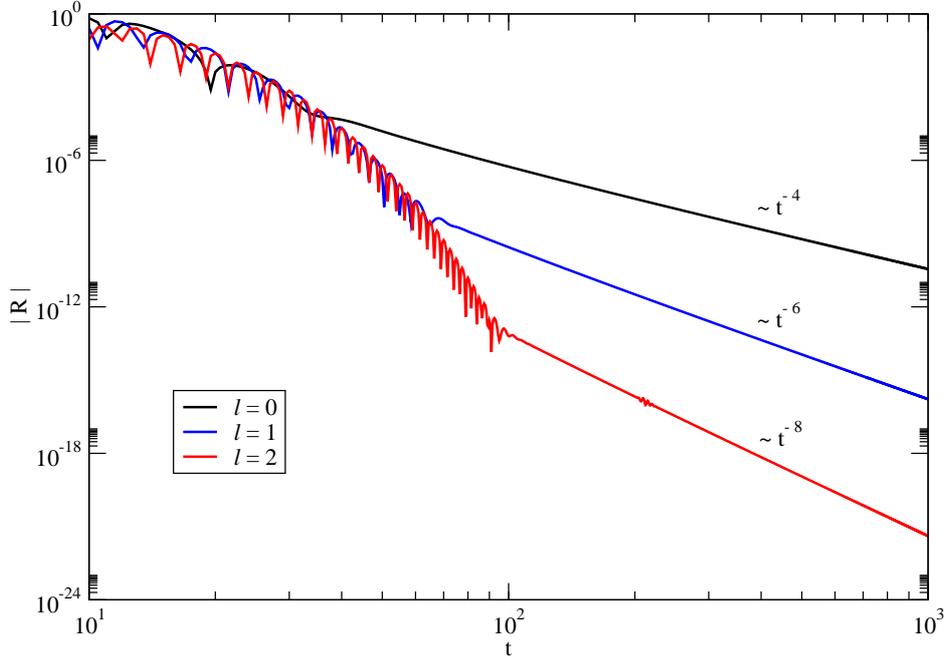}
\end{center}
\caption{Evolução temporal do campo escalar dado por $R(r,t)$ em
$r_{*}=0$ com $C=1$ e $h=1$.}
\label{figevSM2}
\end{figure}

A função de onda que representa a evolução do campo escalar não-massivo $\Psi $ no exterior do BN-CFM é dada por
\begin{eqnarray}
\label{resultSM01}
\Psi(r,t,\theta,\phi)=\frac{R(r,t)}{r}Y_{lm}(\theta,\phi)\ .
\end{eqnarray}
O comportamento de $R(r,t)$, que é solução da parte radial da
equação de Klein-Gordon, no exterior do BN-SM é apresentado na figura \ref{figevSM2} para valores de
$\ell=0,1,2$.
\newpage

Como podemos observar na figura acima, a evolução do campo escalar neste buraco negro também possui três regimes distintos. Inicialmente o campo apresenta uma fase de trasiente. Logo após temos uma fase quasi-normal, sendo atenuada exponencialmente. Após essa fase o campo apresenta um decaimento na forma de lei de potências. Igualmente para o BN-SM podemos escrever o comportamento das caudas utilizando o formalismo de Ching.

Para o BN-SM teremos $\alpha=4$ de modo que as caudas da perturbação escalar decaem como
\begin{eqnarray}
\label{resultSM2b}
t^{-(2\ell+4)}\qquad &\textrm{quando}& \ell \neq 0\\
\nonumber\\
t^{-4}\qquad &\textrm{quando}& \ell = 0
\end{eqnarray}


A figura (\ref{figevSM2}) mostra que o campo escalar decai com o tempo o que nos diz que o BN-SM para esse valore de $C$ é estável quando submetido a perturbações escalares.\\
\\
\\
\\
Os gráficos com a evolução dos campos eletromagnéticos e
gravitacionais dos BN-CFM e BN-SM ainda estão sendo feitos e serão
apresentados no artigo \cite{elcio} a ser submetido.

\newpage

\chapter{Conclusões}
{\it E no fim da jornada percebemos que novos caminhos florescem
 à nossa frente. Seguiremos }\\
 \\
O formalismo desenvolvido para o estudo de perturbações de buracos negros em
$(3+1)$ dimensões pode ser aplicado, sem grandes dificuldades técnicas, para o estudo perturbações de buracos negros em branas com {\it bulk} do tipo $AdS_{5}$.
Observamos que métricas que não satisfazem a condição $g_{tt}=-\frac{1}{g_{rr}}$, geram dificuldades no desacoplamento das equações que governam a evolução dos campos no exterior dos buracos negros, especialmente nos casos onde a paridade é $(-1)^{\ell}$, ou seja, nos casos polares.

De modo geral os buracos negros tipo-CFM e tipo-SM mostraram-se estáveis quando submetidos a perturbações escalares, mesmo quando os valores de $\beta$ e $C$ geravam regiões negativas nos potenciais efetivos. Atribuímos esse comportamento ao fato de que os poços de potenciais eram rasos quando comparados aos máximos dos potenciais.

Uma análise mais ampla sobre o conjunto de parâmetros $M$, $C$, $h$, $\ell$, $\beta$ é necessária para estimar-se melhor as condições limite dos buracos negros e assim poder realizar declarações mais precisas sobre a estabilidade da classe de buracos negros estudados.

Quanto a termodinâmica, ambos os buracos negros se mostraram em acordo com o limite superior da entropia $S_{m}$ de um corpo absorvido por eles. Isso reforça um pouco mais a universalidade desse limite uma vez que ele independe dos parâmetros do {\it bulk}.

A evolução do campo escalar no exterior do Buraco Negro-CFM e do Buraco Negro-SM apresentou comportamento semelhante na fase quasi-normal aos buracos negros em $(3+1)$ dimensões. O atenuamento exponencial das oscilações são observados. O comportamento das caudas mostraram-se semelhantes. 

Para o buraco negro tipo-CFM, as caudas decaem como uma lei de potência do tipo $t^{-(2\ell+3)}$. Apesar do potencial $V_{esc}^{CFM}$ apresentar correções quando comparado ao potencial de Schwarzschild  $V_{esc}^{Sch}$ as caudas decaem da mesma maneira.

Para o buraco negro tipo-SM, as caudas também decaem como lei de potência do tipo $t^{-(2\ell+4)}$. O comportamento das caudas dependem fortemente dos parâmetros $C$, $h$ e $\ell$.

O estudo das componentes polares das perturbações eletromagnéticas e gravitacionais, assim como novas propriedades termodinâmicas e o estabelecimento de freqüências quasi-normais para esses e outros buracos negros sobre a brana, serão alvo de investigações futuras.

\newpage

\appendix
\chapter{Apêndices}
Os apêndices deste trabalho constituem-se de, notações e covenções
adotadas ao longo do trabalho assim como alguns cálculos mais
detalhados que julgamos ser relevantes para um maior
esclarecimento dos assuntos abordados.
\section{Notações e Convenções}
\label{sec:NC}

Certos tensores, tais como, o tensor de Ricci $R_{ij}$ e o tensor
de Einstein $G_{ij}$, podem apresentar conveções de sinal
distintas se compararmos livros ou artigos diferentes. O cuidado
em sermos consistentes com uma única convenção de sinal adotada
deve ser mantido para evitarmos erros ou conclusões equivocadas. A convenção adotada neste trabalho segue abaixo.

Letras latinas minúsculas $(i,j)$ representam o espaço-tempo
4-dimensional e correm de $0\dots 3$. Letras gregas minúsculas
$\mu,\nu$ representam o espaço-tempo 5-D e correm de $0\dots
4$, salvo quando especificado outra notação.

Unidades geométricas são assumidas, onde temos $c=\hbar=k=1$. A assinatura das métricas utilizadas será $(-,+,+,+)$.

A notação de Einstein é assumida nesse trabalho, ou seja, para
índices repetidos que significam soma sobre todo intervalo, a somatória é omitida.
\begin{eqnarray}
A^{j}&=&\sum_{i=0}^{3}\ R_{i}R^{ij}\nonumber\\
\nonumber\\
A^{j}&=&R_{0}R^{0j}+R_{1}R^{1j}+R_{2}R^{2j}+R_{3}R^{3j}=R_{i}R^{ij}
\end{eqnarray}
A conexão $\Gamma^{\ b}_{ij}$ é dada por
\begin{eqnarray}
\label{apI1}
\Gamma^{\ b}_{ij}= \frac{1}{2}g^{ba}\left(\frac{\partial g_{ia}}{\partial
x^{j}}+\frac{\partial g_{ja}}{\partial x^{i}}-\frac{\partial g_{ij}}{\partial
x^{a}}\right).
\end{eqnarray}
A derivada covariante de um tensor $V_{ij}$ que é representada por
$(\ ;\ )$ é definida como
\begin{eqnarray}
\label{apI2}
 V_{ij;a}=V_{ij,a}-\Gamma^{\ b}_{ia}\ V_{bj}-\Gamma^{\ b}_{ja}\ V_{ib}\ .
\end{eqnarray}
A vírgula $(\ ,\ )$ denota uma derivada ordinária sobre o tensor
\begin{eqnarray}
\label{apI3}
 V_{ij,a}=\frac{\partial V_{ij}}{\partial x^{a}}\ .
\end{eqnarray}
O tensor de Ricci $R_{ij}$ adota a seguinte convenção de sinal
\begin{eqnarray}
\label{apI4} R_{ij}= \Gamma^{a}_{ia,j}\ -\Gamma^{a}_{ij,a}+\Gamma^{b}_{ia}\
\Gamma^{a}_{jb} - \Gamma^{b}_{ij}\ \Gamma^{a}_{ab}\ .
\end{eqnarray}
\newpage
\section{Decomposição tensorial de $h_{ij}$}
\label{sec:Dectensorial}
Um estudo do momento angular sobre um espaço-tempo (3+1) dimensões
esfericamente simétrico pode ser feito, se realizarmos rotações
sobre a variedade 2-dimensional formada por $r=const$ e $t=const$.
Tal estudo é necessário para sabermos como se comporta a parte
angular de nosso problema. Sobre tais rotações ao redor da origem
a componentes de $h_{ij}$ se transformam de maneira particular.
Nesse caso ao rotacionarmos nosso referencial $S$ até um
referencial $S'$ os versores dos sistemas se transformarão como
\begin{eqnarray}
\label{Apdecom1}
 \hat{r}\longrightarrow \hat{r}',\quad \hat{t}\longrightarrow \hat{t}', \quad
\hat{\theta}\longrightarrow \hat{\theta}', \quad
\hat{\phi}\longrightarrow \hat{\phi}'.
\end{eqnarray}
Como $r$ e $t$ são constantes $ \hat{r}= \hat{r}'$ e $
\hat{t}=\hat{t}'$. Conseqüentemente
\begin{eqnarray}
\label{Apdecom2}
 \frac{\partial r}{\partial r'}=1, \qquad
\frac{\partial t}{\partial t'}=1.
\end{eqnarray}
Um tensor de ordem 2 quando rotacionado se transforma como
\begin{eqnarray}
\label{Apdecom3}
h_{ij}=\Lambda_{i}^{\phantom{i}{a}}\Lambda_{j}^{\phantom{j}{b}}h_{ab}',\nonumber\\
\nonumber\\
h_{ij}=\frac{\partial x'^{a}}{\partial x^{i}}\frac{\partial
x'^{b}}{\partial x^{j}}\ h_{ab}',
\end{eqnarray}
Adotaremos a seguinte notação para as coordenadas
\begin{eqnarray}
\label{Apdecom4} x^{0}=t, \quad x^{1}=r, \quad x^{2}=\theta, \quad
x^{3}=\phi.
\end{eqnarray}
Desta forma podemos calcular como cada componente de $h_{ij}$ se
transforma.\\
As componentes $h_{00}$, $h_{11}$ e $h_{01}$ se transformam como
escalares pois satisfazem a seguinte condição
\begin{eqnarray}
\Phi(r,t,\theta,\phi)=\Phi'(r',t',\theta',\phi')
\end{eqnarray}
Tomemos $h_{00}$ como exemplo
\begin{eqnarray}
\label{Apdecom5} h_{00}=\frac{\partial x'^{a}}{\partial
x^{0}}\frac{\partial x'^{b}}{\partial x^{0}}\
h_{ab}'&=&\frac{\partial x'^{0}}{\partial x^{0}}\frac{\partial
x'^{0}}{\partial x^{0}}\ h_{00}'\nonumber\\
\nonumber\\
h_{00}&=&h_{00}'
\end{eqnarray}
Isso se deve ao fato de que as coordenadas $x'^{2}$ e $x'^{3}$
dependem apenas de $x^{2}$ e $x^{3}$ uma vez que $r$ e $t$ são
mantidas constantes.\\
As componentes ($h_{02},h_{03}$) e ($h_{12},h_{13}$) se
transformam como 2 vetores, ou seja, satisfazem a condição
\begin{eqnarray}
V_{i}=\frac{\partial x'^{a}}{\partial x^{i}}\ V_{a}.
\end{eqnarray}
Tomemos ($h_{02},h_{03}$) como exemplo de um vetor com componentes $(V_{2},V_{3})$.\\
A componente $h_{02}$ será
\begin{eqnarray}
\label{Apdecom6}
h_{02}&=&\frac{\partial x'^{a}}{\partial
x^{0}}\frac{\partial x'^{b}}{\partial x^{2}}\ h_{ab}',\nonumber\\
\nonumber\\
h_{02}&=& \frac{\partial x'^{0}}{\partial x^{0}}\frac{\partial
x'^{2}}{\partial x^{2}}\ h_{02}'+\frac{\partial x'^{0}}{\partial
x^{0}}\frac{\partial
x'^{3}}{\partial x^{2}}\ h_{03}', \nonumber\\
\nonumber\\
h_{02}&=&\frac{\partial x'^{2}}{\partial
x^{2}}h_{02}'+\frac{\partial x'^{3}}{\partial x^{2}}h_{03}'.
\end{eqnarray}
A componente $h_{03}$ será
\begin{eqnarray}
\label{Apdecom7}
h_{03}&=&\frac{\partial x'^{a}}{\partial
x^{0}}\frac{\partial
x'^{b}}{\partial x^{3}}\ h_{ab}',\nonumber\\
\nonumber\\
h_{03}&=&\frac{\partial x'^{0}}{\partial x^{0}}\frac{\partial
x'^{2}}{\partial x^{3}}\ h_{02}'+\frac{\partial x'^{0}}{\partial
x^{0}}\frac{\partial
x'^{3}}{\partial x^{3}}\ h_{03}'\nonumber,\\
\nonumber\\
h_{03}&=&\frac{\partial x'^{2}}{\partial
x^{3}}h_{02}'+\frac{\partial x'^{3}}{\partial x^{3}}h_{03}'.
\end{eqnarray}
Podemos observar que $h_{02}$ corresponde à componente $V_{2}$ e
$h_{3}$ à componente $V_{3}$ do nosso vetor.
\begin{displaymath}
\left(\begin{array}{c}
V_{2}\\
V_{3}
\end{array}\right)=\
\left(\begin{array}{cc}
\alpha & \beta\\
\gamma & \sigma
\end{array}\right)
\ \left(\begin{array}{c}
V'_{2}\\
V'_{3}
\end{array}\right)
\end{displaymath}
Portanto se transformam como um vetor sobre uma rotação.

As componentes ($h_{22},h_{33},h_{23},h_{32}$) se tranformam como
um tensor de $2^{a}$ ordem $2\times 2$, ou seja satisfazem a
seguinte condição
\begin{eqnarray}
h_{ij}=\frac{\partial x'^{a}}{\partial x^{i}}\frac{\partial
x'^{b}}{\partial x^{j}}\ h_{ab}',
\end{eqnarray}
Tomemos como exemplo a componente $h_{22}$ do nosso tensor
\begin{eqnarray}
\label{Apdecom8}
h_{22}&=&\frac{\partial x'^{a}}{\partial
x^{2}}\frac{\partial x'^{b}}{\partial x^{2}}\ h_{ab}'\nonumber\\
\nonumber\\
h_{22}&=& \frac{\partial x'^{2}}{\partial x^{2}}\frac{\partial
x'^{2}}{\partial x^{2}}\ h_{22}'+\frac{\partial x'^{2}}{\partial
x^{2}}\frac{\partial x'^{3}}{\partial x^{2}}\
h_{23}'+\frac{\partial x'^{3}}{\partial x^{2}}\frac{\partial
x'^{2}}{\partial x^{2}}\ h_{32}'+\nonumber\\
\nonumber\\
&+&\frac{\partial x'^{3}}{\partial x^{2}}\frac{\partial
x'^{3}}{\partial x^{2}}\ h_{33}'
\end{eqnarray}
Portanto $h_{22}$ se transforma como uma componente de um tensor
de $2^{a}$ ordem.

Agora já sabemos como cada componente do tensor $h_{ij}$ se
transforma.

Assim podemos escrever uma matriz $h$ como
\begin{displaymath}
h=\left(\begin{array}{cccc}
E&E&V_{2}&V_{3}\\
E&E&V_{2}&V_{3}\\
V_{2}&V_{2}&T_{22}&T_{23}\\
V_{3}&V_{3}&T_{32}&T_{33}
\end{array}\right)\ .
\end{displaymath}
Desta forma podemos separar $h_{ij}$ de modo geral como a
multiplicação de duas funções, cada uma dependendo de $r,t$ e
$\theta, \phi$ respectivamente
\begin{eqnarray}
h_{ij}=\sum_{l=0}^{\infty}\ \sum_{m=-l}^{l}\ \sum_{n=1}^{10}
C_{lm}^{n}(t,r)(Y_{lm}^{n})_{ij}(\theta,\phi)\ .
\end{eqnarray}
\newpage

\section{Construção dos Harmônicos Esféricos Tensoriais}
\label{sec:ConsHET}
A construção dos harmônicos esféricos tensoriais pode ser
realizada se utilizarmos os harmônicos esféricos escalares
$Y^{lm}(\theta,\phi)$ como base e operarmos com gradientes e
``pseudogradientes" sobre ele de modo a obtermos vetores  e
tensores.

Nesta construção obteremos objetos com paridade distintas
$(-1)^{\ell}$ e $(-1)^{\ell+1}$. Esta distinção nos diz que
teremos 2 tipos de ondas geradas por nossa perturbação.

A parte escalar do nosso tensor é dada obviamente por
\begin{eqnarray}
\label{Apcon1} \phi^{lm}=const\ Y^{lm}(\theta,\phi)\ .
\end{eqnarray}
Este termos pertence a uma onda de paridade $(-1)^{\ell}$ e
momento angular $\ell$ cuja a projeção sobre o eixo $z$ é $m$.

A construção dos vetores é dada pela operação do gradiente e do
pseudogradiente em $Y_{lm}$. Desta operação obteremos dois vetores
com paridades opostas
\begin{eqnarray}
\label{Apcon2} \psi^{lm}_{\phantom{lm}{,i}}&=&const\
\frac{\partial}{\partial
 x^{i}}Y^{lm},\qquad \ \ \ paridade (-1)^{\ell};\\
\nonumber\\
\phi^{lm}_{\phantom{lm}{,i}}&=&const\
\epsilon_{i}^{\phantom{i}{a}}\frac{\partial}{\partial
 x^{a}}Y^{lm}, \qquad paridade (-1)^{\ell+1}.
\end{eqnarray}
Aqui os índices $i$ e $a$ correm sobre os valores 2 e 3, quando
$x^{2}=\theta$ e $x^{3}=\phi$; e $\epsilon_{i}^{\phantom{i}{a}}$ é
um tensor totalmente anti-simétrico que representa as quantidades
$\epsilon_{2}^{\phantom{2}{2}}=\epsilon_{3}^{\phantom{3}{3}}=0$;
$\epsilon_{2}^{\phantom{2}{3}}=-1/sen\theta$ e
$\epsilon_{3}^{\phantom{3}{2}}=sen\theta$.

Por fim construiremos três tensores fundamentais. Eles são obtidos quando
utilizamos a derivação covariante sobre $Y^{lm}$, quando aplicamos
a métrica da esfera $\gamma_{ij}=g_{ij}/r^2$ sobre $Y^{lm}$ e
quando operamos com o pseudogradiente sobre o tensor obtido na
primeira operação. São eles
\begin{eqnarray}
\psi_{\phantom{lm}{ia}}^{lm}&=&const\ Y^{lm}_{\phantom{lm}{;ia}}\ ,
\qquad \qquad \qquad \ \ \ \ \ \ \ paridade (-1)^{\ell}\\
\nonumber\\
\phi^{lm}_{\phantom{lm}{ia}}&=&const\ \gamma_{ia}\ Y^{lm}\ ,\qquad
\qquad \qquad \ \ \ \ paridade (-1)^{\ell}\\
\nonumber\\
\chi_{\phantom{lm}{ia}}^{lm}&=&\frac{1}{2}\ const\
[\epsilon_{i}^{\phantom{i}{b}}
\psi^{lm}_{\phantom{lm}{ba}}+\epsilon_{a}^{\phantom{a}{b}}
\psi^{lm}_{\phantom{lm}{bi}}]\ .\qquad paridade (-1)^{\ell+1}\
\end{eqnarray}
onde as componentes da métrica sobre a esfera são dadas por
$\gamma_{22}=1$, $\gamma_{23}=\gamma_{32}=0$ e
$\gamma_{33}=sen\theta$.
Desta forma fica completa a construção dos Harmônicos Esféricos Tensoriais.

\section{Comentários a respeito do comportamento da coordenada tartaruga $r_{*}$}
\label{sec:coordtartaruga}
Será de interesse ao nosso estudo sobre as propriedades da evolução de campos, alguns comentários sobre o comportamento assintótico da coordenada tartaruga $r_{*}$.

Tomemos uma métrica esfericamente simétrica dada por
\begin{eqnarray}
\label{Aptar01}
ds^2=-A(r)dt^2+B(r)dr^2+r^2(d\theta^2+sen\theta^2d\phi^2).
\end{eqnarray}
A existência de um horizonte de eventos nesse espaço-tempo implica em um zero da função $A(r)$. Seja $r_{h}$ um zero de $A(r)$.\\
Definindo a função $h(r)$ como
\begin{eqnarray}
\label{Aptar02}
h(r)=\sqrt{\frac{A(r)}{B(r)}}\ ,
\end{eqnarray}
percebemos que o ponto $r=r_{h}$ também é um zero da função $h(r)$.

A coordenada tartaruga $r_{*}$ pode ser definida como
\begin{eqnarray}
\label{Aptar03}
r_{*}=\int \frac{1}{h(r)}dr\ .
\end{eqnarray}
De modo geral , a função $r_{*}(r)$ é monotonicamente crescente. Se calcularmos a derivada de $r_{*}$ veremos que
\begin{eqnarray}
\label{Aptar05}
\frac{dr_{*}(r)}{dr}= \frac{d}{dr}\left[\int \frac{1}{h(r)}dr\right]=\frac{1}{h(r)}\ .
\end{eqnarray}
Analisando as funções $A(r)$ e $B(r)$ vemos que, para o intervalo de interesse $(\ ]r_{h},\infty[\ )$, temos
\begin{eqnarray}
\label{Aptar06}
A(r)>0,\qquad B(r)>0 \qquad \Longrightarrow  h(r)>0\ .
\end{eqnarray}
Desta forma a derivada de $r_{*}$ é crescente
\begin{eqnarray}
\label{Aptar07}
\frac{dr_{*}(r)}{dr}>0\ .
\end{eqnarray}
O comportamento de $r_{*}$ próximo ao horizonte de eventos pode ser estudado se soubermos de que tipo são os zeros da função $h(r)$ em $r=r_{h}$.

Vamos supor que $h(r_{h})=0$ é um zero simples, ou seja, que $h(r)$ pode ser escrita como
\begin{eqnarray}
\label{Aptar08}
h(r)=(r-r_{h})P(r)\ ,
\end{eqnarray}
onde $P(r)$ é um polinômio de $r$ e respeite a condição $P(r_{h})\neq 0$.
Sob essas condições  temos
\begin{eqnarray}
\label{Aptar09}
r_{*}(r)=\int \frac{dr}{(r-r_{h})P(r)}\ .
\end{eqnarray}
Se estivermos próximos o suficiente do horizonte de eventos podemos escrever a equação (\ref{Aptar09}) como
\begin{eqnarray}
\label{Aptar10}
r_{*}(r)=\frac{1}{P(r_{h})}\int \frac{dr}{(r-r_{h})}=\frac{1}{P(r_{h})}ln(r-r_{h})\ .
\end{eqnarray}
Lembrando que a gravidade superficial em $r=r_{h}$ é dada por
\begin{eqnarray}
\label{Aptar11}
\kappa_{h}=\frac{1}{2}\left. \frac{dh(r)}{dr}\right|_{r=r_{h}}=\frac{P(r_{h})}{2}\ ,
\end{eqnarray}
podemos reescrever a equação (\ref{Aptar10}) próximo do horizonte de eventos como
\begin{eqnarray}
\label{Aptar12}
r_{*}(r)=\frac{1}{2\kappa_{h}}ln(r-r_{h})\ .
\end{eqnarray}
Neste caso a função $r_{*}$ diverge logaritmicamente quando $r\rightarrow r_{h}$ sendo
\begin{eqnarray}
\label{Aptar13}
\lim_{r\rightarrow r_{h}}r_{*}(r)\rightarrow -\infty
\end{eqnarray}
Neste limite é possível inverter a função $r_{*}(r)$ de modo que a forma analítica de $r(r_{*})$ será
\begin{eqnarray}
\label{Aptar14}
r=r_{h}+e^{2\kappa_{h}r_{*}}\ .
\end{eqnarray}
Agora se supusermos que $h(r_{h})=0$ é um zero duplo, ou seja, que $h(r)$ pode ser escrita como
\begin{eqnarray}
\label{Aptar15}
h(r)=(r-r_{h})^2Q(r)\ ,
\end{eqnarray}
onde $Q(r_{h})\neq 0$, teremos
\begin{eqnarray}
\label{Aptar16}
\kappa_{h}=0,
\end{eqnarray}
\begin{eqnarray}
\label{Aptar16b}
r_{*}(r)=\frac{1}{Q(r_{h})}\int \frac{dr}{(r-r_{h})}\propto -\frac{1}{r-r_{h}}.
\end{eqnarray}
Desta forma, próximo ao horizonte teremos
\begin{eqnarray}
\label{Aptar17}
r_{*}=-\frac{1}{2Q(r_{h})(r-r_{h})}\ .
\end{eqnarray}
Neste caso a função $r_{*}(r)$ diverge como uma lei de potência quando $r\rightarrow r_{h}$.\\
Invertendo $r_{*}(r)$ teremos
\begin{eqnarray}
\label{Aptar18}
r(r_{*})=r_{h}-\frac{1}{2Q(r_{h})r_{*}}\ .
\end{eqnarray}

\section{Cálculo do desacoplamento das equações $\delta R_{12}$, $\delta R_{10}$ e $\delta R_{13}$.}
\label{sec:desacoplamento}
As equações $\delta R_{12}$, $\delta R_{10}$ e $\delta R_{13}$ podem ser obtidas em \cite{Chandra} e são dadas por
\begin{eqnarray}
\label{Apdesac01}
R_{10}=\frac{e^{-2\psi-\mu_{2}-\mu_{3}}}{2}\left[\left(e^{3\psi-\nu-\mu_{2}+\mu_{3}}Q_{20}\right)_{,2}+\left(e^{3\psi-\nu-\mu_{3}+\mu_{2}}Q_{30}\right)_{,3}\right]=0;\\
\nonumber\\
R_{12}=\frac{e^{-2\psi-\nu-\mu_{3}}}{2}\left[\left(e^{3\psi+\nu-\mu_{2}-\mu_{3}}Q_{32}\right)_{,3}-\left(e^{3\psi-\nu+\mu_{3}-\mu_{2}}Q_{02}\right)_{,0}\right]=0;\\
\nonumber\\
R_{13}=\frac{e^{-2\psi-\nu-\mu_{2}}}{2}\left[\left(e^{3\psi+\nu-\mu_{3}-\mu_{2}}Q_{23}\right)_{,2}-\left(e^{3\psi-\nu-\mu_{3}+\mu_{2}}Q_{03}\right)_{,0}\right]=0.
\end{eqnarray}
onde $R_{ij}$ corresponde a $\delta R_{ij}$.

O fator multiplicativo de cada equação pode ser ignorado uma vez que todas as equações são iguais a zero.

Assim se fizermos a seguinte operação
\begin{eqnarray}
\label{Apdesac02}
\delta R_{10,0}=\delta R_{12,2}+\delta R_{13,3}
\end{eqnarray}
teremos uma identidade de modo que a equação $\delta R_{10}$ pode ser escrita como uma combinação da outras duas equações.

Tomando então as derivadas apropriadas teremos
\begin{eqnarray}
\label{Apdesac03}
\left(e^{3\psi-\nu-\mu_{2}+\mu_{3}}Q_{20}\right)_{,2,0}&+&\left(e^{3\psi-\nu-\mu_{3}+\mu_{2}}Q_{30}\right)_{,3,0}=\nonumber\\
\nonumber\\
\left(e^{3\psi+\nu-\mu_{2}-\mu_{3}}Q_{32}\right)_{,3,2}&-&\left(e^{3\psi-\nu+\mu_{3}-\mu_{2}}Q_{02}\right)_{,0,2}+\nonumber\\
\nonumber\\
+\left(e^{3\psi+\nu-\mu_{3}-\mu_{2}}Q_{23}\right)_{,2,3}&-&\left(e^{3\psi-\nu-\mu_{3}+\mu_{2}}Q_{03}\right)_{,0,3}
\end{eqnarray}
Lembrando que
\begin{eqnarray}
\label{Apdesac04}
Q_{AB}=q_{A,B}-q_{B,A}\ ,\qquad Q_{A0}=q_{A,0}-\omega_{,A}\ ,\qquad Q_{0A}=\omega_{,A}-q_{A,0}\ ,
\end{eqnarray}
e aplicando essas condições  na equação (\ref{Apdesac03}) teremos
\begin{eqnarray}
\label{Apdesac05}
\left(e^{3\psi-\nu-\mu_{2}+\mu_{3}}Q_{20}\right)_{,2,0}&+&\left(e^{3\psi-\nu-\mu_{3}+\mu_{2}}Q_{30}\right)_{,3,0}=\nonumber\\
\nonumber\\
=\left(e^{3\psi+\nu-\mu_{2}-\mu_{3}}Q_{32}\right)_{,3,2}&+&\left(e^{3\psi-\nu+\mu_{3}-\mu_{2}}Q_{20}\right)_{,2,0}-\nonumber\\
\nonumber\\
-\left(e^{3\psi+\nu-\mu_{3}-\mu_{2}}Q_{32}\right)_{,3,3}&+&\left(e^{3\psi-\nu-\mu_{3}+\mu_{2}}Q_{30}\right)_{,3,0}\nonumber\\
\nonumber\\
0&=&0.
\end{eqnarray}


\end{document}